\UseRawInputEncoding
\documentclass[prb,preprint,showpacs,preprintnumbers,amsmath,amssymb]{revtex4}

\usepackage{graphicx}
\usepackage{dcolumn}
\usepackage{bm}
\usepackage{float}

\begin{document}


\title{A closer look at how symmetry constraints and the spin-orbit coupling shape the electronic structure of Bi(111)}

\author{Marisol Alc{\'a}ntara Ortigoza}
\email{Marisol.AlcantaraOrtigoza@ucf.edu}




\author{Talat S. Rahman}
\email{Talat.Rahman@ucf.edu}

\affiliation{Department of Physics, Tuskegee University\\
Tuskegee Institute, Alabama 36088, USA
}%

\affiliation{Department of Physics, University of Central Florida\\
Orlando, Florida 32816, USA
}%



\date{\today}

\begin{abstract}

Fully relativistic density-functional-theory calculations of Bi(111) thin films are analyzed to revisit  interpretations made about their electronic structure and the corresponding extrapolations to explain that of macroscopic samples. Our calculated band structure of a 39-bilayer film ($\sim$~15~nm) reproduces satisfactorily three independent  measurements obtained for macroscopic samples and shows that $\sim$9-nm films are thick enough to describe the band structure of Bi(111) within experimental resolution. Our results indicate that the two split surface-state metallic branches along the $\overline{\Gamma M}$ direction do not overlap with the bulk band at the zone boundary but lie within the A7-distortion-induced conduction-valence band gap. We  show that (1) neither the existence of the metallic surface states nor their observed splitting is related to  inversion \emph{asymmetry}  and (2) the spin texture observed in such states cannot be attached to the lifting of the Kramers degeneracy. Thus, such splitting is not of the Rashba-type. We hence provide grounds to identify the large splitting of the metallic branches as a $m_j=\pm1/2$--$m_j=\pm3/2$ splitting. We also propose that the  spin texture observed for the metallic branches may  only occur because the essentially unaltered strong covalent bonds retained by Bi(111) surface atoms cannot afford magnetic polarization.  Moreover, we emphasize that  degeneracy at the $M$-point of the SBZ of Bi(111) ---implied by the translational symmetry of the surface and apparently not satisfied under the assumption that the two split metallic branches are two singlets of opposite magnetic  quantum number--- is  satisfied irrespectively of the presence of inversion symmetry centers, even if not detectable by measurements. We also show that the  magnetic-moment discontinuity at $M$ ---also implied by the same assumption--- does not exist, which also explain why the measured spin-polarization of the metallic branches vanishes near $M$.   Furthermore, we investigate the Rashba effect on the band structure of Bi(111) due to three different types of structural/electronic perturbations to reveal the actual lifting of the Kramers degeneracy. We discuss that depending on the magnitude of the perturbation imposed on a film, the magnitude of the splitting and the localization of the Rashba-split states may change dramatically.

\end{abstract}

\maketitle

\section{\label{sec:intro}introduction}

\subsection{\label{subsec:introinterest} The renewed interest in Bi}

The singular properties and applications of Bismuth sprouting since 1911 make it one of the most remarkable and enlightening elements of the periodic table: It is the heaviest non-radioactive element. It has the lowest thermal conductivity after mercury and is the element with largest diamagnetism. Bi allowed an "early" discovery of the Seebeck, the de Hass-van Alphen, the Shubnikov-de Haas, and the Nernst effects, all of which are inherently present in all metals but were more challenging to observe. Bismuth was in fact the first metal whose Fermi surface was experimentally identified~\cite{shoenberg} and provided the basis to determine that of any metal. Bi also set the grounds for the still ongoing rational search and optimization of thermoelectric materials. Furthermore, Bi and Bi$_{1-x}$Sb$_x$ ($x<0.06$) were among the first materials for which the  excitonic-insulator phase was considered and searched.~\cite{jerome1967,brandt1972,miura1982} Bulk Bi was also the first non-superconducting  material (at least in the ordered phase) that was found to be superconductor for structurally bulk-like nanoparticles.~\cite{weitzel1991} Yet today, Bi remains  a prime subject of intense investigation just as more than 100 years ago.

The interest in Bi has revived at various research fronts. 
Of particular interest to the present investigation is that its surfaces are declared to be much better conductors than the bulk.~\cite{hengsberger,ast1prl} Furthermore, and  uncoincidentally perhaps,~\cite{weitzel1991,hofmann} Bi bicrystals and nanoparticles display a superconducting phase ($T_c = 5K$) not observed in ordered bulk Bi down to 50 mK.~\cite{weitzel1991,vossloh1998,muntyanu2008,mingliang2008}
Moreover, because the spin-orbit coupling is stronger for heavier  elements, lifting of the Kramers degeneracy  has been invoked recently to explain the two metallic surface-state branches of Bi(111), by assuming and inherent lack of inversion symmetry because of the Bi-vacuum interface.~\cite{chulkov2004,chulkov2006,chulkov2007,chulkov2008}

\subsection{\label{subsec:introbulk} Current state of knowledge of the geometric and electronic structure of bulk Bi}

The remarkable electronic properties of Bi are tightly related to its  crystal structure,
the so-called A7 or $\alpha$As structure, which has the space group $R\overline{3}m$.
The primitive cell of the A7 structure is a rhombohedron that contains two atoms. The A7 structure largely resembles a simple cubic structure (SC) with a slight distortion along the body diagonal: the so-called A7 distortion. In bulk Bi, each atom
has three equidistant nearest-neighbor atoms and three equidistant next-nearest neighbors,
slightly further away. This structure thus renders A7-materials as composed by bilayers (BLs) stacked perpendicular the c-axis or trigonal axis (see Fig.~\ref{fig:bulkstrhex}).
The three
nearest neighbors of a Bi atom lie within the same bilayer (BL), whereas its three next-nearest neighbors are located in the adjacent BL.
As a result, every atom in an A7-material has
three equivalent "short" and "strong" covalent-like bonds and three equivalent "long" and "weak"  metallic-like bonds.
A detailed description of the above rhombohedral and hexagonal representation can be found elsewhere.~\cite{bi111ph}

Why Group--V elements arrange themselves in the A7 structure was not understood until
Jones proposed that the A7-distortion creates a band gap almost everywhere in the Brillouin zone (BZ)
that stabilizes the structure. We shall refer to this \emph{partial} gap as the A7-distortion \emph{pseudogap}.~\cite{peierls}  The so-called Jones-Peierls mechanism explaining the A7 structure has been amply supported by
measurements and calculations (see Ref.~\onlinecite{bi111ph}).
In fact, it will be important for the later discussion
to recall that  the A7-distortion \emph{pseudogap}
makes all A7 materials semimetals and exists independently of the strength of the spin-orbit coupling (SOC)  and its effects on the electronic band structure.

\subsection{\label{subsec:introelectronsurface} Current state of knowledge of the geometric and electronic structure of Bi(111)}

As a result of the A7 distortion, bulk Bi can be considered as formed by puckered bilayers perpendicular to the trigonal axis (see Figures~\ref{fig:bulkstrhex} and ~\ref{fig:surfstr}).
The covalent \emph{intra-bilayer} bonds within each bilayer are much stronger than the metallic-like \emph{inter-bilayer} bonds. This
explains why Bi crystals easily cleave along the (111) plane, giving rise to the rhombohedral Bi(111) surface --- also known as the pseudocubic-(111) or the hexagonal-(0001) surface (In the following, however, we confine ourselves to the rhombohedral notation only).

The first investigation of the low-index Bi surfaces  was performed by Jona in 1967 via low-energy electron spectroscopy.~\cite{jona}
He found that (1) cleaving  Bi perpendicular to the trigonal direction  can be achieved easily and (2)  the resulting Bi(111) surface does not reconstruct.~\cite{jona,notejona,sun2006}
Jona \emph{et al.} also reported that Bi(111) is quite inert to oxygen adsorption ---unlike the other two low-index surfaces. He ascribed the easy cleavage, the bulk-like surface termination, and the inertness of Bi(111)  to the fact that cleavage of Bi along the trigonal direction  breaks only the long and weak metallic bonds rather than the  short and strong covalent bonds, thus yielding a surface terminated in a bilayer and with no true dangling bonds.~\cite{jona} These assertions were later confirmed by comparing low-energy electron diffraction (LEED) intensities against dynamical LEED calculations~\cite{chulkov2005} and will be relevant to argue later that macroscopic Bi single-crystals may preserve inversion symmetry with respect to the trigonal axis when cleaved perpendicular to it.

In order to understand  the electronic structure of Bi(111), it is useful to define and distinguish between the A7-distortion pseudogap and the spin-orbit pseudogaps. The A7-distortion pseudogap is  around the Fermi level ($E_F$) and separated the conduction and valence bands. Its existence (though not its magnitude) is  independent of the SOC, inasmuch as it is the A7-pseudogap what makes all A7 materials semimetals. On the other hand, Bi(111) has also SOC pseudogaps, which are known  to lie inside the valence band (unlike those of arsenic,~\cite{dresselhaus} for example) and to be caused solely by the SOC, as suggested  by earlier Density-Functional Theory (DFT) calculations of bulk Bi (see Fig.~\ref{fig:bzelecstr}).~\cite{gonze1990}

Computational investigations on the electronic band structure of the (111) surface of Group-V semimetals started as early as 1976 with the  work of Anishchik \emph{et al}.~\cite{anishchik}
Their  calculations, using a Green-function formalism in the tight-binding representation,~\cite{anishchik}  apparently yielded  pseudogaps within the (111)-projected bulk \emph{p}-band of  Group-V semimetals and  surface \emph{p}-states inside of those pseudogaps. The latter, however, were not associated with the SOC and were not necessarily metallic or outside the valance \emph{p}-band. It is  important to note also that, in this investigation the appearance of surface states strictly demanded the existence of some surface relaxation.~\cite{anishchik} A similar argument was invoked more recently by Ast and H{\"o}chst.~\cite{hofmann} However, we shall see that the relaxation of the surface neither implies nor affects significantly the metallic surface states. In fact, the next and more insightful finding around the band structure of Bi(111)  came not from the surface itself but from one of the earliest fully relativistic pseudopotential calculations of the electronic band structure of bulk Bi by Golin.~\cite{golin} These calculations did not tackle any implications upon the surfaces' band structure whatsoever. Yet Golin's calculations \emph{per se} hinted that, in addition to the A7-distortion pseudogap around $E_F$, the (111)-projected band-structure of bulk Bi might have other "pseudogaps" around the zone center (the $\Gamma$ point) of the surface Brilloiun zone (see Fig.~\ref{fig:surfstr}(c))  because of the spin-orbit splitting of the valence \emph{p}-states along the $\overline{\Gamma T}$ line of the bulk BZ (see Fig.~\ref{fig:bzelecstr}).

The spin-orbit pseudogaps predicted by Golin's calculations and their implications on the existence of surface states inside them was the focus of several surface photoemission experiments.
Jezequel \emph{et al}.~\cite{jezequel}  were first to observe  surface states on Bi(111) via angle-resolved photoemission spectroscopy (ARPES). Moreover, by applying the same computational method of Anishchik but explicitly  including the spin-orbit coupling in the Hamiltonian, they showed that the energy and wave-vector range of the calculated spin-orbit pseudogaps  were consistent with the observations (except for a measured surface state at 3~eV) and  resolve that these states have indeed a \emph{p}-character.
Until now, however, to our knowledge, the projected  band structure of bulk Bi has not been calculated from first principles
to corroborate the existence and position of such gaps and  compare with these ARPES experiments.

Soon after the work of Jezequel \emph{et al.}, Patthey \emph{et al}.~\cite{patthey}  detected that Bi(111) has an enhanced metallicity with respect to  bulk Bi.  Later photoemission experiments aimed  determining the band-structure of Bi(111) with better accuracy.~\cite{tanaka,thomas} However, it was only until Hengsberger \emph{et al}.\cite{hengsberger} focused on probing energy scales of only few hundred meV below the Fermi energy, that high-resolution photoemission studies at 12 and 300~K made patent the existence of the two metallic branches of surface states from the zone boundary (the $M$ point) to the midpoint of the $\overline{\Gamma M}$ line,  spanning regions  closer to the Fermi level than those previously explored (though somehow predicted by the preliminary calculations of Jezequel~\cite{jezequel}).
An important difference between the measurements of Hengsberger \emph{et al.} and earlier ones is that the latter  focused on finding surface states within the \emph{spin-orbit pseudogaps}, whereas Hengsberger's \emph{et al.} scanned for   surface states  in the \emph{A7-distortion pseudogap}.
The two-dimensional Fermi surface near $\Gamma$  was  later investigated by Ast and H\"ochst.~\cite{ast1prl,ast2}  Thus, their results together with those of Hengsberger \emph{et al}. resolved the band structure of Bi(111) all along the $\overline{\Gamma M}$ direction of the surface Brillouin zone (Fig.~\ref{fig:bzelecstr}).

\subsection{\label{subsec:intrometallicbranches} Current state of knowledge about the two surface-state metallic branches of  Bi(111)}

Some doubts have been cast upon the reliability of Hengsberger's \emph{et al}. measurements close to  $M$.~\cite{hofmann} The reason is that  tight-binding-approximation calculations of the bulk-projected bands (see e.g. Ref.~\onlinecite{chulkov2004,liu1995}) suggest that the energy range within which Hengsberger \emph{et al}. found the metallic surface states overlaps with the bulk band well before reaching the $M$-point. If so, then the experiment could not have allowed surface states to be distinguished from bulk states around the zone edge ($M$).
Nevertheless, independent ARPES measurements around the $M$-point for a macroscopic sample by Ast and H\"ochst  confirmed the findings of Hengsberger \emph{et al}.~\cite{AstMpoint2003}
Remarkably, despite the presumed constraints on the extent of the two metallic branches around the $M$-point implied by tight-binding calculations of the bulk band structure, a fitting of Lorenzian line shapes to the  energy distribution of the ARPES intensity
---considering the Fermi function cutoff and removing a Shirley background--- confirm that the emission intensity of the two metallic surface-state branches is weak but still detectable around the zone edge and display a gap of $\sim$~37-meV at the $M$-point.~\cite{AstMpoint2003}

Further attention has been drawn to the two surface-state metallic branches of the Bi(111) surface of macroscopic samples and thin films for over a decade.  One aim of the most recent investigations has been to understand the origin of the energy splitting between them.~\cite{chulkov2004,chulkov2005,hofmann,chulkov2006,chulkov2007,chulkov2008} However, before going any further with the details, it may be helpful to clarify the terminology to which we shall adhere. Lifting of the Kramers degeneracy because of the lack of  inversion-symmetry centers \emph{and} the ubiquitous SOC is known in the literature (including the works on Bi(111)~\cite{rashba10}) as Rashba splitting,~\cite{rashba3,rashba10,rashba1,rashba2,rashba4,rashba5,rashba6,rashba7,rashba8,rashba9}
in regards to the lifting of the twofold spin degeneracy of the electronic states of systems if they lack inversion symmetry and the SOC is taken into account.~\cite{rashba1984}
In order to understand it, we can start by considering systems whose Hamiltonian is conservative and whose geometric structure has an inversion center. Such systems have at least two symmetries: One is the time-reversal symmetry, which guarantees that for any electronic state defined by $\epsilon(\bm{k},\uparrow)$, there is a degenerate state with $\epsilon(-\bm{k},\downarrow)$=$\epsilon(\bm{k},\uparrow)$. The second one is the inversion symmetry, which implies that for each state defined by $\epsilon(\bm{k},\uparrow)$, there is a degenerate state with $\epsilon(-\bm{k},\uparrow)$=$\epsilon(\bm{k},\uparrow)$. Therefore, if a system satisfies both symmetries, its electronic states  are at least doubly degenerate. Specifically, by combining the above two conditions, one gets that $\epsilon(\bm{k},\uparrow)$=$\epsilon(\bm{k},\downarrow)$ for each $\bm{k}$-vector. The latter is known as the \emph{Kramers degeneracy}~\cite{hofmann,grundmann2010} and it is also often referred as the "spin degeneracy" because the two states have same $\bm{k}$-vector but opposite spin. However, one must be careful with the latter designation because ir implicitly assumes a framework in which the spin-orbit coupling  is neglected.
That being said, the Rashba-type splitting is the \emph{lifting of the Kramers degeneracy (1) because of an external and tunable perturbation  breaks the inversion symmetry and (2) by virtue of the inherent presence of the SOC} (Though wurtzite materials which  inherently have no inversion symmetry also display Rashba/Dresselhaus splitting).~\cite{rashba1984} For example, the Rashba splitting model was initially applied to understand  systems with no inversion center either because of  the action of an external electric field,~\cite{marques,bangert} heterojunctions,~\cite{rashba1984,andrada,Ohkawa} but it may also take place in the presence of reactive substrates, surface passivation, or surface erosion.
Also, since the Rashba-splitting exists because of the SOC, it becomes most noticeable  for the electronic states
of heavy elements such as Bi.
Naturally, because the spin and angular momentum are no longer good
quantum numbers, the Rashba-type splitting breaks the magnetic quantum number degeneracy $\pm|m_j|$ rather than the "spin degeneracy."

Figure~\ref{fig:soc} is a qualitative diagram of the effect of the spin-orbit coupling  on \emph{p}-levels.~\cite{krupin2004}  Namely, were the SOC  negligible, all bands would be at least doubly degenerate at any k-point and all six states would be degenerate at $\Gamma$ (Fig.~\ref{fig:soc}(a)). The SOC, on the other hand, leads to a splitting of the $p_{j=3/2}$ and $p_{j=1/2}$ levels even at $\Gamma$, but leaves the
$\pm|m_j|$ degeneracy for systems with an \emph{inversion center} (Fig.~\ref{fig:soc}(b)). The latter case, in particular, is commonly used to represent the splitting of the bands of semiconductors.~\cite{winkler,grundmann2010} On the other hand, in a lattice without any inversion center, the Kramers or $m_j$ degeneracy ($\pm|m_j|$) is lifted at all k-points (Fig.~\ref{fig:soc}(c)), except  $\Gamma$ (as required by the time-reversal symmetry and regardless of the lack of inversion symmetry). That is the Rashba or so-called 'spin' splitting.

Coming back to Bi(111), it was initially concluded that the two metallic surface-states branches result from the removal of the spin (magnetic-quantum-number) degeneracy.~\cite{chulkov2004,chulkov2006} In turn, as mentioned before, lifting of the Kramers degeneracy is always attached to the lack of an inversion-symmetry center of the system plus the inherent SOC.~\cite{chulkov2004,chulkov2006}
The first ARPES investigation of the two metallic surface-state branches on thin films was performed using Si(111) as a substrate. The conclusion ---aided by band-structure calculations of free-standing Bi(111) films---  was that the weak interaction between the Bi(111) surface and the Si(111) substrate does not lift the spin degeneracy and the states behave as if the film preserved space-inversion symmetry.~\cite{chulkov2006}  Thus, at this point, it was  accepted that the splitting between the two metallic bands exists in the presence of inversion symmetry and that each of the two branches are "spin-degenerate."
At the same time, contradictorily, it was
stated again that the good agreement between experiment and DFT calculations shows that the surface states of the Bi(111) films display a Rashba-type SOC splitting.~\cite{chulkov2006}
Later on, the same authors have acknowledged that  non-supported Bi(111) films, or films supported by weakly-interacting substrates such as Si(111), behave as if they preserved inversion symmetry and for that reason the  degeneracy of the surface states is not lifted.~\cite{chulkov2007,chulkov2008} Thus, the spin-orbit coupling is identified  as a sufficient and necessary condition for the splitting. And yet, they again ratify the notion that there are two metallic surface-state branches rather than a single one because of the splitting caused by the Rashba effect (which is understood as that lifting of the Kramers degeneracy because of lack of inversion symmetry~\cite{rashba1,rashba2,rashba3,rashba4,rashba5,rashba6,rashba7,rashba8,rashba9,rashba10}).~\cite{chulkov2007,chulkov2008}
We shall see  that assuming that the measured splitting is of the Rashba-type  brings out several inconsistencies about the origin of the two metallic surface-states branches.

In addition to the above-mentioned time-reversal symmetry ---which guarantees that all states must be doubly degenerate at $\Gamma$--- the two metallic branches are also expected to be degenerate at the $M$-point because of the two-dimensional translational symmetry of the surface. The latter exists irrespectively of whether or not there is an inversion-symmetry center.
Specifically, on the surface layer (or any other layer below) there are six equivalent directions (Fig.~\ref{fig:surfstr}b) and correspondingly, for example, there are six equivalent $M$ points on the surface Brillouin zone (Fig.~\ref{fig:surfstr}c). This implies that $\epsilon(-\bm{M},\uparrow)=\epsilon(\bm{M},\uparrow)$. Clearly, the latter equation combined with the time-reversal symmetry implies that all states at $M$ must be doubly degenerate.
Nevertheless, the supposedly-Rashba-split metallic surfaces-state branches turned out to split all the way up to $M$, according to measurements and calculations.~\cite{hengsberger,AstMpoint2003,chulkov2006,chulkov2008} In other words, the expected degeneracy at $M$ in fulfillment of the translational symmetry is \emph{apparently} not satisfied.~\cite{chulkov2008} Surprisingly, the seeming violation to the translational symmetry was attached to the fact that the clean and relaxed Bi(111) slabs used in calculations (a) \emph{do} have  inversion symmetry\cite{chulkov2006} or (b) are sufficiently thin so that the two surfaces interact,~\cite{chulkov2004,chulkov2008} for which they do not represent an ideal semi-infinite surface,~\cite{chulkov2008} notwithstanding the splitting at all other intermediate points along the $\overline{\Gamma M}$ line has been from the start attributed to \emph{the lack of inversion symmetry at the surface} (Rashba splitting).~\cite{chulkov2004,chulkov2006,chulkov2008} Another inconsistency is that the latter assertion antagonizes two categorically different aspects. Namely, finiteness of the thickness of the slab apparently counteracts the in-plane translational symmetry. The seeming problem has thus been resolved by adsorbing hydrogen on one side of a fairly thin Bi(111) slab.~\cite{chulkov2004,chulkov2008}
Thus, the calculated band structure that  is currently considered to simulate that of a macroscopic Bi(111) sample is the band structure of one side of a film  only 4-nm thick, with broken inversion-symmetry via the adsorption of atomic H  on the other side of the film, and whose metallic surface states are strongly localized all the way up to the $M$-point (contrary to available measurements).~\cite{chulkov2008}

A closer look to the above line of reasoning begs three questions: (1) Is  H-adsorption needed to originate the splitting along $\overline{\Gamma M}$ shown in Refs.~\onlinecite{chulkov2004,chulkov2008}, to break the inversion symmetry, to impose the degeneracy at $M$,~\cite{chulkov2008} or to decouple the already weakly coupled sides of a 10-BL Bi slab?~\cite{chulkov2004,chulkov2008} or rather to better simulate a semi-infinite surface~\cite{chulkov2008}?; (2) Is it reasonable that one must break a symmetry (inversion symmetry) to fulfill another (translational symmetry)~\cite{chulkov2008}?; (3) Why shall we expect the two metallic branches to meet at $M$ ---either for thin or macroscopic films--- if it has been accepted that each  metallic branch corresponds to two degenerate states~\cite{chulkov2006} (which would automatically satisfy the degeneracy at $M$)?
In particular, although (1) the \emph{only} reason to demand the double degeneracy at $M$ is the translation symmetry and (2) it had already been accepted that the two bands are magnetic-quantum-number degenerate, at least for non-strongly perturbed finite films~\cite{chulkov2006},
the degeneracy at $M$ has been further searched and its absence justified.~\cite{chulkov2008,rashba10}
For example,  it has been concluded that the expected degeneracies at $\Gamma$ and $M$ are in fact \emph{removed} by the strong interactions between the upper and the lower surface of a finite film,~\cite{chulkov2008,rashba10} even though the the time-reversal and translational symmetries (or any other symmetry) cannot  be contingent on such interactions.
Furthermore, it was reported  that by increasing the separation
between the upper and lower surfaces (i.e., increasing the film thickness), the gap between the two branches at $M$ progressively decreases. The gap was in fact estimated to decrease by a factor 2 if the
number of layers is doubled.
The latter estimation puts forth that even films as thick as 16-nm (40~BLs)  misrepresent a macroscopic sample.
But moreover, since the only argument to expect a degeneracy at $M$ is the translational symmetry, it apparently implies that the in-plane translational symmetry  is fulfilled only in the limit of infinitely thick slabs. Otherwise, we have to ask ourselves: Why is there is a concern about the behavior of the metallic surface states at zone edge? and How is the splitting or degeneracy at $M$ relevant to the origin of the two metallic bands measured at $M$?

Spin-resolved ARPES (SARPES) measurements have shown that the two metallic branches  display  opposite magnetic-quantum-number polarization in thin-films.~\cite{chulkov2007} These experiments reaffirmed the notion that the Rashba-effect lifts the Kramers degeneracy, giving rise to two branches of opposite magnetic quantum number.~\cite{chulkov2007}
So, one may ask, how is it possible that the spin texture is observable if the bands are  degenerate states of opposite magnetic-quantum-number? The reason is that
 the two degenerate bands composing  each branch are localized on opposite sides of a film.~\cite{chulkov2008}
That is actually not surprising; the time-reversal and the translational symmetries do not demand that the two degenerate bands must be localized in the same spatial region.
Thus, in practice, each side of a thin film  bears two \emph{non-degenerate} bands of opposite magnetic-moment polarization. As such, dependently on the situation, one could use the entanglement of the states of the two sides of the slab or completely ignore the states of one side. Of course, in the experiments so far performed, the measurements correspond to only on one side of the films. Thus, from that point of view \emph{alone}, it is not possible to determine whether the gap between the spin-texture states corresponds to the Rashba splitting or not. Nevertheless, we shall see that the significance of the splitting or degeneracy at the $M$ point is that, if two spin-textured bands do not meet at $M$, then that indicates that the measurements do not correspond to the lifting of the spin or Kramers degeneracy.
But then, the question arises: why does the spin texture of one branch is opposite to that of the other branch   if the two metallic branches are not correlated (i.e., they do not correspond to split Kramers pairs)? Moreover,  it remains to resolve why the magnetic moment of each single band on a given side of the slab is double-valued and discontinuous at the $M$-point (see Fig.4(b) of Ref.~\onlinecite{chulkov2007}).

The  significance of the splitting at $M$   for the origin of the splitting everywhere else along $\overline{\Gamma M}$ can be also appreciated in a more recent SARPES investigation of  inversion-asymmetric (supported) Bi films .~\cite{rashba10} The unexpected splitting at $M$  was explained again by the overlapping of the metallic states with the bulk band but also linked  to the evanescence of the spin polarization near $M$.
The metallic states are  assumed to extend inside the film because of the overlapping with the bulk band, reflect specularly at the film/substrate interface and create quantum-well states, which in turn yield energetically separated bands at the $M$-point.
Then, the states become non-spin-polarized near $M$ simply because  \emph{only in this region} the splitting has no relation to the Rashba effect.  Even more worrying is that the states are also said to somehow become "electronically inversion symmetric" only near $M$.~\cite{rashba10}
As such, the splitting of the metallic bands is now believed to arise from an exotic combination of an inversion-asymmetry-driven Rashba-splitting (away from $M$) and a non-inversion-asymmetry-driven parity-effect splitting (near $M$).~\cite{rashba10}
It is true that the coupling between the two sides of the film  enhances the splitting between the metallic states at $M$ and that their increasing delocalization for decreasing wave vectors must reduce their spin polarization.
Nevertheless, there is neither bulk band for thin films nor energy overlapping with any bulk-like state. Moreover, we shall see that the only overlapping that totally quenches the spin polarization of the metallic bands is that with its own Kramers pair because the Kramers degeneracy is hardly lifted in films supported on weakly interacting substrates.

Because of the considerations exposed in the above introduction, we believe it is timely to revise the current interpretation of the electronic band structure of Bi(111) and, in particular, of its metallic surface-state branches. We shall see that all experimental measurements and previous computational results can be explained coherently and straightforwardly. In this work, we shall address four main issues: (1) What is the origin and character of the splitting of the two metallic branches? (2) What exactly is the role of the time-reversal, inversion and translation symmetry in shaping the band structure of Bi(111)? (3) How is the splitting or degeneracy at $M$ relevant to the origin of the two metallic bands measured at $M$? and (4) How does the Rashba splitting look like in Bi(111)? Can we tune it?

The rest of the manuscript is organized as follows: Section~\ref{sec:comp} describes the computational details of our investigation. Section~\ref{sec:resgeo} summarizes our results concerning the geometric structure of bulk Bi and Bi(111) films for various thicknesses and  the effect of the H-adsorption perturbation. Section~\ref{sec:reselec} contains our results for the band structure of Bi(111) for various thicknesses and examines the various aspects related to the electronic structure of Bi(111) that we have discussed above, including the effect of H adsorption. Section~\ref{sec:resrashba} describes the Rashba splitting of the two metallic branches of Bi(111) achieved by three different perturbations and analyzes how the magnitude of the perturbation correlates with that of the Rashba splitting and with the spatial localization of the surface states. Section~\ref{sec:recross} analyzes the density of electronic states as a function of thickness to resolve quantitatively whether the concentration of charge carriers is localized at the surface is sufficiently high to dominate the transport properties for thin films. Section~\ref{sec:discussion} discusses the implications of our results. Finally, Section~\ref{sec:concl} summarizes our main results and the conclusions drawn from them.

\section{\label{sec:comp}COMPUTATIONAL DETAILS}

Our investigation is based on fully relativistic periodic-supercell DFT calculations of the electronic structure of bulk Bi and Bi(111).
Our calculations are performed
 within the local density approximation (LDA). For the exchange and correlation terms,
we use the Ceperley and Alder~\cite{ceperly} functional. The  generalized gradient approximation with the Perdew-Burke-Ernzerhof~\cite{r56} functionals was also applied to obtain the structural properties of bulk Bi and used for comparison with the LDA functional.

The electron-ion interaction is treated within the pseudo-potential (PP) approach.~\cite{r60}
We use projector-augmented-waves  pseudo-potentials.~\cite{paw}
As valence electrons, we have taken into account only the  $6s^2$ and $6p^3$ electrons of Bi.
The Kohn-Sham orbitals are expanded in plane waves (PW) as implemented in the Vienna \emph{ab initio} simulation package (VASP).~\cite{vasp}
The SOC  is implemented by the developers as described in Ref.~\onlinecite{vaspmanual}. Notice that in this scheme, the inversion symmetry is not assumed
within the calculation even if the structure has inversion symmetry so that the Kramers degeneracy is \emph{not} assumed but derived. Moreover, all k-points in the grid are used in the calculation.

We have used  the both rhombohedral and hexagonal unit cells of  Bi (which contain two and six atoms per unit cell, respectively) to describe the bulk, and the hexagonal one~\cite{hofmann} to describe the Bi(111) surface.
The Bi(111) surface was modeled with a slab having on atom per layer (1$\times$1 in-plane periodicity) and  39 bilayers.
In the calculations involving the surface, a vacuum layer  of 14~{\AA}
separate the periodic images of the slab  to avoid interaction between them.

For  integrations over the  Brilloiun zone, that of the bulk (in the hexagonal representation) and surface were sampled by 2592 and 432 irreducible k-points, respectively, which are selected while ignoring  inversion symmetry.
The integrations use a Gaussian broadening  with a smearing parameter of 0.1 eV.
The Kohn-Sham orbitals were  expanded in a plane-wave basis set with a maximum kinetic energy  of 163 eV.
The kinetic energy for the augmentation of charges is cut off at 265.4 eV.
The positions of all atoms in the slab were optimized until the  forces on each atom and each direction was smaller than  6$\times$10$^{-3}$ eV/\AA. For this purpose, the conjugate-gradient algorithm has been applied. The total energy for each ionic configuration sampled within the relaxation process was convergent up to 2$\times$10$^{-5}$ eV.

Preliminary  calculations of the  lattice parameters, volume and cohesive energy of  bulk Bi show
that (1) semicore \emph{d}-electrons of Bi influence these properties negligibly; (2) the generalized gradient approximation using the Perdew-Burke-Ernzerhof~\cite{r56} functional does not outperform LDA since
significantly overestimates the length of the basis vectors and thus nearest-neighbor distances and the supercell volume; (3)  denser k-point grids to sample the bulk Brillouin zone
do not yield variations in the electronic  properties of bulk Bi; (4) doubling the energy cutoff for plane waves does not affect the electronic  properties of bulk Bi.

For the calculations concerning the Bi(111) surface, we consider slabs of various thicknesses, with inversion symmetry and without it.
The inversion symmetry of a relaxed Bi(111) slab is broken by (1) placing hydrogen at 2~{\AA} of Bi, as in Ref.~\onlinecite{chulkov2004}, and at 1.86~{\AA}, which is the distance resulting after relaxation; (2) an inter-bilayer perturbation ---achieved by increasing  by 0.3~{\AA} the  distance between the first two bilayers of  one side of the slab; and (3) an intra-bilayer perturbation ---achieved by increasing by 0.1~{\AA} the  distance between the two layers forming the first bilayer of one side of the slab.

The XCrySDen software~\cite{xcrysden} is used to obtain a schematic representation
of the structure of the bulk Bi and Bi(111).

\section{\label{sec:resgeo}Results: Geometric structure of bulk Bi, clean Bi(111)  and H/Bi(111)}

\subsection{\label{subsec:strucbi111}Structure of  Bi(111)}

The Bi(111) surface is a triangular lattice (see Fig.~\ref{fig:surfstr}(b)) perpendicular to the trigonal axis. There are four  parameters in A7-materials characterizing the bilayered structure that facilitate the analysis of the properties of bulk Bi and, in particular, of  Bi(111). These parameters are the distance between first nearest neighbors $d_{1NN}$,  the distance between second nearest neighbors $d_{2NN}$, the intra-bilayer distance $d{ij}$, and the inter-bilayer distance $d{jk}$.
The bilayered structure of Bi underlying the Bi(111) surface is shown in Fig.~\ref{fig:surfstr}(a). As shown by the color code and the labels, the bilayered structure repeats itself only every six planes. The vertical spacing between the layers forming a bilayer, the \emph{intra-bilayer distance}, is denoted by $d_{ij} = d_{12}$, $d_{34}$, etc., and is related to the strong, short and covalent bonds existing among first nearest neighbors. In turn, the vertical spacing between two bilayers, the \emph{inter-bilayer distance}, is denoted by $d_{jk} = d_{23}$, $d_{45}$, etc., and is related to the weak, long and metallic bonds existing among second nearest neighbors. A complete analysis of the structure of Bi(111) will be provided elsewhere. Here it will be important to mention only that  the changes in $d_{ij}$ and $d_{jk}$ within the four topmost bilayers are well converged for slabs of 10 or more bilayers (up to 39~BLs). Moreover, the bulk-like interlayer spacings is recovered around the fifth bilayer, in the sense that at such distance from the surface, changes of $d_{ij}$ and $d_{jk}$ with respect to the bulk values are less than 0.1\%.

\subsection{\label{subsec:strucHbi111}Effect of H-adsorption on Bi(111) geometric structure}

In  this subsection, we present the results for H/Bi(111), which clearly is an \emph{inversion asymmetric} film. We limit ourselves to investigate  a hydrogen monolayer (ML) adsorbed atop on one side of a Bi(111) film of (a) 6~BLs, a thickness which in the absence of H is insufficient to recover  bulk-like $d_{ij}$ and $d_{jk}$, and (b) 10 and 12~BLs,  thickness which in the absence of H are sufficient to recover bulk-like $d_{ij}$ and $d_{jk}$.
These calculations are performed in order to analyze whether H-adsorption can be expected to help  model the geometric structure of a \emph{semi-infinite} Bi(111) slab.
Namely, as mentioned in Section~\ref{subsec:introelectronsurface},  it has been assumed in previous DFT calculations that  it is possible to simulate  the Bi(111) surface of a \emph{semi-infinite} system  by using a  film of  10~BLs with a monolayer of atop H placed $\sim$2{\AA} away from one side of the Bi film, the H-host surface.~\cite{chulkov2004,chulkov2008} In this way, the other side of the slab, the \emph{clean side}, is expected to be a reliable model of the   Bi(111) surface of a \emph{semi-infinite} system or, for that sake, of any macroscopic sample. Hence, our goal here is to find out whether the clean side of H-adsorbed \emph{inversion-asymmetric} films indeed displays  the converged structural properties of the clean inversion-symmetric films reported in the previous subsection.

The first aspect that calls the attention is that a monolayer of atop H has been fixed at a particular (and arbitrary) distance from the surface, $\sim$~2\AA.~\cite{chulkov2004,chulkov2008} We thus aim to find out whether, with respect to the Bi-H radius of interaction, a distance of $\sim$2\AA is  long or short so as to represent a small or large perturbation to the Bi slab, especially to the clean side of it. One cannot obtain the structure of a Bi slab when the H ML is exactly 2{\AA} away unless one fixes  the position of both the topmost Bi surface layer and H.
Nevertheless, should one monolayer of H, situated at $\sim$2{\AA} from the surface, be a small perturbation, then  either (1) 2{\AA} is too long with respect to the relaxed distance or (2) the  effect of one H ML on the structure of the clean side is  mild. Our results for the totally relaxed H/Bi(111) films displayed in Table~\ref{tab:surfrx2} indicate the opposite:  (1) The relaxed distance  between  H  and the topmost Bi layer is actually not too far to 2{\AA} (1.86{\AA}) and
(2) changes in the interlayer distances around the relaxed clean side indicate that the perturbation induced by  H is considerably strong and of long range. Specifically, as shown in Table~\ref{tab:surfrx2}, even for 12-BL films, H-adsorption  causes that the inter-layer distances adopted around the "clean" side  deviate significantly from the converged distances that were thoroughly tested for inversion-symmetric slabs of up to 39~BLs. For example,   $\Delta d_{12}$ for the clean side of the three H-adsorbed films is $\sim$-1.9\%, whereas the convergent value for clean inversion-symmetric films is $\sim$-1.6\%. Therefore, because the position of the atoms and their charge density entail each other, one cannot expect that a monolayer of H  will render a charge density ---and thus a geometric structure--- which is in any way a good approximation of that of a semi-infinite slab, no matter whether one keeps the positions of all atoms fixed at their converged values or not.

\section{\label{sec:reselec}Results: Electronic structure of bulk Bi and Bi(111) thin films}

\subsubsection{\label{subsec:ssrx} The effect of structural relaxation on the metallic states of Bi(111)}

It is worth to mention first that the metallic surface states along $\overline{\Gamma M}$  are quite insensitive to the relaxation of the structure, as long as the inversion symmetry is preserved. Namely, the electronic structure of a relaxed film does not differ significantly from that of a bulk-terminated film. Also, if the inversion symmetry of a Bi(111) film is broken by hydrogen adsorption on one of its surfaces,  the electronic structure of the metallic states energy does not depend on whether the system is relaxed or not. Specifically, the band structure of H/Bi(111) in which H is placed  2~{\AA} away from the Bi surface and the forces arising upon adsorption are not relaxed (as in Ref.~\onlinecite{chulkov2004}) is not significantly different from that in which H is 1.86~{\AA} away from  Bi surface  after a full relaxation of the forces. In contrast, we shall see later the noticeable differences between  the metallic surface states of an inversion-symmetric clean film   and those of an inversion-asymmetric slab in which even minor perturbations break the inversion symmetry.

\subsubsection{\label{subsec:ssblk} \emph{Ab initio} projected band-structure of bulk Bi: The A7-distortion and SOC surface states }

The effect of the SOC on the electronic structure of bulk Bi has long been known from \emph{ab initio} calculations through the early work of Golin,~\cite{golin} and more recently from Gonze \emph{et al.} (see Fig.~\ref{fig:bzelecstr}).~\cite{gonze1990} Nonetheless,  the \emph{projected} band-structure of bulk Bi on the SBZ of Bi(111) has so far only been obtained  from tight-binding calculations. In fact, even \emph{ab initio} studies of Bi(111) have  used the tight-binding  band structure of the bulk to draw conclusions about that of Bi(111) and about the effects of the SOC on it.~\cite{hofmann,chulkov2004} Thus, we shall first analyze the \emph{ab initio} and fully relativistic projected band-structure of bulk Bi on the SBZ of Bi(111) in order to introduce and discuss the surface states of Bi(111) in a consistent manner.

Figure~\ref{fig:bpdosss}(top) shows  the projected band structure of bulk Bi along the $\overline{\Gamma M}$ and $\overline{\Gamma K}$ directions of the SBZ of Bi(111) in the energy range within which the spin-orbit coupling is expected to give rise to gaps and surface states. It  shows that there are indeed gaps at  the regions where Jezequel \emph{et al}. observe  surface states for Bi(111) via ARPES.~\cite{jezequel} Specifically, there are gaps from $\Gamma$ to $\frac{1}{3} K$ spanning from -0.6 to -0.8 eV and from -1.4 to -2.8 eV; and from $\frac{3}{4} K$ to $K$ spanning from -1.8 to -2.8 eV, that are consistent with Jezequel's \emph{et al}. measurements.~\cite{jezequel} Furthermore, our calculation of the band structure of a 39-BL film with inversion symmetry and including  SOC  confirms the existence of those states, as shown in Figure~\ref{fig:bpdosss}(bottom).  Comparison between the top and bottom plots reveals that there are surface states arising \emph{only} if the SOC is taken into account and in regions gapped \emph{only} because of the SOC. These surface states are thus indeed "spin-orbit coupling surface states". In contrast, the pair of metallic surface-state branches (denoting that it is not a single band) around $E_F$ appear within the "A7-distortion" gap and regardless of whether or not the SOC is taken into account. In other words, the origin of both the conduction-valence band gap and the metallic surface-state branch(es)  does not lie in the SOC.

Note that for this inversion-symmetric film of up to 15.5~nm, with and without SOC, the branches are at least doubly degenerate. Importantly, at least for inversion-symmetric films, the two Kramers degenerate bands composing each metallic branch  are primarily localized on  opposite sides of the film. Thus, the SOC alone does not lift the Kramers degeneracy anywhere inside the SBZ.
Namely, although the metallic branches appear separated all along $\overline{\Gamma M}$, each metallic branch in Fig.~\ref{fig:bpdosss}(bottom) is composed of two degenerate bands (In other words, there are actually four metallic surface-state bands). Because the film is inversion-symmetric and the bands are doubly degenerate, that means that the splitting  arises without requiring inversion asymmetry and without breaking the Kramers degeneracy, which demonstrates that the SOC splitting is not of the Rashba type. Moreover, it is clear for the same reasons that, despite the splitting, the degeneracy at $M$ is guaranteed  for inversion-symmetric films.

Figure~\ref{fig:bpdosss}(bottom) also shows that the splitting of the metallic branches exist all along between the $\overline{\Gamma M}$ midpoint to $M$ \emph{even if the SOC is ignored altogether and in the presence of inversion symmetry} for films our 15.5-nm film. In fact, the splitting between the two metallic branches is about two times larger near $M$ without SOC than with SOC (it might really overlap with the non-SOC bulk band of bulk Bi). For example, the splitting at $M$ without SOC for the 15.5nm film can be as large as 90~meV, yet it bears no relation to the SOC or the Rashba effect but to the coupling between the two side of a film, as noted in Ref.~\onlinecite{rashba10}. Figure~\ref{fig:bpdosss}(bottom) also shows that the splitting of the metallic branches  from $\Gamma$ to the midpoint of the $\overline{\Gamma M}$ line arises only once the SOC is included.
From the midpoint of $\overline{\Gamma M}$ to $M$, the SOC significantly increases the energy of both metallic branches, for which they do not overlap with the bulk band. Interestingly,  the splitting between the metallic branches that exists near $M$  in the absence of SOC is reduced once the SOC is included. However, it is not clear whether this reduction is caused by the directly by the SOC or indirectly because of the energy increase.

\subsubsection{\label{subsec:ssevo} Splitting and delocalization of the metallic surface states along $\overline{\Gamma M}$ as a function of film thickness}

Figure.~\ref{fig:surfelec} displays the band-structure of Bi(111) around $E_F$ and including the SOC, for an inversion-symmetric film of  39~BLs. The evolution and convergence of the dispersion will be provided elsewhere. Here it will  be important to mention only that  21~BLs are enough to converge the dispersion of the two metallic branches  along the $\overline{\Gamma M}$ line.

Note in Fig.~\ref{fig:surfelec} that our fully relativistic DFT calculations indicate that the width of the A7-distortion gap of bulk Bi around  at $M$ is $\sim$120~meV and spans from 15 to 135~meV below $E_F$, in stark contrast to the $\sim$20--30-meV gap predicted by tight-binding calculations (spanning at most from 20 to 50~meV below $E_F$).~\cite{chulkov2004,AstMpoint2003} In addition, contrary to the belief  that the surface states of macroscopic samples overlap with the bulk band (based on tight-binding calculations~\cite{chulkov2004,liu1995}),~\cite{rashba10,hofmann,AstMpoint2003} our calculations for slabs thicker than 8~nm ($\sim$~21 BL) show that the metallic branches do not overlap  with the bulk bands at or in the vicinity of  $M$. This is important because, as mentioned in subsection~\ref{subsec:intrometallicbranches}, the alleged overlapping has served as a premise to conclude on the behavior of the metallic branches at $M$  and as an explanation for the incapability of ARPES  to confirm  the expected degeneracy at $M$ in macroscopic samples. Namely, even for ultra-thin films, it has been argued that the observations are rather surface resonances that result from the hybridization of the metallic surface state  with quantum well states~\cite{chulkov2006}  and are, hence, so delocalized that the spin-splitting weakens.~\cite{chulkov2007} We shall address this issue in subsection~\ref{subsec:magmom_discon}. In the meantime, we must  highlight as well that our results in Fig.~\ref{fig:surfelec} also indicate that  ARPES measurements around  $M$~\cite{hengsberger,AstMpoint2003} are within the gap and, thus, are not surface resonances (hybridization of surface states with quantum well states~\cite{chulkov2006}) but surface states.
Therefore, our results indicate that ARPES measurements of the metallic branches around $M$ are reliable despite their low intensity because there is no  overlapping with the bulk band. The intensity is certainly weak because the metallic branches become more delocalized around $M$, whereas the  energy of the X-rays used in ARPES to study surfaces is fixed and chosen to sample the least amount of subsurface atoms.~\cite{AstMpoint2003,Shah1979,Damascelli2004} Nonetheless, integration of  partial intensities over a higher (or lower) photon energies might be useful to  resolve more clearly the dispersion of macroscopic samples without the concern that part of the intensity comes from bulk states.

Importantly, there is good agreement between our calculations for inversion-symmetric films and the ARPES measurements of Hengsberger \emph{et al}.~\cite{hengsberger} and Ast and H\"ochst (Fig.1(a) of Ref.~\onlinecite{AstMpoint2003}) for macroscopic slabs regarding the dispersion and  the fact that the two bands do not meet at the $M$ point. This agreement supports that, should the crystal display inversion asymmetry, (1) the splitting along  $\overline{\Gamma M}$ of the metallic branches takes place regardless of inversion symmetry and (2) the gap at $M$ between the two metallic branches may comes up  not only for simulated inversion-symmetric thin films but also for macroscopic samples and supported thin films. We shall denote the energy gap at $M$ as $\Delta E$ and discuss it in subsection~\ref{subsec:ssatm}.

\subsubsection{\label{subsec:SOCSIA} Evolution of the metallic branches along the $\overline{\Gamma M}$ as a function of the SOC strength in an inversion-symmetric and -asymmetric film}

We have seen that the two degenerate bands of each branch are localized on the opposite surfaces of the slab. However, only one of them is investigated experimentally. In order to understand the effect of the SOC and the inversion symmetry on the band-structure of Bi(111) films, and compare with experimental observations, we show in Fig.~\ref{fig:socsia}(Top) the evolution of the splitting between and degree of localization  of the metallic branches of a 24-BL Bi(111) with inversion symmetry as a function of the strength of the SOC and on a single chosen side of the slab. In order to do that, we must distinguish between the two surfaces of the slab. We will refer to them as top and bottom surface. The states located on, say, the top surface are then highlighted with red dots whose diameter represents to what extent the state is localized on the surface. The Kramers pair at each $k$-point (located in the bottom surface) cannot be appreciated because it is degenerate with the bands located on the top surface.

As expected from available measurements and previous calculations,~\cite{AstMpoint2003,chulkov2008} both  metallic branches bear some localization weight on any of the two surfaces of a film. As expected from  subsection\ref{subsec:ssblk}, the splitting between the two metallic branches increases almost all along $\overline{\Gamma M}$ and decreases near $M$ with  increasing strength of the SOC. Fig.~\ref{fig:socsia}(Top) also shows that with  increasing strength of the SOC, the localization of the metallic branches on the topmost bilayer slightly decreases from $\Gamma$ to the midpoint of $\overline{\Gamma M}$, slightly increases from the midpoint of $\overline{\Gamma M}$ to $M$, and strongly increases along $\overline{\Gamma K}$.

Let us now  consider the effect of the SOC in a slab without inversion symmetry.  Fig.~\ref{fig:socsia}(Bottom) displays  the evolution of the metallic branches of a 24-BL Bi(111) without inversion symmetry as a function of the strength of the SOC. In this case,  the symmetry is broken by increasing  the inter-bilayer distance $d_{jk}$ between the bottom bilayer and the rest of the film. Again, the red dots indicate how much the metallic states are localized on the top bilayer.  By comparing Figs.~\ref{fig:socsia}(Top) and (Bottom), one can see that the dispersion of the bands  on the top bilayer and their dependency on the strength of the SOC are nearly the same with and without inversion symmetry. In addition, can see that, without inversion symmetry,  the metallic branches display  extra splittings. In the absence of the SOC, the branches resulting from the extra  splitting are still doubly degenerate. Hence inversion asymmetry alone does not lift the Kramers degeneracy either. However,  with broken symmetry and once the SOC is included, the obtained splitting  finally lifts the Kramers degeneracy and reveals the four states composing the two metallic branches.  That is indeed the lifting of the Kramers degeneracy, the spin or Rashba splitting. Still,  Kramers pairs are on opposite sides of the slab. Therefore, once again, the   splitting found for the top bilayer  bears no relation to the lack of inversion symmetry or the lifting of the Kramers degeneracy (Rashba effect). Importantly, Fig.~\ref{fig:socsia}(Bottom) shows that in this inversion-asymmetric film, the resulting Rashba splitting does satisfy the translational-symmetry-required degeneracy at $M$.  We shall come back to analyze inversion-asymmetric slabs when we discuss the Rashba splitting in Bi(111) in section~\ref{sec:resrashba}.

\subsubsection{\label{subsec:originsplit} Origin of the two metallic bands along the $\overline{\Gamma M}$}

We have seen that  the Kramers degeneracy is not lifted by the SOC alone and hence the observed splitting between the metallic bands is not of the Rashba type.  The question is, what is then the origin of the splitting?

Ast and H\"ochst have proposed  strong surface relaxations to explain the origin of the metallic surface states of Bi(111), but it has long been clear that these are not dramatic.~\cite{hofmann} Moreover,  the surface states appear regardless whether the surface is relaxed or not (subsection~\ref{subsec:ssrx}). Hofmann in turn has evoke the relation between  the existence of surface states and dangling bonds, later acknowledging that  Bi(111) has no dangling bonds. He then proposes the spin-orbit interaction as the origin of the surface states. However,  our results in Fig.~\ref{fig:bpdosss} for a 39-BL slab demonstrate that  the presence of two bands (and not one) is not even necessarily attached to the spin-orbit-coupling. Thus, based on the fact that the metallic surface states exist regardless of the SOC and/or the lack of inversion symmetry, we can say that the surface states of Bi(111) are first of all  a pair of Shockley states arising because of the abrupt termination of the periodic potential for  the highly delocalized $p$-states of Bi. It is true though that the splitting is enhanced along the  $\overline{\Gamma M}$ direction by the spin-orbit coupling. However, around the $M$-point, the delocalized character of these states plays an important role,~\cite{rashba10} for which the splitting exists regardless of the SOC for thin films.

Still, what does the SOC split then? We propose that the splitting does not lift the Kramers degeneracy and yet  it is caused by the SOC, and mainly around $\Gamma$, because it is a $m_j=\pm1/2$--$m_j=\pm3/2$ splitting, as described in Fig.~\ref{fig:soc} in subsection~\ref{subsec:intrometallicbranches}. In the following, we  provide some arguments and indirect evidence that the two metallic branches (accounting for a total of four bands) have a $m_j=\pm1/2$ and $m_j=\pm3/2$ character, respectively, particularly from $\Gamma$ to the k-point along $\overline{\Gamma M}$ where they reach highest energy.

The states shown in Fig.~\ref{fig:bpdosss}(top) appear as  purely \emph{p}-states of  bulk Bi if projected on spherical harmonics (\emph{s}-states have much lower energy ---$\sim$8eV below $E_F$, as shown in Fig.~\ref{fig:bzelecstr}). However, it would be necessary  to project the wavefunction onto the eigenfunctions of the $\bm{J^2}$ and $J_z$ operators to obtain their main total-angular-momentum and magnetic-quantum-number character, which is a computational task that we could not afford. Nonetheless, by analyzing  the splittings of the bulk bands along $\overline{\Gamma T}$ and $\overline{\Gamma KX}$ occurring upon inclusion of the SOC (Fig.~\ref{fig:bzelecstr}) and comparing them with those sketched in Fig.~\ref{fig:soc}(b), it is possible to identify that along those directions the lower single \emph{p}-band must have a $j=1/2$ character and the upper pair of \emph{p}-bands must have a $j=3/2$ character. The latter in turn must split into $m_j=\pm1/2$- and $m_j=\pm3/2$-states.
Therefore, one can conclude that in Fig.~\ref{fig:bpdosss}(top), the projected bulk bands around $\Gamma$ of the SBZ that span from $\sim$2 to $\sim$3~eV  correspond to the $j=1/2$-states, whereas the upper occupied bands  (those up to $\sim$1.5eV below $E_F$) correspond to $j=3/2$-states that split into $m_j=\pm1/2$- and $m_j=\pm3/2$-states (even though in other regions of the BZ the energy ordering of the $j=1/2$ and $j=3/2$ may be swapped so that the two contributions overlap in that region~\cite{golin}).
One can thus speculate that along $\overline{\Gamma M}$ of the SBZ (and particulary around $\Gamma$), the bulk bands up to $\sim$1.5eV below $E_F$ must be in large part  states of a $m_j=\pm1/2$ and $m_j=\pm3/2$ character that overlap energetically for the most part once projected on the SBZ, whereas those below -1.5~eV  have a $j=1/2$ character. In line with these arguments, one can see that the surface states that peel off from the lower band ($j=1/2$) is a single band but the surface states that peel off from the upper band (overlapping of $m_j=\pm1/2$ and $m_j=\pm3/2$) come in pairs (see Fig.~\ref{fig:bpdosss}(bottom)). This includes the pair surface-state metallic bands. Thus,  the surface states derived from these three major bands ($j=1/2$, $m_j=\pm1/2$- and $m_j=\pm3/2$), which  do not overlap with anything else, might acquire a relatively pure ${j,m_j}$ character.

The $m_j=\pm1/2$--$m_j=\pm3/2$ splitting is known in semiconductors physics as the splitting between the light-hole (LH) and heavy hole (HH) branches  (\emph{the LH-HH splitting}), in reference to the effective masses attached to  carriers because of their dispersion.~\cite{grundmann2010,winkler} In the case of the metallic branches of Bi(111), the high-energy metallic branch has a larger dispersion than the low-energy branch. We would thus say that the high-energy metallic branch (light) corresponds to lighter carriers than the low-energy metallic branch (heavy). In line with this analogy (with the band-structure of semiconductors), there is also the anisotropy of the charge density. Namely, the charge density associated to  light and heavy bands is anisotropic with respect to the direction perpendicular to the surface: while
the corresponding in-plane ($xy$) charge density is larger for a HH-band ($m_j=\pm1/2$) than for a LH-band ($m_j=\pm3/2$),
the charge density along the direction perpendicular to the surface ($z$) is  larger for a LH-band than for a HH-band.~\cite{rajeevMIT} Indeed, as shown in Fig.~\ref{fig:pldosani}, we find that, along $\overline{\Gamma M}$, the projected LDOS $p_{xy}$ attached to the low-energy metallic branch (heavy) is larger than that of the high-energy branch (light), whereas
the projected LDOS $p_{z}$ attached to the  high-energy branch (light) is larger than that of the low-energy metallic branch (heavy).

Up to now, we have just provided similarities between the two split metallic branches of Bi(111) and the  LH-HH splitting found in bulk semiconductors, suggesting that high- and low-energy branches have a $m_j=\pm1/2$ and $m_j=\pm3/2$ character, respectively. However, we now turn to the local magnetic moment of the two metallic branches of Bi(111), which can gauge the $m_j=\pm1/2$ and $m_j=\pm3/2$ character of the two metallic branches of Bi(111).
The total magnetic moment is zero for both branches because of the Kramers degeneracy, but we can focus on only one of two degenerate bands of each metallic branch, say, that with positive magnetization. Namely, we will apply that
for a quantization axis parallel to the normal to the surface ($z$), states with a predominant $m_j=+1/2$-character must have a larger contribution to the in-plane magnetization than states with a predominant $m_j=+3/2$-character.
We  use the $m_y$ magnetization because it dominates the in-plane magnetization all along $\overline{\Gamma M}$. Indeed, as shown in Figure~\ref{fig:magy}, the low-energy metallic branch (heavy) has a larger in-plane magnetization all along $\Gamma M$ than the high-energy branch (light), indicating that the low-energy  branch has a predominant $m_j=+1/2$-character and the high-energy branch has a predominant $m_j=+3/2$-character. In summary, our results suggest that the reason for which the splitting of the two metallic branches is certainly the result of the SOC  but it does necessarily lift the  Kramers degeneracy is that the splitting correspond to the SOC-driven $m_j=\pm1/2$--$m_j=\pm3/2$ splitting.

\subsubsection{\label{subsec:originspin} Origin of the spin polarization of the two metallic bands along the $\overline{\Gamma M}$}

Spin-resolved ARPES measurements for ultra-thin films have shown that the two metallic branches  display  opposite magnetic-quantum-number polarization in thin-films.~\cite{chulkov2007,rashba10} On the other hand, we have shown that the spin texture observed in such states cannot be attached to the lifting of Kramers degeneracy. We must then explain why the  measurements for each branch give opposite magnetic-moment polarization if the two metallic branches are in principle not correlated; \emph{i.e.}, they are not split Kramers pairs.

We  propose that the  spin texture observed for the metallic branches occurs only because the surface cannot afford to be magnetic.
Namely,
covalent bonds are formed at the expense of quenching spin pairing. Bi atoms, for example, have three unpaired electrons and nevertheless Bi is non-magnetic in the bulk phase precisely because  the three covalent bonds that each atom makes  lower the energy more than retaining any spin pairing.
Therefore, the net magnetization of the surface bilayer is expected to be zero because
the surface has no dangling bonds (no covalent bonds are broken) and hence each surface atom retains essentially unaltered the relatively strong covalent bonds that any bulk atom has.
Thus, just as in the bulk phase, the covalent bonds of surface atoms  cannot afford magnetic polarization.
That being said, let us now analyze the observed spin texture of the metallic bands. Although the SOC does not lift the Kramers degeneracy, the Kramers pairs (which have opposite magnetic quantum number)
cannot quench each other their spin polarization because, as mentioned earlier, they
are primarily located on opposite sides of the slab. So, for example, if the high-energy band of the top bilayer yields a spin-up polarization, the only possibility to cancel the net magnetization at the surface is that the low-energy branch acquires the opposite spin polarization.

\subsubsection{\label{subsec:magmom_discon} The magnetic moment discontinuity at $M$  }

So far, we have shown that the splitting at $M$ between the two metallic bands lying on a single side of a film is not in conflict with the translational  symmetry in inversion-symmetric or inversion-asymmetric films because, as we have seen, (1) the splitting found on a single side of a film does not correspond to the lifting of the Kramers degeneracy and (2) the two-fold degeneracy at $M$ is guaranteed for inversion-symmetric and even asymmetric films.
There could be, however, a subtle issue with the splitting at $M$ for thin films. Namely, spin-resolved ARPES measurements on ultra-thin films (7~BLs) have shown that the two metallic branches  display  opposite magnetic-quantum-number polarization  and this has been confirmed by DFT calculations.~\cite{chulkov2007}
Thus, even though the two-fold degeneracy at $M$ is strictly satisfied,  there is an apparent non-physical discontinuity and indetermination of the magnetic moment one either side of the film (top or bottom) at the $M$-point for each metallic band. This magnetic discontinuity and indetermination at $M$ can be appreciated if one focuses the attention on one side of the slab, as illustrated in Fig.~\ref{fig:magmom_discon} (see also Fig.4(b) of Ref.~\onlinecite{chulkov2007}).

On the other hand, as mentioned in subsection~\ref{subsec:intrometallicbranches},  SARPES measurements on Bi(111) ultra-thin films also show that the measured net spin polarization of each metallic band vanishes near $M$. Namely, both spin-up and spin-down intensities are detected and become comparable at $k$-points near $M$.~\cite{chulkov2007,rashba10}
In that work, the evanescence of the spin polarization near $M$ is understood by simply declaring that
\emph{only in this region} the splitting does derive from the Rashba effect. Yet, the opening premise of the above  argument is that the surface states overlap with the bulk band~\cite{rashba10} (or bulk-like quantum-well states~\cite{chulkov2007}). Nevertheless, we have shown that there is neither bulk band for such thin films nor  overlapping of the metallic surface states with any bulk-like state (see Fig.~\ref{fig:surfelec}). The question is then, why does the spin polarization vanish? In this section, we shall resolve the  non-physical discontinuity and indetermination of the magnetic moment  to put our understanding of the metallic branches at the   $M$ point on firm grounds, and at the same time show that the solution to this issue provides  insight on the vanishing of the spin polarization obtained from the above spin-resolved ARPES measurements.

Figure~\ref{fig:magmom_evo} provides the contribution of the two surface bilayers (top and bottom) of an inversion-symmetric film to the magnetic moment associated with the high- and low-energy metallic branches. Because experimentally only one surface has been sampled, we  concentrate only on the "top" bilayer,  though the same arguments apply for the other bilayer. As expected from experiment, the high- and low-energy branches have opposite magnetic-moment polarization. As shown in the figure, the top bilayer hosts mainly the the high-energy band H1 and the low-energy band L2. Also, for the most part along the $\overline{\Gamma M}$ line, the high-energy band H1 has "up" polarization (Fig.~\ref{fig:magmom_evo}(a)), whereas the low-energy band L2 has "down" polarization (Fig.~\ref{fig:magmom_evo}(b)). In agreement with experiment,~\cite{chulkov2007} the high-energy branch has a smaller magnetic moment than the low-energy branch and the magnetic polarization decreases as $k$ approaches to $M$. The latter happens for the same reason that the ARPES intensity decreases around the $M$-point. Namely, the metallic surface-state branch becomes increasingly delocalized and so does the magnetic moment. Most importantly, note that, close to $M$, it is no longer true that the top bilayer gets magnetic-moment contributions only from the H1 and L2 bands since their Kramers degenerate pairs, H2 and L1, respectively, also contribute. Yet the contributions of H2 and L1 have opposite polarization. The contributions of the H2 and L1 bands further reduce the magnetic moment polarization of each band. So that  the magnetic moment of each band is exactly zero at $M$. This result hence implies that there is no magnetic-moment discontinuity or indetermination. It remains, however, to understand whether the exact cancelation of the magnetic moment at $M$ is a general result attached to the time-reversal symmetry.

Regarding the observed  evanescence of the spin polarization near $M$,~\cite{rashba10} we conclude that (1) the experiment detects "spin-up" and "spin-down" intensities around $M$ and \emph{at the same energy} (within experimental resolution) because,  at least around $M$, spin-resolved ARPES has actually sampled the two states composing each Kramers pair (which at other regions of the $\overline{\Gamma M}$ line are on
opposite sides of the film) and (2) the interaction of the Bi film with the substrate causes a negligible Rashba splitting around $M$, as we shall discuss in more detail in section~\ref{sec:resrashba}. In summary, the fact that the spin polarization vanishes at $M$ not only precludes the magnetic-moment discontinuity and indetermination but  is also a signature that the Kramers degeneracy is hardly lifted.

\subsubsection{\label{subsec:ssatm} Significance  and evolution as a function of thickness of the gap at $M$ between the metallic branches}

Despite  the splitting at $M$ between the two metallic bands lying on a single side of a film is not in conflict with the translational  symmetry in inversion-symmetric or inversion-asymmetric films, it is interesting to look into the evolution of gap $\Delta E$ as a function of thickness, in order to address the  issue of whether the metallic branches meet or not for inversion-symmetric films. Fig.~\ref{fig:surfelec} shows that the gap $\Delta E$ between the two metallic branches  decreases as a function of thickness, in agreement with previous calculations.~\cite{chulkov2006,chulkov2008}  It has been estimated that $\Delta E$ decreases as  $\frac{1}{N_B}$, which suggest that the two bands are degenerate only in the limit of an infinite number of layers. Accordingly, the converged band structure can be attained only for semi-infinite slabs.~\cite{chulkov2008} We, nevertheless,  do not find  evidence that the bands should meet in such limit. A non-linear regression  fitting of our data (Fig.~\ref{fig:convgap}) indicates that although $\Delta E$ varies as $\frac{1}{N_B}$ within ${N_B}=6$ and ${N_B}=39$, there is an addition constant factor $C=40$~meV ($\Delta E = \frac{B}{N_B}-C$) that invalidates that the two bands are degenerate only in the limit of an infinite number of layers. Our fitting in Fig.~\ref{fig:convgap} is of course limited to the range of the data but it shows that there are no grounds to conclude on the behavior of thicker slabs. Thus, while it is true that the splitting around $M$ is related to the interaction between the two Bi surfaces, the two branches might not meet even at the macroscopic scale if the gap remains once the two sides of the film are decoupled, as indicated so far by ARPES measurements.~\cite{AstMpoint2003,hengsberger} Or, for example, the bands might  meet at $N_B \sim$72~BLs (see Fig.~\ref{fig:convgap}) and then behave differently for thicker slabs if the two sides of the film remain coupled.

We have not pursued a fully relativistic calculation for $N_B\sim$72~BLs for several reasons. First of all, it is unfortunately an extremely demanding calculation. But more importantly, we have shown already that the key aspects to understand ARPES measurements of Bi(111) metallic bands is not buried in their behavior at the $M$ point.   Nevertheless, the thickness at which the two states become essentially decoupled may have still a physical  significance, particularly if such thickness  is finite, as suggested by the extrapolation in Fig.~\ref{fig:convgap}.

We have found, at least for calculations without SOC, that there might be indeed a crossover thickness between the ultra-thin regime ---in which the two sides of the film are coupled--- and a possible regime in which they are not coupled, which in turn might determine  the behavior of the metallic bands.
Preliminary calculations \emph{without SOC}  for $N_B=39$ and $78$~BLs, show that for inversion-asymmetric slabs, if the two sides are separated by 15.5~nm or less, the metallic branches split and the Kramers pair of each branch live on opposite sides of the film coupled (as we have seen in all the above discussion) but, if the two sides are  31~nm apart and are thus nearly degenerate, the Kramers pair of each metallic branch live on the same side of the film. Therefore, in order to fully understand the band structure of macroscopic samples it remains to see whether the SOC reallocates each Kramers pair on opposite sides of the slab or not and whether it increases the splitting of the metallic branches (as it does around $\Gamma$) or leaves it unaffected. Moreover, if the two bands composing each Kramers pair are located on opposite sides of the film only for  films as thin as 15~nm or less, then  the  entanglement of the Kramers pair may  only exists in this regime.

\subsubsection{\label{subsec:hydro} The effect of hydrogen on the metallic states of Bi(111): deviations from the convergent shape of the dispersion}

As mentioned in  subsection~\ref{subsec:intrometallicbranches}, previous investigations of the electronic structure have resorted to place a  monolayer of atomic hydrogen on one side of a thin (11~BLs) Bi(111) slab with the aim of
imposing a degeneracy at $M$,~\cite{chulkov2008}  decoupling the two sides of a 10-BL Bi slab~\cite{chulkov2004,chulkov2008} and/or   simulating a semi-infinite surface.~\cite{chulkov2008}
Therefore, the calculated band structure currently considered to simulate that of a macroscopic Bi(111) sample is that of the "clean side" of a 11-BL Bi(111) film   with broken inversion-symmetry via the adsorption of atomic H  on the other side of the film, the hydrogen "host" surface. We have seen that no reinforcement of this symmetry is  required, for which artificially breaking the symmetry of the slab is not justified \emph{a priori}. But moreover, we shall see that H-adsorption is a very strong perturbation for a slab of only 11~BLs, as it yields a band structure that is actually more deviated from the  calculated converged and experimental ones than the same 11~BLs slab without hydrogen. This result can be anticipated from the analysis of the structural relaxations in subsection~\ref{subsec:strucHbi111}, which showed that H not only breaks the inversion symmetry of Bi(111) films but is  able to significantly change the geometric structure of the "clean" side that is $\sim$~5~nm apart.

In order to investigate whether adsorbing hydrogen on one side of a film is meaningful to simulate the band structure of the (111) surface of macroscopic samples~\cite{chulkov2004} and/or useful to reach the convergent band structure of a semi-infinite slab, we have  obtained the dispersion of the metallic surface-state branches around $E_F$ of the "clean" surface of inversion-asymmetric Bi(111) films that are adsorbed with one monolayer atop H at 2~{\AA} from the "host" surface. We have investigated three  thicknesses: 6, 12, and 21 BLs. Our results are shown in Figure~\ref{fig:surfelecH} and compared with  the metallic surface states of the Bi(111) surface of an inversion-symmetric film of 39~BLs (equivalent to $\sim$15.5~nm, the thickest slab we tackled).

Our results show  that breaking the inversion symmetry by H adsorption  does not help to converge the dispersion of the metallic branches along  $\overline{\Gamma M}$. On the contrary, it makes it worse. It  causes that the two  bands reach  the $M$ at a much lower energy than the convergent range. That indeed predicts that the two metallic branches of a macroscopic slab indeed overlap with the bulk band. Moreover, H-passivation for films of 12~BLs or less perturbs the dispersion at regions  where the bands of the corresponding inversion-symmetric clean slabs were already converged (\emph{i.e} regions that are  no longer affected by an increase of the thickness). Specifically,  by comparing  the dispersion of slabs of 7 and 27~BLs (Fig.~\ref{fig:surfelec}), one sees that 7~BLs are enough to reproduce the hole-pocket that appears within the second fifth of the $\overline{\Gamma M}$ line and also the dispersion of the upper surface band within the first fifth of  $\overline{\Gamma M}$. In turn, H on a 6-BL film (Fig.~\ref{fig:surfelecH}) changes significantly the diameter of the hole-pocket and even deviates the dispersion of the upper surface state within the first fifth of $\overline{\Gamma M}$. Then, for a clean 12-BL film, both hole-pockets are well converged and everything within the first three fifths of $\overline{\Gamma M}$. Upon H-passivation, however, the small hole-pocket is shrunken and the diameter of the bigger pocket is also slightly shrunken from the left side. For a clean 21-bilayer slab, the convergence is perfect up to the resolution of the figure within the first four fifths of the $\overline{\Gamma M}$ yet, even at  that thickness, the effect of H  does not completely vanish  within the latter  segment. More importantly, our calculations shows that the band structure of 12~BLs adsorbed on one side by H also strongly deviates from  experimental results, particularly from the midpoint of $\overline{\Gamma M}$ to $M$.~\cite{hengsberger,AstMpoint2003} Specifically, the comparison between measurements and calculations is worsened in both the energy of the surface states and in localization: Both metallic branches influenced by H indeed overlap with the bulk band situated below the A7-distortion gap, 140~meV below $E_F$, whereas the experiment shows that the two branches ---though not clearly resolved near $M$--- never cross. The H-influenced band structure of Bi(111) also misrepresents the band structure of macroscopic samples in terms of localization because a 12-BL film with H  yields metallic surface states that  are  strongly localized all the way up to the $M$-point (see \emph{e.g} Fig. of Ref.~\onlinecite{chulkov2008}), which is also not representative of the experiments performed on macroscopic samples,~\cite{AstMpoint2003,hengsberger} as discussed in subsection~\ref{subsec:ssevo}. Moreover, the influence of H on the "clean" surface is such that quenches the surface states arising in the SOC gap at $\sim$0.7~eV below $E_F$. In summary, the band structure of a Bi(111) film of 21~BLs (or less) influenced by H adsorption is a much worse approximation of the calculated converged and experimental band structure than that of a clean inversion-symmetric 12-BL film.

Note that the metallic bands of the "clean" surface of H/Bi(111) in Fig.~\ref{fig:surfelecH} \emph{do} split because of the Rashba effect; there is a total depletion of the metallic states of the H-host surface. In fact, one can see that the bands indeed meet at the $M$ point. We shall discuss the Rashba splitting in more detail the next section.

\section{\label{sec:resrashba}Results: Rashba splitting of Bi(111) metallic states due to three perturbations}

We have shown that the origin of the observed splitting between the two metallic branches corresponds to  the lifting of the $m_j=\pm1/2$--$m_j=\pm3/2$ degeneracy by the SOC. This does not mean that the Rashba splitting is not possible but that it is just not intrinsic to Bi(111) or to surfaces \emph{per se}. In this section we shall discuss the Rashba splitting of the two metallic branches of Bi(111) produced  by three different perturbations to the inversion symmetry.

\subsubsection{\label{subsec:rashbastr} The Rashba splitting by two structural perturbations}

To test the true Rashba-type splitting and compare with experiments on ultrathin films,~\cite{chulkov2006} we have broken the inversion symmetry of a 12~BL Bi slab via (1) increasing by 0.3~{\AA} the relaxed inter-bilayer distance between the surface and subsurface bilayers of one side of the slab ($d_{23}$); and (2) increasing  by 0.1~{\AA} the relaxed inter-layer distance of the surface bilayer ($d_{12}$). We have chosen these perturbations because the Si(111) substrate of the Bi films in the experiments of Refs.~\onlinecite{chulkov2007,rashba10} most likely  causes a  weak perturbation on the Bi film  because (1) the grown Bi film has not dangling bonds and (2) Bi atoms have slightly higher electronegativity that the  atoms of the semiconducting Si substrate, for which no significant ionic or covalent bond between the Si substrate and the Bi film is expected.

The resulting electronic structure after each structural perturbation is shown in Figure~\ref{fig:interintrabi} and compared with that of an inversion-symmetric film. As expected, both structural perturbations cause  that the two metallic branches split and display a total of four non-degenerate bands.
That is indeed the Rashba splitting because it lifts the  Kramers degeneracy.
Not surprisingly, the metallic surface-state branches of the perturbed surface are much less  sensitive to the perturbation of the weak inter-bilayer distance (Fig.~\ref{fig:interintrabi}(top)) than to the perturbation of the relatively strong covalent intra-bilayer distance (Fig.~\ref{fig:interintrabi}(bottom)). In turn, this correlates with the fact that the Rashba splitting is larger in the latter case.

There are several remarks that ought to be done about the effect of these perturbations. Namely, note that (1) although both of the above structural perturbations clearly break inversion symmetry, they do not empty the metallic states of the perturbed surface and thus does not affect significantly the metallic bands of the other side; (2) comparison between  the metallic bands of the unperturbed surface of the inversion-\emph{asymmetric} slab and those of the surface of the inversion-symmetric slab (see \emph{e.g} Fig.~\ref{fig:interintrabi}(bottom)) shows that they are almost identical; (3) in both inversion-asymmetric films, the resulting Rashba splitting certainly satisfies the degeneracy at $M$, required by translational symmetry; (4) Kramers pairs lie mostly on opposite sides of thin films, as mentioned before.
These four points are important to explain the experimental observations  in Refs.~\cite{chulkov2006,chulkov2007,rashba10}.
Specifically, for a perturbation as those shown in Fig.~\ref{fig:interintrabi}, the $M$-degeneracy would be undetectable  because (1) the Kramers pair of each measured band is primarily localized on the other side and (2)  near $M$ ---\emph{i.e.}, the only small region where the two Kramers pairs could be detected one a single side of the film (see subsection~\ref{subsec:magmom_discon}), the Rashba splitting is too small to be resolved with the available resolution. In fact, that explains the measurements in Refs.~\onlinecite{chulkov2007,rashba10}:  two states of opposite spin polarization are detected at the same energy (within experimental resolution).

Importantly as well, an experiment  performed on the unperturbed surface of a film bearing such kind of perturbations on the other side would not be able to  (1) distinguish whether the sample has inversion symmetry or not, certainly not based on the splitting at $M$ and (2) detect the Rashba splitting. In other words,  the splitting of the metallic branches is not of the Rashba type neither far from $M$ (as proposed in Ref.~\onlinecite{rashba10}) nor close to it. Moreover, it is erroneous to attach inversion symmetry to systems in which the degeneracy at $M$ is not seen.~\cite{rashba10} Therefore, the idea that interaction between the two side of a film makes the metallic branches "electronically inversion symmetric"~\cite{rashba10} is not well based. Our calculations suggest that as long as the  perturbation let the states of the perturbed surface to be occupied, the experiment will keep detecting the splitting ($\Delta E(N_B)$) at the $M$, even though the degeneracy at $M$ is satisfied as shown in Fig.~\ref{fig:interintrabi}.  Interestingly, though, the Rashba splitting for  relatively mild perturbations could be measured if the two sides of a perturbed film were investigated. We shall see, however, that this condition is not necessary if the perturbation has a different nature, as we shall see in the following subsection.

\subsubsection{\label{subsec:rashbahydro} The Rashba splitting by H-adsorption perturbation}

Remarkably, the
earliest DFT work to understand the band structure of  Bi(111) of clean macroscopic samples~\cite{ast1prl} obtained ---not an intrinsic lifting of the spin degeneracy of the states of Bi(111)--- but the strong Rashba splitting,  \emph{specific to the depletion of the metallic surface states on the "H-host" surface of ultra-thin films}, of  the only metallic surface-state Kramers pair that remains occupied after H adsorption (which is situated on the "clean" surface of H/Bi(111)). Indeed, unlike the doubly-degenerate  metallic bands of the clean and inversion-symmetric Bi(111), the only two occupied metallic bands (see Fig.~\ref{fig:surfelecH}) that are on  the "clean" side of the H/Bi(111) film are singly degenerate.
We say that this Rashba splitting is specific to the depletion of the surface states from one side of the film and not to H because other species might have the same effect. In fact, note that the Rashba splitting may be large but it is apparently always circumscribed within  the limits of the $m_j=\pm1/2$--$m_j=\pm3/2$ splitting.

Thus, adsorption of H in Refs.~\cite{chulkov2004,chulkov2008} and in Fig.~\ref{fig:surfelecH} certainly not merely breaks the inversion symmetry, as other ways of breaking the symmetry do not produce such a  strong effect  (see also subsection~\ref{subsec:hydro}). On the contrary, the structural perturbations described earlier, although significantly splitting the magnetic-quantum-number (spin) degeneracy, cause so little change in the electronic structure of the unperturbed surface  that  experiments cannot tell whether the film has or not inversion symmetry. H-adsorption also not merely decouples the two sides of a film, as macroscopic Bi samples~\cite{AstMpoint2003} have surely the two sides essentially decoupled, and yet display metallic bands whose dispersion is closer to that of our inversion-symmetric 15-nm film. H-adsorption does break the symmetry and "decouples" (if we may call it that way) the two sides of a thin film, but such "decoupling" is actually a total depletion of the metallic states of the H-host surface. H essentially takes charge from the host Bi(111) surface, for which  the metallic states of H-host surface become unoccupied and pushed above $E_F$. Of course, in this profoundly different scenario of inversion asymmetry, the way the system satisfies  translation symmetry (the $M$ degeneracy) is also fundamentally different. Namely, unlike the  inversion asymmetries that do not involve charge donation, sharing or polarization,  the H-adsorption-induced asymmetry depletes the metallic states of the host surface, for which (1) there is only one split Kramers pair occupied and (2) it is not formed on opposite sides of the slab but on the same side. Therefore, for this type of inversion asymmetry, one should really expect to find the degeneracy at $M$ via ARPES on a single side of the film. These results suggest that, depending on the nature of the perturbation that breaks the symmetry, the splitting of the Kramers pairs may be found on opposite sides of the slab (as in the structural perturbations in subsection~\ref{subsec:rashbastr}) or on the same side (if there is charge depletion of the surface states).

This way of producing  Rashba splitting is  not alien to those used in semiconductors. The charge depletion of the H-host surface creates a scenario similar to, for example, that in which an inversion charge-density layer is created on the surface of semiconductors  by an external electric fields~\cite{marques,bangert} while the other side undergoes charge depletion.
Therefore, should such strong perturbation and the charge the reallocation within the film be possible by mere chemisorption, it could be more interesting to study such systems than pristine Bi(111), in part also because in  H/Bi(111) films, the only one Kramers pair occupied is localized on the "clean" side and there might be a stronger  localization of the metallic states on the clean bilayer of a H/Bi(111) film that in the topmost bilayer of an inversion-symmetric Bi(111) film, as mentioned earlier.
In fact, notice that the trend observed in Fig.~\ref{fig:surfelecH} indicates that a sufficiently thick film with  inversion asymmetry caused by H-adsorption should  give states strictly lying in the gap and well localized all along  $\overline{\Gamma M}$.

\section{\label{sec:recross} The electronic density of surface carriers}

As shown in Fig.~\ref{fig:bpdosss}, the largest contribution to the density of states (DOS) of bulk Bi around $E_F$ comes from  states around $\Gamma$. For this reason the interpretation of ARPES measurements of  the metallic surface states around this area is a delicate issue. Namely, even  X-rays of $\sim$~20~eV  kick electrons into the  vacuum that mostly come from layers as deep as $\sim$1nm and probably deeper.~\cite{Shah1979,Damascelli2004} Thus the largest contribution to ARPES measurements around $\Gamma$ may come, not from surface states, but from  states  buried in subsurface layers. For example, while ARPES measurements  have been able to detect many  electronic bands from Bi(111) ultrathin films,~\cite{chulkov2006} not all of them are localized in the first bilayer.

Despite the above reservations, based on ARPES measurements have suggested that the carrier concentration (holes) at the surface ---created by the presence of the metallic surface states--- could be high enough  so as to dominate the transport properties of films whose thickness is less than $\sim$~30~nm ($\sim$~70-80 BLs)   ---even if only the Fermi surface elements around $\Gamma$ are taken into account.~\cite{ast1prl} One reason supporting this prediction may be that the contribution from bulk states to the DOS at  $E_F$  is reduced for thin films  because their density becomes sparser with decreasing thickness (see Fig.~\ref{fig:surfelec}). Another reason is that the surface carriers velocity at $E_F$ is significantly lower than that of the bulk carriers, for which the local density of states (LDOS) of surface atoms at $E_F$ should be larger than that of an atom in the bulk.~\cite{hofmann} In order to confirm the latter assertion, however, one must calculate the density of electronic states (DOS) around the Fermi level, which takes into account the entire SBZ. Thus, we provide a quantitative  assessment of the concentration of carriers by comparing the LDOS of the surface atom ---obtained from the converged electronic structure of a slab of 39~BLs--- and the LDOS of a bulk atom. Our results in Fig.~\ref{fig:dosatom} show that indeed, the density of  carriers at $E_F$ in a single surface atom is larger than that in a single bulk atom. Our results, nevertheless, indicate that the LDOS of a surface atom is  at most 15 times larger than that of a bulk atom. Thus, at $\sim$~30~nm, the  DOS of at least $\sim$140 bulk-like atoms would be overwhelmingly larger than that of the surface atoms. In fact, as shown in Figure~\ref{fig:dosfilm}(a) the DOS of the topmost bilayer is only $\sim$17\% of the total DOS of a 15.5~nm film.

Still, one could think that the carrier concentration of the topmost bilayer could dominate the  transport properties of thin films only if they are thinner than $\sim$6nm (15 BLs).
In fact, yet another reason to think that the carrier concentration at the surface could dominate the transport properties of ultrathin films might be that the energy of the high-energy metallic branch approaches the Fermi level as the thickness decreases below $\sim$6nm (15 BLs), as suggested in Fig.~\ref{fig:surfelec}. For example, for films thinner than $\sim$2nm (6-7 BLs), the high-energy band is  above the Fermi level for the most part along the $\overline{\Gamma M}$ line, except very close to  $\Gamma$ (see Fig.~\ref{fig:surfelec}).
 A qualitative assessment of the density of  carriers at the surface is actually  not  straightforward for ultrathin films. Namely, while decreasing the thickness causes that  (1) the density of bulk-like states at $E_F$  becomes increasingly sparser around $\Gamma$ and (2) the surface states get significantly closer to the Fermi level (for films thinner than $\sim$6nm) in a relatively large vicinity around $M$, the spatial localization of the metallic  branches at the film surface decreases exactly in that vicinity around $M$ (see \emph{e.g} Fig.2 in Ref.~\onlinecite{chulkov2008}).
Indeed,  Figs.~\ref{fig:dosfilm}(b) and (c)  compare the total electronic DOS of films of thickness of 6 and 3.2~nm versus the LDOS at the topmost bilayer and show that even for films of 3.2nm (8 BLs)  the local DOS of states  at $E_F$ from the topmost bilayer is not more than 20\%.  In other words, according to our calculations, even in 8-BL films and although detectable by ARPES,  $\sim$60\% of the carriers are localized in the six inner bilayers of the film.

The reason why most carriers are not in the surface layer \emph{and} are detectable by ARPES is that
the contribution from subsurface layers is much larger than that from bulk-like atoms. In fact, knowing from the analysis of the geometric structure that bulk-like atoms are found as deep as in the fifth bilayer, one can estimate from Fig.~\ref{fig:dosfilm}(a) that, for a 15.5-nm film,
each of the two surface bilayers,
each of the six subsurface bilayers,
and each of the 31 bulk bilayers
contributes by 0.40($\sim$17\%), $\sim$0.18($\sim$7\%), and 0.02($\sim$0.8\%) states per two atoms per eV, respectively, to the total DOS at $E_F$.
Then, as expected from the  increasing delocalization of the metallic states with decreasing thickness, for films of 3.2~nm, the contribution from each subsurface bilayer increases  and that of each surface bilayers decreases to  $\sim$0.21 and $\sim$0.28 states per two atoms per eV, respectively. In summary, our calculations suggest that the carrier concentration of one surface alone (the topmost bilayer, for instance) cannot outweigh that of the bulk or subsurface bilayers practically at any thickness. Yet the carrier concentration  at $E_F$  from  atoms within a depth of $\sim$1.4~nm can certainly outweigh the contribution of the rest of a film if the latter is thinner than $\sim$18~nm.

\section{\label{sec:discussion}Discussion}

Our results have shown that the association between the spin-orbit coupling and various observations in Bi(111) has been based on uncertain and unproved premises.  Jezequel, for example, first acknowledged that the  existence of the surface states is, as in all covalently-bonded materials, only due to the abrupt termination of the (periodic) crystal potential but at the same time highlighted that the  energy range of the surface states is necessarily governed by the SOC.~\cite{jezequel} Our calculations indicate that  this is  not the case for the metallic surface states of Bi(111) found by Hengsberger \emph{et al}.\cite{hengsberger}
Unfortunately, the "spin-orbit" tag seems to have been  imparted  to the A7-distortion pseudogap around the Fermi level
to the point that its very appearance in the (111)-projected  bulk band and thus that of the  metallic surface states
have been  ascribed to the spin-orbit coupling,~\cite{FuKane2007, Kane2008, roushan} notwithstanding that the  works Gonze~\cite{gonze1990} and Shick~\cite{shick}  clearly  confirmed that the existence of the  gap  around the Fermi level along the $\overline{\Gamma M}$ direction  is independent of the spin-orbit coupling but driven by the $u$-deformation,  reason for which it is present in the same degree in Sb and even in the lightest group-V semimetal, As.

It has been alleged and accepted  that there are two energetically distinct branches and not one is because these two branches result from a strong spin or Rashba-type splitting, i.e., from lifting the spin (or Kramers) degeneracy that occurs in the absence of inversion centers.~\cite{rashba10,chulkov2004,chulkov2008,HasanKane} The splitting has in fact been tagged as a "spin splitting" or "lift of the spin degeneracy" even though the spin  and  orbital momentum are not autonomous and the essence of all these discussions~\cite{chulkov2004,chulkov2006,chulkov2008,hofmann,hsieh,HasanKane}, including those related to the theory of topological insulators, is the  particulary large role of the SOC in heavy atoms such as Bi. But more importantly, although the splitting is dramatically enhanced for short wavelengths  by the SOC, it is patently questionable that it is of the Rashba-type in the light of the experimental evidence and all  computational studies so far.~\cite{chulkov2008,chulkov2006,chulkov2007}  Instead, we have provide some evidence that the splitting corresponds to the SOC $m_j=\pm1/2$--$m_j=\pm3/2$ splitting.

As Rashba pointed out, it is only in the absence of  inversion centers in a crystal that the SOC splits the electronic bands that otherwise would be doubly degenerated (Kramers degeneracy).~\cite{rashba1965} Some materials have crystal structures with no inversion centers. Yet, in a inversion-symmetry material such as Bi, some external factor must break the inversion center of a Bi(111) film. In the case of pristine Bi(111), however, that the lack of inversion symmetry is guaranteed in
macroscopic samples because they are equivalent to system with a  semi-infinite number of layer, and then that the latter is equivalent to a two-sided ultra-thin film having the surface states of one side depleted by H-adsorption~\cite{chulkov2006,chulkov2008} are arguments not supported  by measurements. The first argument fails because, although the surface is not an inversion center, it does not mean that the system is locally asymmetric.~\cite{chulkov2004,chulkov2007}  Namely, the inversion symmetry or any other symmetry refers to the entire system and not to a particular surface.

The SOC-driven Rashba-type splitting for surface states was first reported for the $sp$-derived surface states of Au(111) that are found around the Fermi level.~\cite{LaShell}
The inference that the splitting is of the Rashba-type was based on the idea that creating a surface on Au(111) guarantees automatically the elimination of inversion centers. The latter is not necessarily true, but it is  likely that a sample of Au displaying a (111) surface, which needs to be prepared  by extensive sputter anneal cycles,~\cite{LaShell} has no inversion centers. The situation, however, is  not the same for Bi(111)  because, while its bilayers easily cleave, the intra-bilayer bonds are quite stable.~\cite{jona} This was in fact confirmed in experiments on Bi(111) thin films.~\cite{chulkov2006} Still, it remains to investigate whether the splitting for Au(111) is  a SOC $m_j=\pm1/2$--$m_j=\pm3/2$ splitting.

That the metallic bands of inversion-symmetric films are doubly degenerate and primarily localized on opposite sides of the film is a known~\cite{chulkov2006,chulkov2008} and expected result. Namely, as mentioned in the introduction (Subsection~\ref{subsec:intrometallicbranches}), the inversion symmetry center of the slab and the time-reversal symmetry demand that \emph{all} states in the irreducible surface Brillouin zone must be doubly degenerate. What has not yet been acknowledged is that, because each branch is a pair of degenerate bands, the degeneracy at $M$ --- demanded by the translational symmetry (see Subsection~\ref{subsec:intrometallicbranches})--- is \emph{automatically} satisfied despite the branches  split. Therefore,  neither the time-reversal, the translational or the inversion symmetry requires that these two observed metallic branches meet at $M$ for inversion-symmetric and even for certain inversion-asymmetric slabs, as we showed  in section~\ref{sec:resrashba}. Thus, although the coupling between the two sides of the film  indeed enhances the splitting of the bands at $M$ (see Fig.~\ref{fig:surfelec}), one cannot argue that such coupling lifts an expected degeneracy at $M$,~\cite{rashba10,chulkov2008} because the degeneracy is already satisfied. Moreover, we have shown that an experiment  performed on a inversion-asymmetric  thin film bearing a "mild" perturbation (at least one that does not deplete the metallic states on the other side) would still find the splitting at the $M$ point and would not be able to distinguish whether the sample has inversion symmetry or not. Therefore,  one cannot say that the interaction between the two sides of a film makes the metallic branches "electronically inversion symmetric."~\cite{rashba10}

Let us believe for a moment that the two branches are the result of a Rashba splitting, therefore their splitting must arise from the combination of two factors: the SOC and the lack on inversion symmetry. If that were the case, because of the time-reversal symmetry and the translational in-plane symmetry of the hexagonal surface Brillouin zone, the branches must become degenerate at $\Gamma$ and $M$, respectively. However, computations from slabs with inversion symmetry~\cite{chulkov2008} and measurements~\cite{chulkov2006,chulkov2007,hengsberger} performed in supported thin Bi(111) films (i.e., the inversion symmetry is broken by substrate) or in macroscopic samples do not provide evidence of that. While the calculations and  measurements  do show the splitting, in neither case the expected degeneracy at $M$ is satisfied,~\cite{chulkov2006,chulkov2008,chulkov2007,hofmann} that is, the branches are always energetically separated by a gap $\Delta E$.  Clearly, if the splitting of the bands were of the Rashba-type, that the supposedly Rashba-split bands split all the way up to $M$ would be a violation/contradiction of the patent inversion \emph{and} translational symmetry existing at least in periodic supercell calculations. Rather than addressing that, Koroteev \emph{et al}.~\cite{chulkov2004,chulkov2008} attached the inconsistency between symmetry conditions and results to the fact that the Bi slabs used in calculations preserve the inversion symmetry (i.e., the bands are degenerate) and are thus not representative of an ideal semi-infinite surface. So the inversion asymmetry ---necessary for the spin or Rashba splitting, supposedly giving rise to the splitting of the surface states along $\overline{\Gamma M}$, is in fact never fulfilled in clean inversion-symmetric Bi(111) slabs, and yet  the strong splitting near $\Gamma$ was obtained.

The computational manoeuvre that followed, to apparently satisfy the translational symmetry condition, was to actually break the symmetry   by adsorbing a monolayer of H on one side of the slab.~\cite{chulkov2004,chulkov2006,chulkov2008} Here the concepts of a semi-infinite slab and that of broken inversion symmetry have been  mixed up through consideration of the $M$ point.
Specifically, on the one hand, it was found that increasing the thickness of the slabs (slabs with inversion symmetry) causes the gap between the two surface-state branches at $M$ to decrease.~\cite{chulkov2008} On the other hand, it was indirectly implied that $\Delta E$ vanishes only in the limit of an infinite number of layers (it was stated in Ref.~\onlinecite{chulkov2008} that $\Delta E$ decreases by a factor of 2 when the number of layer is doubled), which incidentally  suggests that the branches would not merge at $M$ even in a macroscopic but finite sample. On the other hand, it was known~\cite{chulkov2004} that the degeneracy at $M$ can be achieved (i.e., $\Delta E$=0)  by adsorbing H. Thus, the logic  applied is to presume H absorbtion as the means to mimic a semi-infinite slab.~\cite{chulkov2008} The equivalency of this two scenarios is rather questionable because the reasons underlying the behavior at $M$ in either case are physically different. Namely, in a  very thick slab the states of the two surfaces are decoupled. However, adsorption of H depletes the states of the H-host surface.

More problematic is the claim that, the supposedly spin-split states   along the $\overline{\Gamma M}$  become non-spin polarized because of the overlap with the bulk band.~\cite{rashba10} First of all, there is no bulk band in thin films. Second, the states do not overlap with anything. Therefore, if the states are still detectable, as they are in ARPES and SARPES, they must preserve their polarization if they are indeed Rashba spin-split bands. Instead, we have shown that bands lying on the "top" surface loose polarization because the Kramers degeneracy is not lifted  and thus the Kramers pair of each band, mostly lying on the "bottom" bilayer, overlaps with  as they gains some localization at the "top" bilayer near $M$. Therefore, the origin of the spin polarization as well as the explanation why it vanishes  must be also address in the discussions of the advances in TIs theory based on spin-resolved ARPES for a Bi$_{(1-x)}$Sb$_x$(111).~\cite{hsieh}

We remark that even though the two branches are resolvable by spin-resolved ARPES,~\cite{chulkov2007} and have opposite $m_j$, none of this contravene  the fact that the Kramers degeneracy can be preserved, and has been "preserved"  at least up to current experimental resolution.
In fact, as shown in our calculations, that the degeneracy at $M$ of the metallic bands has not yet been reported, confirms that the Si-substrate-driven Rashba splitting is weak as the structural perturbation that we have studied and certainly weaker than that produced by the H overlayer.
In this sense, calling this a "spin splitting" in the presence of the Kramers degeneracy (either because of inversion symmetry or a negligible small true Rashba splitting) may not be an adequate to describe this measurements. Specifically, although we may choose to ignore that grown films and sharply cleaved surfaces have two surfaces, the symmetries shaping the band-structure apply   not only to one surface or the surface that is studied experimentally but to both surfaces. Thus, in order to understand the implication of such symmetries (or lack of them)  on the observations, we have to keep in mind both surfaces. For example, for thin-films with inversion symmetry  considering the two sides of the film is essential to  understand how the translation symmetry is fulfilled even in the two metallic branches do not meet at $M$.
Moreover, it is important to be certain whether or not one has achieved  Rashba splitting. For example, if it is  not Rashba splitting, as in the case of the Bi(111) films in Ref.~\onlinecite{rashba10,chulkov2007} the magnetic-moment polarization may simply vanish, \emph{i.e.}, the Kramers degenerate bands cancel each other's spin polarization. In turn, if it is Rashba splitting, then the spin polarization should be preserved, as it is expected for topological insulators.

The efforts to explain the observed splitting at the $M$ point for macroscopic samples and supported ultra-thin films are prodigious.~\cite{hofmann,chulkov2006,chulkov2007,chulkov2008,rashba10} They involves doubts on the interpretation of ARPES measurements at $M$ because of the possible overlapping with the bulk band, hybridization with quantum-well states, creation of other quantum-well states because of the film/substrate interface, the even/odd characterization of the metallic bands as well as their avoided crossing, and even a model of a crystal potential that is somehow inversion-asymmetric  for \emph{in-plane} long-wavelengths states but inversion-symmetric for \emph{in-plane} short-wavelength states. WThe question
 is, Why? What is the significance of the splitting or degeneracy at $M$? The reason is clear: It is  relevant to the origin of the two metallic bands.  Macroscopic samples and supported ultra-thin films must be inversion-asymmetric. However, if the bands are found to split at $M$, that is taken as an indication that  the degeneracy is not broken. That, in turn, means that inversion symmetry is  preserved and, therefore, the splitting is not of the Rashba type. So much so indeed that it has been   acknowledged that near $M$, but only near $M$, the splitting is not of the Rashba type. Nevertheless, we have shown that experiments and calculations coherently complement each other to reveal that all measurements of the metallic bands along $\overline{\Gamma M}$ are likely a $m_j=\pm1/2$--$m_j=\pm3/2$ splitting but certianly not a Rashba splitting.   We have shown that  pristine single Bi crystals and supported ultra-thin films may actually display a tiny Rashba splitting because of weak inversion asymmetries. So, if ARPES experiments do not find that the bands meet at M, it does not means that the inversion symmetry is preserved. It just means that the Rashba-split pair of each band  is not being measured. In other words, it means that what is being measured is not the Rashba splitting.

The  Rashba splitting has not yet been  resolved experimentally on Bi(111), not even in thin films supported on Si(111)~\cite{chulkov2006}
because perhaps it has never been probed and furthermore because of the resolution needed for a weakly interacting substrate. In order to observe the Rashba splitting for such a weak perturbation, the electrons of both sides of the film have to probed  because the $\pm m_j$ degeneracy is between the states lying  on each side of a film.
Finally,  the magnitude of the Rashba splitting depends on the strength and nature of the external perturbation; its magnitude in Bi does not have a well-defined value for each k-point as that of the $m_j=\pm3/2-m_j=\pm1/2$-splitting, yet it  remains to understand whether it is limited by the $m_j=\pm3/2-m_j=\pm1/2$-splitting.

\section{SUMMARY\label{sec:concl} }

We present fully relativistic density-functional-theory calculations of the electronic structure of Bi(111) films to revisit of the interpretation of the metallic surface states of Bi(111) and some of the unresolved issues around thin films and their limit to macroscopic samples.
We  investigate the spin-orbit pseudogaps and surface states in the valence band of the projected  band-structure of bulk Bi and  obtain for the first time some of the actual spin-orbit surface states from first principles.
The evolution of the electronic band structure as a function of thickness indicates that calculations for films as thick as $\sim$9~nm reproduce fairly well the measured band-structure of macroscopic samples within experimental resolution.
Our first-principles calculations show that the metallic surface states near the zone edge (the $M$ point) are not resonances in the sense that they do not overlap with the bulk band at the zone boundary. They lie strictly inside the A7-distortion conduction-valence band gap, which at $M$ is at least 3 times wider than predicted by tight-binding calculations and do not meet at $M$ for films of 15.5~nm. Thus, our results  attest  available  measurements of the metallic surface-state bands on a single surface of a macroscopic sample.~\cite{hengsberger,AstMpoint2003} This also shows that juxtaposing tight-binding calculations of the bulk-projected band against first-principles calculations of the surface states~\cite{chulkov2004,hofmann} has been unsuitable, as well as the conclusions drawn from the assumption that the metallic branches overlap with the bulk band.
Our  calculations together with ARPES experiments~\cite{hengsberger,AstMpoint2003}    show that  the band structure obtained  by adsorbing H on one side of a film only 4~nm (11 BLs)  is not representative of that of a macroscopic sample.
We show that the magnitude of the energy gap $\Delta E$ between the two metallic branches at $M$ as a function of thickness  does not indicate that the two branches become degenerate  in the  limit of  infinite thickness.
On the other hand, we show that, even if the observed branches are  separated by  $\Delta E$,  the degeneracy at $M$ is satisfied, as required by the translation symmetry, for films with or even without inversion symmetry. Thus,  the splitting at $M$ requires no further explanation,~\cite{chulkov2008,rashba10} since  the expected degeneracy may not be accessible in thin films~\cite{chulkov2007} or even macroscopic samples.
Our calculations show that H adsorption does not help the electronic structure to converge to that of thicker slab (of up to 15.5~nm). On the contrary, it rather deviates the  dispersion of the surface states from the found convergent dispersion.
We find that the only reason for which the gap between the two metallic branches at $M$ could have been  worrisome, the magnetic discontinuity and indetermination at $M$, is resolved since the magnetization at $M$ is zero because the Kramers degeneracy is not lifted and the pairs do overlap spatially on the same side of a film near $M$.
We also confirm that  two split metallic branches exists and would be measurable (on a single side of a film) in the presence of inversion symmetry and, for sufficiently thin films, even if the SOC were negligible. The latter splitting is  related to the delocalized character of the surface states since the energy gap between the two metallic  bands at $M$ progressively reduces as the thickness of a film increases. Still, the behavior of the metallic states near $M$ and the effect of the SOC once they are totally decoupled needs further investigation.
However, because the splitting between the two metallic branches, as measured in experiment,  exists in inversion-symmetric films,  the presence of the two metallic bands is  not caused by  a Rashba effect. In other words, it does not lift the Kramers or "spin" degeneracy. In fact, we have shown that if the supposedly "spin"-split metallic bands of Bi(111) experimentally do not merge at the $M$-point~\cite{chulkov2006} proves that, even for supported thin-films, ARPES is simply not detecting the Rashba splitting.
Thus, we show that   considering the two sides of a film is essential to  understand how the translation symmetry is fulfilled even if the two metallic branches do not meet at $M$,    why the magnetic-moment discontinuity is not a concern, and why the magnetic-moment polarization of each side of the film vanishes around $M$.~\cite{chulkov2007}
We show  that the presence of the two metallic surface-state branches can be understood as a SOC-induced $m_j=\pm3/2-m_j=\pm1/2$-splitting.
In addition, we  provide a rationale for the observed  opposite spin-texture of the two  metallic bands, in light of the fact that they cannot be attached to a broken Kramers degeneracy and  that they are not correlated. Specifically,  each surface atom retains essentially unaltered the relatively strong covalent bonds that any bulk atom has, which cannot afford magnetization. Thus the two bands localized at the surface, although  non-correlated, must have opposite spin.
We show that lifting of the Kramers degeneracy (Rashba splitting)   takes place only if the inversion symmetry is broken and its magnitude is dependent upon the strength of the perturbation that breaks it. In other words, the Rashba splitting is, in general, not intrinsic to  surfaces  but to the perturbation imposed on the sample to which they belong. Thus, even though the surface is not an inversion center, the sample may still have inversion symmetry and thus the Kramers degeneracy must hold.
Our calculations also illustrate that the Rashba-splitting  is  in general dependent on the magnitude of the perturbation. We show that for structural perturbations that do not involve charge polarization, donation or sharing, the Rashba splitting is so small that it could not be detected within the current experimental resolution, which  explain the measurements in Refs.~\onlinecite{chulkov2006,chulkov2007}.
In turn, the band structure yielded by H/Bi(111), although not representative of that of pristine Bi(111), is representative  of the Rashba splitting caused by the strong perturbation imparted by H adsorption, which depletes the metallic states of one side of the Bi film.
Finally, we find that, as a result of the variations in the electronic structure as a function of thickness, the highest valence band re-crosses the Fermi level and creates extra electron pockets for films $\sim$3-nm thick. Still, the surface-states bands  do not dominate the DOS at $E_F$ even at this thickness, in part because thickness has a strong effect on the localization of the metallic states. However, we find that the carrier concentration  at $E_F$  of  atoms within a depth of $\sim$1.4~nm can certainly outweigh the contribution of the rest of a film if the latter is thinner than $\sim$18nm.

\begin{acknowledgments}
This work was supported by the Karlsruher Institut f\"ur Technologie (KIT), Germany. Computations
were performed at the Institut f\"ur Festk\"orperphysik of KIT, the Stokes HPCC facility at UCF Institute for Simulation and Training (IST) and local machines at UCF. MAO is indebted to Sergey Stolbov for countless insightful discussions and computational support.

\end{acknowledgments}

\bibliography{bi111_elect}

\begin{thebibliography}{79}
\expandafter\ifx\csname natexlab\endcsname\relax\def\natexlab#1{#1}\fi
\expandafter\ifx\csname bibnamefont\endcsname\relax
  \def\bibnamefont#1{#1}\fi
\expandafter\ifx\csname bibfnamefont\endcsname\relax
  \def\bibfnamefont#1{#1}\fi
\expandafter\ifx\csname citenamefont\endcsname\relax
  \def\citenamefont#1{#1}\fi
\expandafter\ifx\csname url\endcsname\relax
  \def\url#1{\texttt{#1}}\fi
\expandafter\ifx\csname urlprefix\endcsname\relax\def\urlprefix{URL }\fi
\providecommand{\bibinfo}[2]{#2}
\providecommand{\eprint}[2][]{\url{#2}}

\bibitem[{\citenamefont{Shoenberg}(1939)}]{shoenberg}
\bibinfo{author}{\bibfnamefont{D.}~\bibnamefont{Shoenberg}},
  \bibinfo{journal}{Proc. Roy. Soc. A. Math. and Phys. Sci.}
  \textbf{\bibinfo{volume}{170}}, \bibinfo{pages}{341} (\bibinfo{year}{1939}).

\bibitem[{\citenamefont{J{\'e}rome et~al.}(1967)\citenamefont{J{\'e}rome, Rice,
  and Kohn}}]{jerome1967}
\bibinfo{author}{\bibfnamefont{D.}~\bibnamefont{J{\'e}rome}},
  \bibinfo{author}{\bibfnamefont{T.~M.} \bibnamefont{Rice}}, \bibnamefont{and}
  \bibinfo{author}{\bibfnamefont{W.}~\bibnamefont{Kohn}},
  \bibinfo{journal}{Phys. Rev.} \textbf{\bibinfo{volume}{158}},
  \bibinfo{pages}{462} (\bibinfo{year}{1967}).

\bibitem[{\citenamefont{Brandt and Chudinov}(1972)}]{brandt1972}
\bibinfo{author}{\bibfnamefont{N.~B.} \bibnamefont{Brandt}} \bibnamefont{and}
  \bibinfo{author}{\bibfnamefont{S.~M.} \bibnamefont{Chudinov}},
  \bibinfo{journal}{J. Low temperarure Phys.} \textbf{\bibinfo{volume}{8}},
  \bibinfo{pages}{339} (\bibinfo{year}{1972}).

\bibitem[{\citenamefont{Miura et~al.}(1982)\citenamefont{Miura, Hiruma, Kido,
  and Chikazumi}}]{miura1982}
\bibinfo{author}{\bibfnamefont{N.}~\bibnamefont{Miura}},
  \bibinfo{author}{\bibfnamefont{K.}~\bibnamefont{Hiruma}},
  \bibinfo{author}{\bibfnamefont{G.}~\bibnamefont{Kido}}, \bibnamefont{and}
  \bibinfo{author}{\bibfnamefont{S.}~\bibnamefont{Chikazumi}},
  \bibinfo{journal}{Phys. Rev. Lett.} \textbf{\bibinfo{volume}{49}},
  \bibinfo{pages}{1339} (\bibinfo{year}{1982}).

\bibitem[{\citenamefont{Weitzel and Micklitz}(1991)}]{weitzel1991}
\bibinfo{author}{\bibfnamefont{B.}~\bibnamefont{Weitzel}} \bibnamefont{and}
  \bibinfo{author}{\bibfnamefont{H.}~\bibnamefont{Micklitz}},
  \bibinfo{journal}{Phys. Rev. Lett.} \textbf{\bibinfo{volume}{66}},
  \bibinfo{pages}{385} (\bibinfo{year}{1991}).

\bibitem[{\citenamefont{Hengsberger et~al.}(2000)\citenamefont{Hengsberger,
  Segovia, Garnier, Purdie, and Baer}}]{hengsberger}
\bibinfo{author}{\bibfnamefont{M.}~\bibnamefont{Hengsberger}},
  \bibinfo{author}{\bibfnamefont{P.}~\bibnamefont{Segovia}},
  \bibinfo{author}{\bibfnamefont{M.}~\bibnamefont{Garnier}},
  \bibinfo{author}{\bibfnamefont{D.}~\bibnamefont{Purdie}}, \bibnamefont{and}
  \bibinfo{author}{\bibfnamefont{Y.}~\bibnamefont{Baer}},
  \bibinfo{journal}{Eur. Phys. J. B} \textbf{\bibinfo{volume}{17}},
  \bibinfo{pages}{603} (\bibinfo{year}{2000}).

\bibitem[{\citenamefont{Ast and H{\"o}chst}(2001)}]{ast1prl}
\bibinfo{author}{\bibfnamefont{C.~R.} \bibnamefont{Ast}} \bibnamefont{and}
  \bibinfo{author}{\bibfnamefont{H.}~\bibnamefont{H{\"o}chst}},
  \bibinfo{journal}{Phys. Rev. Lett.} \textbf{\bibinfo{volume}{87}},
  \bibinfo{pages}{177602} (\bibinfo{year}{2001}).

\bibitem[{\citenamefont{Hofmann}(2006)}]{hofmann}
\bibinfo{author}{\bibfnamefont{P.}~\bibnamefont{Hofmann}},
  \bibinfo{journal}{Prog. Surf. Sci.} \textbf{\bibinfo{volume}{81}},
  \bibinfo{pages}{191} (\bibinfo{year}{2006}).

\bibitem[{\citenamefont{Vossloh et~al.}(1998)\citenamefont{Vossloh, Holdenried,
  and Micklitz}}]{vossloh1998}
\bibinfo{author}{\bibfnamefont{C.}~\bibnamefont{Vossloh}},
  \bibinfo{author}{\bibfnamefont{M.}~\bibnamefont{Holdenried}},
  \bibnamefont{and} \bibinfo{author}{\bibfnamefont{H.}~\bibnamefont{Micklitz}},
  \bibinfo{journal}{Phys. Rev. B} \textbf{\bibinfo{volume}{58}},
  \bibinfo{pages}{12422} (\bibinfo{year}{1998}).

\bibitem[{\citenamefont{Muntyanu et~al.}(2008)\citenamefont{Muntyanu, Gilewski,
  Nekov, Zaleski, and Chistol}}]{muntyanu2008}
\bibinfo{author}{\bibfnamefont{F.~M.} \bibnamefont{Muntyanu}},
  \bibinfo{author}{\bibfnamefont{A.}~\bibnamefont{Gilewski}},
  \bibinfo{author}{\bibfnamefont{K.}~\bibnamefont{Nekov}},
  \bibinfo{author}{\bibfnamefont{A.}~\bibnamefont{Zaleski}}, \bibnamefont{and}
  \bibinfo{author}{\bibfnamefont{V.}~\bibnamefont{Chistol}},
  \bibinfo{journal}{Sol. State Comm.} \textbf{\bibinfo{volume}{147}},
  \bibinfo{pages}{183} (\bibinfo{year}{2008}).

\bibitem[{\citenamefont{Tian et~al.}(2008)\citenamefont{Tian, Kumar, and
  Chan}}]{mingliang2008}
\bibinfo{author}{\bibfnamefont{M.}~\bibnamefont{Tian}},
  \bibinfo{author}{\bibfnamefont{N.}~\bibnamefont{Kumar}}, \bibnamefont{and}
  \bibinfo{author}{\bibfnamefont{M.~H.~W.} \bibnamefont{Chan}},
  \bibinfo{journal}{Phys. Rev. B} \textbf{\bibinfo{volume}{78}},
  \bibinfo{pages}{045417} (\bibinfo{year}{2008}).

\bibitem[{\citenamefont{Koroteev et~al.}(2004)\citenamefont{Koroteev,
  Bihlmayer, Gayone, Chulkov, Bl{\"u}gel, Echenique, and
  Hofmann}}]{chulkov2004}
\bibinfo{author}{\bibfnamefont{Y.}~\bibnamefont{Koroteev}},
  \bibinfo{author}{\bibfnamefont{G.}~\bibnamefont{Bihlmayer}},
  \bibinfo{author}{\bibfnamefont{J.~E.} \bibnamefont{Gayone}},
  \bibinfo{author}{\bibfnamefont{E.}~\bibnamefont{Chulkov}},
  \bibinfo{author}{\bibfnamefont{S.}~\bibnamefont{Bl{\"u}gel}},
  \bibinfo{author}{\bibfnamefont{P.}~\bibnamefont{Echenique}},
  \bibnamefont{and} \bibinfo{author}{\bibfnamefont{P.}~\bibnamefont{Hofmann}},
  \bibinfo{journal}{Phys. Rev. Lett.} \textbf{\bibinfo{volume}{93}},
  \bibinfo{pages}{046403} (\bibinfo{year}{2004}).

\bibitem[{\citenamefont{Hirahara et~al.}(2006)\citenamefont{Hirahara, Nagao,
  Matsuda, Bihlmayer, Chulkov, Koroteev, Echenique, Saito, and
  Hasegawa}}]{chulkov2006}
\bibinfo{author}{\bibfnamefont{T.}~\bibnamefont{Hirahara}},
  \bibinfo{author}{\bibfnamefont{T.}~\bibnamefont{Nagao}},
  \bibinfo{author}{\bibfnamefont{I.}~\bibnamefont{Matsuda}},
  \bibinfo{author}{\bibfnamefont{G.}~\bibnamefont{Bihlmayer}},
  \bibinfo{author}{\bibfnamefont{E.}~\bibnamefont{Chulkov}},
  \bibinfo{author}{\bibfnamefont{Y.~M.} \bibnamefont{Koroteev}},
  \bibinfo{author}{\bibfnamefont{P.~M.} \bibnamefont{Echenique}},
  \bibinfo{author}{\bibfnamefont{M.}~\bibnamefont{Saito}}, \bibnamefont{and}
  \bibinfo{author}{\bibfnamefont{S.}~\bibnamefont{Hasegawa}},
  \bibinfo{journal}{Phys. Rev. Lett.} \textbf{\bibinfo{volume}{97}},
  \bibinfo{pages}{146803} (\bibinfo{year}{2006}).

\bibitem[{\citenamefont{Hirahara et~al.}(2007)\citenamefont{Hirahara, Miyamoto,
  Matsuda, Kadono, Kimura, Nagao, Bihlmayer, Chulkov, Qiao, Shimada
  et~al.}}]{chulkov2007}
\bibinfo{author}{\bibfnamefont{T.}~\bibnamefont{Hirahara}},
  \bibinfo{author}{\bibfnamefont{K.}~\bibnamefont{Miyamoto}},
  \bibinfo{author}{\bibfnamefont{I.}~\bibnamefont{Matsuda}},
  \bibinfo{author}{\bibfnamefont{T.}~\bibnamefont{Kadono}},
  \bibinfo{author}{\bibfnamefont{A.}~\bibnamefont{Kimura}},
  \bibinfo{author}{\bibfnamefont{T.}~\bibnamefont{Nagao}},
  \bibinfo{author}{\bibfnamefont{G.}~\bibnamefont{Bihlmayer}},
  \bibinfo{author}{\bibfnamefont{E.~V.} \bibnamefont{Chulkov}},
  \bibinfo{author}{\bibfnamefont{S.}~\bibnamefont{Qiao}},
  \bibinfo{author}{\bibfnamefont{K.}~\bibnamefont{Shimada}},
  \bibnamefont{et~al.}, \bibinfo{journal}{Phys. Rev. B}
  \textbf{\bibinfo{volume}{76}}, \bibinfo{pages}{153305}
  (\bibinfo{year}{2007}).

\bibitem[{\citenamefont{Koroteev et~al.}(2008)\citenamefont{Koroteev,
  Bihlmayer, Chulkov, and Bl{\"u}gel}}]{chulkov2008}
\bibinfo{author}{\bibfnamefont{Y.~M.} \bibnamefont{Koroteev}},
  \bibinfo{author}{\bibfnamefont{G.}~\bibnamefont{Bihlmayer}},
  \bibinfo{author}{\bibfnamefont{E.~V.} \bibnamefont{Chulkov}},
  \bibnamefont{and}
  \bibinfo{author}{\bibfnamefont{S.}~\bibnamefont{Bl{\"u}gel}},
  \bibinfo{journal}{Phys. Rev. B} \textbf{\bibinfo{volume}{77}},
  \bibinfo{pages}{045428} (\bibinfo{year}{2008}).

\bibitem[{\citenamefont{Alc{\'a}ntara~Ortigoza
  et~al.}(2014)\citenamefont{Alc{\'a}ntara~Ortigoza, Sklyadneva, Heid, Chulkov,
  Rahman, Bohnen, and Echenique}}]{bi111ph}
\bibinfo{author}{\bibfnamefont{M.}~\bibnamefont{Alc{\'a}ntara~Ortigoza}},
  \bibinfo{author}{\bibfnamefont{I.~Y.} \bibnamefont{Sklyadneva}},
  \bibinfo{author}{\bibfnamefont{R.}~\bibnamefont{Heid}},
  \bibinfo{author}{\bibfnamefont{E.~V.} \bibnamefont{Chulkov}},
  \bibinfo{author}{\bibfnamefont{T.}~\bibnamefont{Rahman}},
  \bibinfo{author}{\bibfnamefont{K.~P.} \bibnamefont{Bohnen}},
  \bibnamefont{and} \bibinfo{author}{\bibfnamefont{P.~M.}
  \bibnamefont{Echenique}}, \bibinfo{journal}{Phys. Rev. B}
  \textbf{\bibinfo{volume}{90}}, \bibinfo{pages}{195438}
  (\bibinfo{year}{2014}).

\bibitem[{\citenamefont{Jona}(1967)}]{jona}
\bibinfo{author}{\bibfnamefont{F.}~\bibnamefont{Jona}}, \bibinfo{journal}{Surf.
  Sci.} \textbf{\bibinfo{volume}{8}}, \bibinfo{pages}{57}
  (\bibinfo{year}{1967}).

\bibitem[{\citenamefont{Peierls}(1991)}]{peierls}
\bibinfo{author}{\bibfnamefont{R.}~\bibnamefont{Peierls}},
  \emph{\bibinfo{title}{More Surprises in Theoretical Physics}}
  (\bibinfo{publisher}{Princeton University Press}, \bibinfo{year}{1991}).

\bibitem[{not()}]{notejona}
\bibinfo{note}{He concluded that Bi(10$\overline{1}$) reconstructs, whereas
  Bi(110) and Bi(111) have in-plane bulk-like structures .~\cite{jona} More
  recent measurements, however, concluded that Bi(10$\overline{1}$) does not
  reconstruct.~\cite{sun2006}}.

\bibitem[{\citenamefont{Sun et~al.}(2006)\citenamefont{Sun, Mikkelsen,
  Fuglsang-Jensen, Koroteev, Bihlmayer, Chulkov, Adams, Hofmann, and
  Pohl}}]{sun2006}
\bibinfo{author}{\bibfnamefont{J.}~\bibnamefont{Sun}},
  \bibinfo{author}{\bibfnamefont{A.}~\bibnamefont{Mikkelsen}},
  \bibinfo{author}{\bibfnamefont{M.}~\bibnamefont{Fuglsang-Jensen}},
  \bibinfo{author}{\bibfnamefont{Y.~M.} \bibnamefont{Koroteev}},
  \bibinfo{author}{\bibfnamefont{G.}~\bibnamefont{Bihlmayer}},
  \bibinfo{author}{\bibnamefont{Chulkov}},
  \bibinfo{author}{\bibfnamefont{D.~L.} \bibnamefont{Adams}},
  \bibinfo{author}{\bibfnamefont{P.}~\bibnamefont{Hofmann}}, \bibnamefont{and}
  \bibinfo{author}{\bibfnamefont{K.}~\bibnamefont{Pohl}},
  \bibinfo{journal}{Phys. Rev. B} \textbf{\bibinfo{volume}{74}},
  \bibinfo{pages}{245406} (\bibinfo{year}{2006}).

\bibitem[{\citenamefont{M{\"o}nig et~al.}(2005)\citenamefont{M{\"o}nig, Sun,
  Koroteev, Bihlmayer, Wells, Chulkov, Pohl, and Hofmann}}]{chulkov2005}
\bibinfo{author}{\bibfnamefont{H.}~\bibnamefont{M{\"o}nig}},
  \bibinfo{author}{\bibfnamefont{J.}~\bibnamefont{Sun}},
  \bibinfo{author}{\bibfnamefont{Y.~M.} \bibnamefont{Koroteev}},
  \bibinfo{author}{\bibfnamefont{G.}~\bibnamefont{Bihlmayer}},
  \bibinfo{author}{\bibfnamefont{J.}~\bibnamefont{Wells}},
  \bibinfo{author}{\bibfnamefont{E.~V.} \bibnamefont{Chulkov}},
  \bibinfo{author}{\bibfnamefont{K.}~\bibnamefont{Pohl}}, \bibnamefont{and}
  \bibinfo{author}{\bibfnamefont{P.}~\bibnamefont{Hofmann}},
  \bibinfo{journal}{Phys. Rev. B} \textbf{\bibinfo{volume}{72}},
  \bibinfo{pages}{085410} (\bibinfo{year}{2005}).

\bibitem[{\citenamefont{Dresselhaus}(1971)}]{dresselhaus}
\bibinfo{author}{\bibfnamefont{M.~S.} \bibnamefont{Dresselhaus}},
  \emph{\bibinfo{title}{The physics of semimetals and narrow-gap
  semiconductors: Proceedings (Supplement No. 1 to the Journal of physics and
  chemistry of solids, v. 32, 1971)}} (\bibinfo{publisher}{Pergamon Press},
  \bibinfo{year}{1971}).

\bibitem[{\citenamefont{Gonze et~al.}(1990)\citenamefont{Gonze, Michenaud, and
  Vigneron}}]{gonze1990}
\bibinfo{author}{\bibfnamefont{X.}~\bibnamefont{Gonze}},
  \bibinfo{author}{\bibfnamefont{J.~P.} \bibnamefont{Michenaud}},
  \bibnamefont{and} \bibinfo{author}{\bibfnamefont{J.~P.}
  \bibnamefont{Vigneron}}, \bibinfo{journal}{Phys. Rev. B}
  \textbf{\bibinfo{volume}{41}}, \bibinfo{pages}{11827} (\bibinfo{year}{1990}).

\bibitem[{\citenamefont{Anishchik et~al.}(1976)\citenamefont{Anishchik,
  Falicov, and Yndurain}}]{anishchik}
\bibinfo{author}{\bibfnamefont{V.}~\bibnamefont{Anishchik}},
  \bibinfo{author}{\bibfnamefont{L.~M.} \bibnamefont{Falicov}},
  \bibnamefont{and} \bibinfo{author}{\bibfnamefont{F.}~\bibnamefont{Yndurain}},
  \bibinfo{journal}{Surf. Sci.} \textbf{\bibinfo{volume}{57}},
  \bibinfo{pages}{375} (\bibinfo{year}{1976}).

\bibitem[{\citenamefont{Golin}(1968)}]{golin}
\bibinfo{author}{\bibfnamefont{S.}~\bibnamefont{Golin}},
  \bibinfo{journal}{Phys. Rev.} \textbf{\bibinfo{volume}{166}},
  \bibinfo{pages}{643} (\bibinfo{year}{1968}).

\bibitem[{\citenamefont{Jezequel et~al.}(1986)\citenamefont{Jezequel, Petroff,
  Pinchaux, and Yndurain}}]{jezequel}
\bibinfo{author}{\bibfnamefont{G.}~\bibnamefont{Jezequel}},
  \bibinfo{author}{\bibfnamefont{Y.}~\bibnamefont{Petroff}},
  \bibinfo{author}{\bibfnamefont{R.}~\bibnamefont{Pinchaux}}, \bibnamefont{and}
  \bibinfo{author}{\bibfnamefont{F.}~\bibnamefont{Yndurain}},
  \bibinfo{journal}{Phys. Rev. B} \textbf{\bibinfo{volume}{33}},
  \bibinfo{pages}{4352} (\bibinfo{year}{1986}).

\bibitem[{\citenamefont{Patthey et~al.}(1994)\citenamefont{Patthey, Schneider,
  and Micklitz}}]{patthey}
\bibinfo{author}{\bibfnamefont{F.}~\bibnamefont{Patthey}},
  \bibinfo{author}{\bibfnamefont{W.~D.} \bibnamefont{Schneider}},
  \bibnamefont{and} \bibinfo{author}{\bibfnamefont{H.}~\bibnamefont{Micklitz}},
  \bibinfo{journal}{Phys. Rev. B} \textbf{\bibinfo{volume}{49}},
  \bibinfo{pages}{11293} (\bibinfo{year}{1994}).

\bibitem[{\citenamefont{Tanaka et~al.}(1999)\citenamefont{Tanaka, Hatano,
  Takahashi, Sasaki, Suzuki, and Sato}}]{tanaka}
\bibinfo{author}{\bibfnamefont{A.}~\bibnamefont{Tanaka}},
  \bibinfo{author}{\bibfnamefont{M.}~\bibnamefont{Hatano}},
  \bibinfo{author}{\bibfnamefont{K.}~\bibnamefont{Takahashi}},
  \bibinfo{author}{\bibfnamefont{H.}~\bibnamefont{Sasaki}},
  \bibinfo{author}{\bibfnamefont{S.}~\bibnamefont{Suzuki}}, \bibnamefont{and}
  \bibinfo{author}{\bibfnamefont{S.}~\bibnamefont{Sato}},
  \bibinfo{journal}{Phys. Rev. B} \textbf{\bibinfo{volume}{59}},
  \bibinfo{pages}{1786} (\bibinfo{year}{1999}).

\bibitem[{\citenamefont{Thomas et~al.}(1999)\citenamefont{Thomas, Jezequel, and
  Pollini}}]{thomas}
\bibinfo{author}{\bibfnamefont{J.}~\bibnamefont{Thomas}},
  \bibinfo{author}{\bibfnamefont{G.}~\bibnamefont{Jezequel}}, \bibnamefont{and}
  \bibinfo{author}{\bibfnamefont{I.}~\bibnamefont{Pollini}},
  \bibinfo{journal}{J. Phys.: Condens. Matter} \textbf{\bibinfo{volume}{11}},
  \bibinfo{pages}{9571} (\bibinfo{year}{1999}).

\bibitem[{\citenamefont{Ast and H{\"o}chst}(2002)}]{ast2}
\bibinfo{author}{\bibfnamefont{C.~R.} \bibnamefont{Ast}} \bibnamefont{and}
  \bibinfo{author}{\bibfnamefont{H.}~\bibnamefont{H{\"o}chst}},
  \bibinfo{journal}{Phys. Rev. B} \textbf{\bibinfo{volume}{66}},
  \bibinfo{pages}{125103} (\bibinfo{year}{2002}).

\bibitem[{\citenamefont{Liu and Allen}(1995)}]{liu1995}
\bibinfo{author}{\bibfnamefont{Y.}~\bibnamefont{Liu}} \bibnamefont{and}
  \bibinfo{author}{\bibfnamefont{R.~E.} \bibnamefont{Allen}},
  \bibinfo{journal}{Phys. Rev. B} \textbf{\bibinfo{volume}{52}},
  \bibinfo{pages}{1556} (\bibinfo{year}{1995}).

\bibitem[{\citenamefont{Ast and H{\"o}chst}(2003)}]{AstMpoint2003}
\bibinfo{author}{\bibfnamefont{C.~R.} \bibnamefont{Ast}} \bibnamefont{and}
  \bibinfo{author}{\bibfnamefont{H.}~\bibnamefont{H{\"o}chst}},
  \bibinfo{journal}{Phys. Rev. B} \textbf{\bibinfo{volume}{67}},
  \bibinfo{pages}{113102} (\bibinfo{year}{2003}).

\bibitem[{\citenamefont{Hirahara et~al.}(2008)\citenamefont{Hirahara, Miyamoto,
  Kimura, Niinuma, Bihlmayer, Chulkov, Nagao, Matsuda, Qiao, Shimada
  et~al.}}]{rashba10}
\bibinfo{author}{\bibfnamefont{T.}~\bibnamefont{Hirahara}},
  \bibinfo{author}{\bibfnamefont{T.}~\bibnamefont{Miyamoto}},
  \bibinfo{author}{\bibfnamefont{A.}~\bibnamefont{Kimura}},
  \bibinfo{author}{\bibfnamefont{Y.}~\bibnamefont{Niinuma}},
  \bibinfo{author}{\bibfnamefont{G.}~\bibnamefont{Bihlmayer}},
  \bibinfo{author}{\bibfnamefont{E.~V.} \bibnamefont{Chulkov}},
  \bibinfo{author}{\bibfnamefont{T.}~\bibnamefont{Nagao}},
  \bibinfo{author}{\bibfnamefont{I.}~\bibnamefont{Matsuda}},
  \bibinfo{author}{\bibfnamefont{S.}~\bibnamefont{Qiao}},
  \bibinfo{author}{\bibfnamefont{K.}~\bibnamefont{Shimada}},
  \bibnamefont{et~al.}, \bibinfo{journal}{New J. Phys.}
  \textbf{\bibinfo{volume}{10}}, \bibinfo{pages}{083038}
  (\bibinfo{year}{2008}).

\bibitem[{\citenamefont{Shikin et~al.}(2008)\citenamefont{Shikin, Varykhalov,
  Prudnikova, Usachov, Adamchuk, Yamada, Riley, and Rader}}]{rashba3}
\bibinfo{author}{\bibfnamefont{A.~M.} \bibnamefont{Shikin}},
  \bibinfo{author}{\bibfnamefont{A.}~\bibnamefont{Varykhalov}},
  \bibinfo{author}{\bibfnamefont{G.~V.} \bibnamefont{Prudnikova}},
  \bibinfo{author}{\bibfnamefont{D.}~\bibnamefont{Usachov}},
  \bibinfo{author}{\bibfnamefont{V.~K.} \bibnamefont{Adamchuk}},
  \bibinfo{author}{\bibfnamefont{Y.}~\bibnamefont{Yamada}},
  \bibinfo{author}{\bibfnamefont{J.~D.} \bibnamefont{Riley}}, \bibnamefont{and}
  \bibinfo{author}{\bibfnamefont{O.}~\bibnamefont{Rader}},
  \bibinfo{journal}{Phys. Rev. Lett.} \textbf{\bibinfo{volume}{100}},
  \bibinfo{pages}{057601} (\bibinfo{year}{2008}).

\bibitem[{\citenamefont{Ishizaka et~al.}(2011)\citenamefont{Ishizaka, Bahramy,
  Murakawa, Sakano1, Shimojima1, Sonobe1, Koizumi, Shin, Miyahara, Kimura
  et~al.}}]{rashba1}
\bibinfo{author}{\bibfnamefont{K.}~\bibnamefont{Ishizaka}},
  \bibinfo{author}{\bibfnamefont{M.~S.} \bibnamefont{Bahramy}},
  \bibinfo{author}{\bibfnamefont{H.}~\bibnamefont{Murakawa}},
  \bibinfo{author}{\bibfnamefont{M.}~\bibnamefont{Sakano1}},
  \bibinfo{author}{\bibfnamefont{T.}~\bibnamefont{Shimojima1}},
  \bibinfo{author}{\bibfnamefont{T.}~\bibnamefont{Sonobe1}},
  \bibinfo{author}{\bibfnamefont{K.}~\bibnamefont{Koizumi}},
  \bibinfo{author}{\bibfnamefont{S.}~\bibnamefont{Shin}},
  \bibinfo{author}{\bibfnamefont{H.}~\bibnamefont{Miyahara}},
  \bibinfo{author}{\bibfnamefont{A.}~\bibnamefont{Kimura}},
  \bibnamefont{et~al.}, \bibinfo{journal}{Nat. Mat.}
  \textbf{\bibinfo{volume}{10}}, \bibinfo{pages}{521} (\bibinfo{year}{2011}).

\bibitem[{\citenamefont{Zhang et~al.}(2010)\citenamefont{Zhang, He, Chang,
  Song, Wang, Chen, Jia, Fang, Dai, Shan et~al.}}]{rashba2}
\bibinfo{author}{\bibfnamefont{Y.}~\bibnamefont{Zhang}},
  \bibinfo{author}{\bibfnamefont{K.}~\bibnamefont{He}},
  \bibinfo{author}{\bibfnamefont{C.~Z.} \bibnamefont{Chang}},
  \bibinfo{author}{\bibfnamefont{C.~L.} \bibnamefont{Song}},
  \bibinfo{author}{\bibfnamefont{L.~L.} \bibnamefont{Wang}},
  \bibinfo{author}{\bibfnamefont{X.}~\bibnamefont{Chen}},
  \bibinfo{author}{\bibfnamefont{J.~F.} \bibnamefont{Jia}},
  \bibinfo{author}{\bibfnamefont{Z.}~\bibnamefont{Fang}},
  \bibinfo{author}{\bibfnamefont{X.}~\bibnamefont{Dai}},
  \bibinfo{author}{\bibfnamefont{W.~Y.} \bibnamefont{Shan}},
  \bibnamefont{et~al.}, \bibinfo{journal}{Nat. Phys.}
  \textbf{\bibinfo{volume}{6}}, \bibinfo{pages}{584} (\bibinfo{year}{2010}).

\bibitem[{\citenamefont{Dil et~al.}(2008)\citenamefont{Dil, Meier, Lobo-Checa,
  Patthey, Bihlmayer, and Osterwalder}}]{rashba4}
\bibinfo{author}{\bibfnamefont{J.~H.} \bibnamefont{Dil}},
  \bibinfo{author}{\bibfnamefont{F.}~\bibnamefont{Meier}},
  \bibinfo{author}{\bibfnamefont{J.}~\bibnamefont{Lobo-Checa}},
  \bibinfo{author}{\bibfnamefont{L.}~\bibnamefont{Patthey}},
  \bibinfo{author}{\bibfnamefont{G.}~\bibnamefont{Bihlmayer}},
  \bibnamefont{and}
  \bibinfo{author}{\bibfnamefont{J.}~\bibnamefont{Osterwalder}},
  \bibinfo{journal}{Phys. Rev. Lett.} \textbf{\bibinfo{volume}{101}},
  \bibinfo{pages}{266802} (\bibinfo{year}{2008}).

\bibitem[{\citenamefont{Studer et~al.}(2009)\citenamefont{Studer, Salis,
  Ensslin, Driscoll, and Gossard}}]{rashba5}
\bibinfo{author}{\bibfnamefont{M.}~\bibnamefont{Studer}},
  \bibinfo{author}{\bibfnamefont{G.}~\bibnamefont{Salis}},
  \bibinfo{author}{\bibfnamefont{K.}~\bibnamefont{Ensslin}},
  \bibinfo{author}{\bibfnamefont{D.~C.} \bibnamefont{Driscoll}},
  \bibnamefont{and} \bibinfo{author}{\bibfnamefont{A.~C.}
  \bibnamefont{Gossard}}, \bibinfo{journal}{Phys. Rev. Lett.}
  \textbf{\bibinfo{volume}{103}}, \bibinfo{pages}{027201}
  (\bibinfo{year}{2009}).

\bibitem[{\citenamefont{Varykhalov et~al.}(2012)\citenamefont{Varykhalov,
  Marchenko, Scholz, Rienks, Kim, Bihlmayer, S\'anchez-Barriga, and
  Rader}}]{rashba6}
\bibinfo{author}{\bibfnamefont{A.}~\bibnamefont{Varykhalov}},
  \bibinfo{author}{\bibfnamefont{D.}~\bibnamefont{Marchenko}},
  \bibinfo{author}{\bibfnamefont{M.~R.} \bibnamefont{Scholz}},
  \bibinfo{author}{\bibfnamefont{E.~D.~L.} \bibnamefont{Rienks}},
  \bibinfo{author}{\bibfnamefont{T.~K.} \bibnamefont{Kim}},
  \bibinfo{author}{\bibfnamefont{G.}~\bibnamefont{Bihlmayer}},
  \bibinfo{author}{\bibfnamefont{J.}~\bibnamefont{S\'anchez-Barriga}},
  \bibnamefont{and} \bibinfo{author}{\bibfnamefont{O.}~\bibnamefont{Rader}},
  \bibinfo{journal}{Phys. Rev. Lett.} \textbf{\bibinfo{volume}{108}},
  \bibinfo{pages}{066804} (\bibinfo{year}{2012}).

\bibitem[{\citenamefont{Hortamani and Wiesendanger}(86)}]{rashba7}
\bibinfo{author}{\bibfnamefont{M.}~\bibnamefont{Hortamani}} \bibnamefont{and}
  \bibinfo{author}{\bibfnamefont{R.}~\bibnamefont{Wiesendanger}},
  \bibinfo{journal}{Phys. Rev. B} \textbf{\bibinfo{volume}{2012}},
  \bibinfo{pages}{235437} (\bibinfo{year}{86}).

\bibitem[{\citenamefont{Moreschini et~al.}(2009)\citenamefont{Moreschini,
  Bendounan, Gierz, Ast, Mirhosseini, H{\"o}chst, Kern, Henk, Ernst, Ostanin
  et~al.}}]{rashba8}
\bibinfo{author}{\bibfnamefont{L.}~\bibnamefont{Moreschini}},
  \bibinfo{author}{\bibfnamefont{A.}~\bibnamefont{Bendounan}},
  \bibinfo{author}{\bibfnamefont{I.}~\bibnamefont{Gierz}},
  \bibinfo{author}{\bibfnamefont{C.~R.} \bibnamefont{Ast}},
  \bibinfo{author}{\bibfnamefont{H.}~\bibnamefont{Mirhosseini}},
  \bibinfo{author}{\bibfnamefont{H.}~\bibnamefont{H{\"o}chst}},
  \bibinfo{author}{\bibfnamefont{K.}~\bibnamefont{Kern}},
  \bibinfo{author}{\bibfnamefont{J.}~\bibnamefont{Henk}},
  \bibinfo{author}{\bibfnamefont{A.}~\bibnamefont{Ernst}},
  \bibinfo{author}{\bibfnamefont{S.}~\bibnamefont{Ostanin}},
  \bibnamefont{et~al.}, \bibinfo{journal}{Phys. Rev. B}
  \textbf{\bibinfo{volume}{79}}, \bibinfo{pages}{075424}
  (\bibinfo{year}{2009}).

\bibitem[{\citenamefont{Bentmann and Reinert}(2013)}]{rashba9}
\bibinfo{author}{\bibfnamefont{H.}~\bibnamefont{Bentmann}} \bibnamefont{and}
  \bibinfo{author}{\bibfnamefont{F.}~\bibnamefont{Reinert}},
  \bibinfo{journal}{New J. Phys.} \textbf{\bibinfo{volume}{15}},
  \bibinfo{pages}{115011} (\bibinfo{year}{2013}).

\bibitem[{\citenamefont{Bychkov and Rashba}(1984)}]{rashba1984}
\bibinfo{author}{\bibfnamefont{Y.~A.} \bibnamefont{Bychkov}} \bibnamefont{and}
  \bibinfo{author}{\bibfnamefont{E.~I.} \bibnamefont{Rashba}},
  \bibinfo{journal}{JETP Lett.} \textbf{\bibinfo{volume}{39}},
  \bibinfo{pages}{78} (\bibinfo{year}{1984}).

\bibitem[{\citenamefont{Grundmann}(2010)}]{grundmann2010}
\bibinfo{author}{\bibfnamefont{M.}~\bibnamefont{Grundmann}},
  \emph{\bibinfo{title}{The Physics of Semiconductors: An Introduction
  Including Nanophysics and Applications}} (\bibinfo{publisher}{Springer
  Science \& Business Media}, \bibinfo{year}{2010}).

\bibitem[{\citenamefont{G.E. and L.J.}(1982)}]{marques}
\bibinfo{author}{\bibfnamefont{M.}~\bibnamefont{G.E.}} \bibnamefont{and}
  \bibinfo{author}{\bibfnamefont{S.}~\bibnamefont{L.J.}},
  \bibinfo{journal}{Surf. Sci.} \textbf{\bibinfo{volume}{113}},
  \bibinfo{pages}{131–136} (\bibinfo{year}{1982}).

\bibitem[{\citenamefont{Bangert et~al.}(1974)\citenamefont{Bangert,
  v.~Klitzing, and Landwehr}}]{bangert}
\bibinfo{author}{\bibfnamefont{E.}~\bibnamefont{Bangert}},
  \bibinfo{author}{\bibfnamefont{K.}~\bibnamefont{v.~Klitzing}},
  \bibnamefont{and} \bibinfo{author}{\bibfnamefont{G.}~\bibnamefont{Landwehr}},
  \emph{\bibinfo{title}{Proceedings of the Twelfth International Conference on
  the Physics of Semiconductors}} (\bibinfo{publisher}{Vieweg+Teubner Verlag},
  \bibinfo{year}{1974}).

\bibitem[{\citenamefont{de~Andrada~e Silva
  et~al.}(1997)\citenamefont{de~Andrada~e Silva, La~Rocca, and
  Bassani}}]{andrada}
\bibinfo{author}{\bibfnamefont{E.~A.} \bibnamefont{de~Andrada~e Silva}},
  \bibinfo{author}{\bibfnamefont{G.~C.} \bibnamefont{La~Rocca}},
  \bibnamefont{and} \bibinfo{author}{\bibfnamefont{F.}~\bibnamefont{Bassani}},
  \bibinfo{journal}{Phys. Rev. B} \textbf{\bibinfo{volume}{55}},
  \bibinfo{pages}{16293} (\bibinfo{year}{1997}).

\bibitem[{\citenamefont{Ohkawa and Uemura}(1974)}]{Ohkawa}
\bibinfo{author}{\bibfnamefont{F.~J.} \bibnamefont{Ohkawa}} \bibnamefont{and}
  \bibinfo{author}{\bibfnamefont{Y.}~\bibnamefont{Uemura}},
  \bibinfo{journal}{J. Phys. Soc. of Jpn.} \textbf{\bibinfo{volume}{37}},
  \bibinfo{pages}{1325} (\bibinfo{year}{1974}).

\bibitem[{\citenamefont{Krupin}(2004)}]{krupin2004}
\bibinfo{author}{\bibfnamefont{O.}~\bibnamefont{Krupin}}, Ph.D. thesis,
  \bibinfo{school}{Freie Universit{\"a}t Berlin, Universit{\"a}tsbibliothek}
  (\bibinfo{year}{2004}),
  \urlprefix\url{http://www.diss.fu-berlin.de/diss/servlets/MCRFileNodeServlet/FUDISS_derivate_000000003802/}.

\bibitem[{\citenamefont{Winkler}(2003)}]{winkler}
\bibinfo{author}{\bibfnamefont{R.}~\bibnamefont{Winkler}},
  \emph{\bibinfo{title}{Spin–Orbit Coupling Effects in Two-Dimensional Electron
  and Hole Systems}} (\bibinfo{publisher}{Springer-Verlag, Heidelberg},
  \bibinfo{year}{2003}).

\bibitem[{\citenamefont{Ceperley and Alder}(1980)}]{ceperly}
\bibinfo{author}{\bibfnamefont{D.~M.} \bibnamefont{Ceperley}} \bibnamefont{and}
  \bibinfo{author}{\bibfnamefont{B.~J.} \bibnamefont{Alder}},
  \bibinfo{journal}{Phys. Rev. Lett.} \textbf{\bibinfo{volume}{45}},
  \bibinfo{pages}{566} (\bibinfo{year}{1980}).

\bibitem[{\citenamefont{Perdew et~al.}(1996)\citenamefont{Perdew, Burke, and
  Ernzerhof}}]{r56}
\bibinfo{author}{\bibfnamefont{J.~P.} \bibnamefont{Perdew}},
  \bibinfo{author}{\bibfnamefont{K.}~\bibnamefont{Burke}}, \bibnamefont{and}
  \bibinfo{author}{\bibfnamefont{M.}~\bibnamefont{Ernzerhof}},
  \bibinfo{journal}{Phys.\ Rev.\ Lett.} \textbf{\bibinfo{volume}{77}},
  \bibinfo{pages}{3865} (\bibinfo{year}{1996}).

\bibitem[{\citenamefont{Payne et~al.}(1992)\citenamefont{Payne, Teter, Allan,
  Arias, and Joannopoulos}}]{r60}
\bibinfo{author}{\bibfnamefont{M.}~\bibnamefont{Payne}},
  \bibinfo{author}{\bibfnamefont{M.}~\bibnamefont{Teter}},
  \bibinfo{author}{\bibfnamefont{D.}~\bibnamefont{Allan}},
  \bibinfo{author}{\bibfnamefont{T.}~\bibnamefont{Arias}}, \bibnamefont{and}
  \bibinfo{author}{\bibfnamefont{J.}~\bibnamefont{Joannopoulos}},
  \bibinfo{journal}{Rev.\ Mod.\ Phys.} \textbf{\bibinfo{volume}{64}},
  \bibinfo{pages}{1045} (\bibinfo{year}{1992}).

\bibitem[{\citenamefont{Kresse and Furthmüller}(1996)}]{vasp}
\bibinfo{author}{\bibfnamefont{G.}~\bibnamefont{Kresse}} \bibnamefont{and}
  \bibinfo{author}{\bibfnamefont{J.}~\bibnamefont{Furthmüller}},
  \bibinfo{journal}{Comput. Mat. Sci.} \textbf{\bibinfo{volume}{6}},
  \bibinfo{pages}{15} (\bibinfo{year}{1996}).

\bibitem[{\citenamefont{Kresse and Joubert}(1999)}]{paw}
\bibinfo{author}{\bibfnamefont{G.}~\bibnamefont{Kresse}} \bibnamefont{and}
  \bibinfo{author}{\bibfnamefont{J.}~\bibnamefont{Joubert}},
  \bibinfo{journal}{Phys. Rev. B 59} \textbf{\bibinfo{volume}{59}},
  \bibinfo{pages}{1758} (\bibinfo{year}{1999}).

\bibitem[{vas()}]{vaspmanual}
\emph{\bibinfo{title}{Vasp manual}},
  \eprint{http://cms.mpi.univie.ac.at/vasp/vasp.pdf}.

\bibitem[{\citenamefont{Kokalj}(1999)}]{xcrysden}
\bibinfo{author}{\bibfnamefont{A.}~\bibnamefont{Kokalj}}, \bibinfo{journal}{J.
  Mol. Ghraphics Modelling} \textbf{\bibinfo{volume}{17}}, \bibinfo{pages}{176}
  (\bibinfo{year}{1999}), \eprint{Code available from
  http://www.xcrysden.org/.}

\bibitem[{\citenamefont{Lide}(2004)}]{cohBi}
\bibinfo{editor}{\bibfnamefont{D.~R.} \bibnamefont{Lide}}, ed.,
  \emph{\bibinfo{title}{CRC Handbook of Chemistry and Physics, 85th Edition}}
  (\bibinfo{publisher}{Taylor \& Francis}, \bibinfo{year}{2004}).

\bibitem[{\citenamefont{Tamt{\"o}gl et~al.}(2010)\citenamefont{Tamt{\"o}gl,
  Mayrhofer-Reinhartshuber, Balak, Ernst, and Rieder}}]{tamtogl}
\bibinfo{author}{\bibfnamefont{A.}~\bibnamefont{Tamt{\"o}gl}},
  \bibinfo{author}{\bibfnamefont{M.}~\bibnamefont{Mayrhofer-Reinhartshuber}},
  \bibinfo{author}{\bibfnamefont{N.}~\bibnamefont{Balak}},
  \bibinfo{author}{\bibfnamefont{W.~E.} \bibnamefont{Ernst}}, \bibnamefont{and}
  \bibinfo{author}{\bibfnamefont{K.~H.} \bibnamefont{Rieder}},
  \bibinfo{journal}{J. Phys.: Condens. Matter} \textbf{\bibinfo{volume}{22}},
  \bibinfo{pages}{304019} (\bibinfo{year}{2010}).

\bibitem[{\citenamefont{Seah and Dench}(1979)}]{Shah1979}
\bibinfo{author}{\bibfnamefont{M.~P.} \bibnamefont{Seah}} \bibnamefont{and}
  \bibinfo{author}{\bibfnamefont{W.~A.} \bibnamefont{Dench}},
  \bibinfo{journal}{Surf. Interface Anal.} \textbf{\bibinfo{volume}{2}},
  \bibinfo{pages}{1} (\bibinfo{year}{1979}).

\bibitem[{\citenamefont{Damascelli}(2004)}]{Damascelli2004}
\bibinfo{author}{\bibfnamefont{A.}~\bibnamefont{Damascelli}},
  \bibinfo{journal}{Physica Scripta} \textbf{\bibinfo{volume}{T109}},
  \bibinfo{pages}{61–74} (\bibinfo{year}{2004}).

\bibitem[{\citenamefont{Ram and Orlando}(2003)}]{rajeevMIT}
\bibinfo{author}{\bibfnamefont{R.}~\bibnamefont{Ram}} \bibnamefont{and}
  \bibinfo{author}{\bibfnamefont{T.}~\bibnamefont{Orlando}},
  \emph{\bibinfo{title}{6.730 physics for solid-state applications}}
  (\bibinfo{year}{2003}).

\bibitem[{\citenamefont{Fu and Kane}(2007)}]{FuKane2007}
\bibinfo{author}{\bibfnamefont{L.}~\bibnamefont{Fu}} \bibnamefont{and}
  \bibinfo{author}{\bibfnamefont{C.~L.} \bibnamefont{Kane}},
  \bibinfo{journal}{Phys. Rev. B} \textbf{\bibinfo{volume}{76}},
  \bibinfo{pages}{045302} (\bibinfo{year}{2007}).

\bibitem[{\citenamefont{Teo et~al.}(2008)\citenamefont{Teo, Fu, and
  Kane}}]{Kane2008}
\bibinfo{author}{\bibfnamefont{J.~C.~Y.} \bibnamefont{Teo}},
  \bibinfo{author}{\bibfnamefont{L.}~\bibnamefont{Fu}}, \bibnamefont{and}
  \bibinfo{author}{\bibfnamefont{C.~L.} \bibnamefont{Kane}},
  \bibinfo{journal}{Phys. Rev. B} \textbf{\bibinfo{volume}{78}},
  \bibinfo{pages}{045426} (\bibinfo{year}{2008}).

\bibitem[{\citenamefont{Roushan}(2011)}]{roushan}
\bibinfo{author}{\bibfnamefont{P.}~\bibnamefont{Roushan}},
  \emph{\bibinfo{title}{Phd. thesis}} (\bibinfo{year}{2011}).

\bibitem[{\citenamefont{Shick et~al.}(1999)\citenamefont{Shick, Ketterson,
  Novikov, and Freeman}}]{shick}
\bibinfo{author}{\bibfnamefont{A.~B.} \bibnamefont{Shick}},
  \bibinfo{author}{\bibfnamefont{J.~B.} \bibnamefont{Ketterson}},
  \bibinfo{author}{\bibfnamefont{D.~L.} \bibnamefont{Novikov}},
  \bibnamefont{and} \bibinfo{author}{\bibfnamefont{A.~J.}
  \bibnamefont{Freeman}}, \bibinfo{journal}{Phys. Rev. B}
  \textbf{\bibinfo{volume}{60}}, \bibinfo{pages}{15484} (\bibinfo{year}{1999}).

\bibitem[{\citenamefont{Hasan and Kane}(2010)}]{HasanKane}
\bibinfo{author}{\bibfnamefont{M.~Z.} \bibnamefont{Hasan}} \bibnamefont{and}
  \bibinfo{author}{\bibfnamefont{C.~L.} \bibnamefont{Kane}},
  \bibinfo{journal}{Rev. Mod. Phys.} \textbf{\bibinfo{volume}{82}},
  \bibinfo{pages}{3045} (\bibinfo{year}{2010}).

\bibitem[{\citenamefont{Hsieh et~al.}(2009)\citenamefont{Hsieh, Xia, Wray,
  Qian, Pal, Dil, Osterwalder, Meier, Bihlmayer, Kane et~al.}}]{hsieh}
\bibinfo{author}{\bibfnamefont{D.}~\bibnamefont{Hsieh}},
  \bibinfo{author}{\bibfnamefont{Y.}~\bibnamefont{Xia}},
  \bibinfo{author}{\bibfnamefont{L.}~\bibnamefont{Wray}},
  \bibinfo{author}{\bibfnamefont{D.}~\bibnamefont{Qian}},
  \bibinfo{author}{\bibfnamefont{A.}~\bibnamefont{Pal}},
  \bibinfo{author}{\bibfnamefont{J.~H.} \bibnamefont{Dil}},
  \bibinfo{author}{\bibfnamefont{J.}~\bibnamefont{Osterwalder}},
  \bibinfo{author}{\bibfnamefont{F.}~\bibnamefont{Meier}},
  \bibinfo{author}{\bibfnamefont{G.}~\bibnamefont{Bihlmayer}},
  \bibinfo{author}{\bibfnamefont{C.~L.} \bibnamefont{Kane}},
  \bibnamefont{et~al.}, \bibinfo{journal}{Science}
  \textbf{\bibinfo{volume}{323}}, \bibinfo{pages}{919} (\bibinfo{year}{2009}).

\bibitem[{\citenamefont{Rashba}(1965)}]{rashba1965}
\bibinfo{author}{\bibfnamefont{{\'E}.~I.} \bibnamefont{Rashba}},
  \bibinfo{journal}{Soviet Phys. Uspekhi} \textbf{\bibinfo{volume}{7}},
  \bibinfo{pages}{823} (\bibinfo{year}{1965}).

\bibitem[{\citenamefont{LaShell et~al.}(1996)\citenamefont{LaShell, McDougall,
  and Jensen}}]{LaShell}
\bibinfo{author}{\bibfnamefont{S.}~\bibnamefont{LaShell}},
  \bibinfo{author}{\bibfnamefont{B.~A.} \bibnamefont{McDougall}},
  \bibnamefont{and} \bibinfo{author}{\bibfnamefont{E.}~\bibnamefont{Jensen}},
  \bibinfo{journal}{Phys. Rev. Lett.} \textbf{\bibinfo{volume}{77}},
  \bibinfo{pages}{3419} (\bibinfo{year}{1996}).

\bibitem[{\citenamefont{D\'iaz-S\'anchez
  et~al.}(2007)\citenamefont{D\'iaz-S\'anchez, Romero, and Gonze}}]{diaz}
\bibinfo{author}{\bibfnamefont{L.~E.} \bibnamefont{D\'iaz-S\'anchez}},
  \bibinfo{author}{\bibfnamefont{A.~H.} \bibnamefont{Romero}},
  \bibnamefont{and} \bibinfo{author}{\bibfnamefont{X.}~\bibnamefont{Gonze}},
  \bibinfo{journal}{Phys. Rev. B} \textbf{\bibinfo{volume}{76}},
  \bibinfo{pages}{104302} (\bibinfo{year}{2007}).

\bibitem[{\citenamefont{Zitter}(1971)}]{zitter}
\bibinfo{author}{\bibfnamefont{R.~N.} \bibnamefont{Zitter}},
  \emph{\bibinfo{title}{The physics of semimetals and narrow-gap
  semiconductors: Proceedings (Supplement No. 1 to the Journal of physics and
  chemistry of solids, v. 32, 1971)}} (\bibinfo{publisher}{Pergamon Press},
  \bibinfo{year}{1971}).

\bibitem[{\citenamefont{Wyckoff}(1960)}]{wyckoff1960}
\bibinfo{author}{\bibfnamefont{R.~W.~G.} \bibnamefont{Wyckoff}},
  \emph{\bibinfo{title}{Crystal Structure, Vol. 1}}
  (\bibinfo{publisher}{Interscience Publishers, New York},
  \bibinfo{year}{1960}).

\bibitem[{\citenamefont{Cucka and Barrett}(1962)}]{cucka1962}
\bibinfo{author}{\bibfnamefont{P.}~\bibnamefont{Cucka}} \bibnamefont{and}
  \bibinfo{author}{\bibfnamefont{C.~S.} \bibnamefont{Barrett}},
  \bibinfo{journal}{Acta Crystallographica} \textbf{\bibinfo{volume}{15}},
  \bibinfo{pages}{865} (\bibinfo{year}{1962}).

\bibitem[{\citenamefont{Barrett}(1959)}]{barret1959}
\bibinfo{author}{\bibfnamefont{C.~S.} \bibnamefont{Barrett}},
  \bibinfo{journal}{Australian J. Phys.} \textbf{\bibinfo{volume}{13}},
  \bibinfo{pages}{209} (\bibinfo{year}{1959}).

\bibitem[{\citenamefont{Goetz and Hergenrother}(1932)}]{goetz1932}
\bibinfo{author}{\bibfnamefont{A.}~\bibnamefont{Goetz}} \bibnamefont{and}
  \bibinfo{author}{\bibfnamefont{R.~C.} \bibnamefont{Hergenrother}},
  \bibinfo{journal}{Phys. Rev.} \textbf{\bibinfo{volume}{40}},
  \bibinfo{pages}{137} (\bibinfo{year}{1932}).

\bibitem[{\citenamefont{James}(1921)}]{james1921}
\bibinfo{author}{\bibfnamefont{R.~W.} \bibnamefont{James}},
  \bibinfo{journal}{Phil. Mag. Series 6} \textbf{\bibinfo{volume}{42}},
  \bibinfo{pages}{193} (\bibinfo{year}{1921}).

\bibitem[{\citenamefont{Tamt{\"o}gl et~al.}(2013)\citenamefont{Tamt{\"o}gl,
  Kraus, Mayrhofer-Reinhartshuber, Campi, Bernasconi, Benedek, and
  Ernst}}]{tamtogl2013}
\bibinfo{author}{\bibfnamefont{A.}~\bibnamefont{Tamt{\"o}gl}},
  \bibinfo{author}{\bibfnamefont{P.}~\bibnamefont{Kraus}},
  \bibinfo{author}{\bibfnamefont{M.}~\bibnamefont{Mayrhofer-Reinhartshuber}},
  \bibinfo{author}{\bibfnamefont{D.}~\bibnamefont{Campi}},
  \bibinfo{author}{\bibfnamefont{M.}~\bibnamefont{Bernasconi}},
  \bibinfo{author}{\bibfnamefont{G.}~\bibnamefont{Benedek}}, \bibnamefont{and}
  \bibinfo{author}{\bibfnamefont{W.~E.} \bibnamefont{Ernst}},
  \bibinfo{journal}{Phys. Rev. B} \textbf{\bibinfo{volume}{87}},
  \bibinfo{pages}{035410} (\bibinfo{year}{2013}).

\bibitem[{\citenamefont{Jeavons and Saunders}(1969)}]{AsA7Bigood}
\bibinfo{author}{\bibfnamefont{A.~P.} \bibnamefont{Jeavons}} \bibnamefont{and}
  \bibinfo{author}{\bibfnamefont{G.~A.} \bibnamefont{Saunders}},
  \bibinfo{journal}{Proc. Roy. Soc. A.} \textbf{\bibinfo{volume}{310}},
  \bibinfo{pages}{415} (\bibinfo{year}{1969}).

\end{thebibliography}


\newpage

\clearpage

\begin{table*}[ht]
\caption{Structural rhombohedral and hexagonal parameters describing bulk Bi (see text and Fig.~\ref{fig:bulkstrhex}); volume of the rhombohedral supercell (V), first and second nearest-neighbor distances ($d_{1NN}$ and $d_{2NN}$); the two bulk-like vertical inter-layer separations along the trigonal axis ($d_{ij}$ and $d_{jk}$). The results of the present work are compared with  measurements at liquid helium and room temperature (T), as well as with previous calculations.}
 \centering
 \begin{ruledtabular}
 \begin{tabular}{c|ccc|cccccc}
               & \multicolumn{3}{c|}{Calculations} & \multicolumn{6}{c}{Experiments} \\
               &\multicolumn{2}{c}{This work}&Ref.~\onlinecite{diaz}&Ref.~\onlinecite{zitter}&Ref.~\onlinecite{wyckoff1960}&Ref.~\onlinecite{cucka1962}&Ref.~\onlinecite{barret1959}&Ref.~\onlinecite{goetz1932}&Ref.~\onlinecite{james1921}\\
               & GGA   & LDA    & LDA  &      &     & T=4.2K   & T=4.2K    & T=300K  & \\
 \hline
 $a_r$(\AA)    &4.85   &4.68    &4.69  &4.75  &4.75 & -     &4.73   & -      & 4.71-4.77\\
 $\alpha$($^o$)&57.30  &58.05   &57.57 &57.27 &57.24& -     &57.31  & -      & 57.20, 58.12 \\
 $u$           &0.233  &0.237   &0.234 &0.233 &0.237&0.234  &0.234  &0.235   & 0.237 \\
 $d_1/d_2$     &0.87   &0.90    &0.88  &0.87  &0.90 &0.88   &0.88   &0.89    & 0.90\\
 V({\AA}$^3$)  &75.82  &69.24   &69.06 &71.02 &70.78&-      &70.05  & -      & 70.55, 73.30\\
 $a_h$(\AA)    &4.65   &4.54    &4.52  &4.55  &4.55 &-      &4.53   &  -     & 4.57, 4.63\\
 $c_h$(\AA)    &12.12  &11.63   &11.71 &11.87 &11.86&11.80  &11.81  &11.84   & 11.69, 11.84\\
 $c_h/a_h$     &2.61   &2.56    &2.59  &2.61  &2.61 &-      &2.60   &  -     & 2.56\\
 $d_{1NN}$(\AA)&3.13   &3.08    &3.05  &3.06  &3.11 &-      &3.06   &  -     &3.11, 3.15\\
 $d_{2NN}$(\AA)&3.63   &3.46    &3.50  &3.55  &3.48 &-      &3.51   &  -     & 3.47, 3.52\\
 $d_{ij}$(\AA) &1.608   &1.615    &1.58  &1.58  &1.67 &1.59   &1.59   &1.62    &1.66\\
 $d_{jk}$(\AA) &2.432   &2.262    &2.33  &2.38  &2.28 &2.34   &2.35   &2.33    &2.28 \\
\end{tabular}
\end{ruledtabular}
\label{tab:bulkstr}
\end{table*}%

\begin{table*}[ht]
\caption{Converged changes (\%) in the interlayer spacing, $\Delta d_{i,i+1}$, of Bi(111) films with respect to those in bulk Bi obtained in this work, previous calculations and  low-energy electron diffraction experiments (LEED). The experimental values at T~=~0~K are values obtained from an extrapolation and linear fit based on measurements obtained at T=140~K.~\cite{chulkov2005}}
 \centering
 \begin{ruledtabular}
 \begin{tabular}{c|ccccc|cc}
                 &  \multicolumn{5}{c|}{Calculations} & \multicolumn{2}{c}{LEED Experiment~\cite{chulkov2005}} \\
                 & 10--39 BLs        & 6-12 BLs            & 7 BLs                   &  6 BLs                     &  6 BLs                     & \multicolumn{2}{c}{} \\
                 & LDA (This work)  & LDA~\cite{bi111ph}   & LDA~\cite{chulkov2005}  &  LDA~\cite{chulkov2008}   &  GGA~\cite{tamtogl2013}      &  T=140K       & T=0K     \\
\hline
$\Delta d_{12}$(\%)  & -1.57            &  -1.62/-1.56         & +0.6                  &  +0.9                      &   -0.83                     & +0.5$\pm$1.1  & +1.2$\pm$2.3\\
$\Delta d_{23}$(\%)  & +1.31            &  1.37/1.32           & +6.2                  &  +6.5                      &   +3.13                     & +1.9$\pm$0.8 & +2.6$\pm$1.7\\
$\Delta d_{34}$(\%)  & -0.87            &   -0.87/-0.87        &   -                     &  -                        &     -                      &  0.0$\pm$1.1 &      -    \\
$\Delta d_{45}$(\%)  & +0.13            &   0.13/0.13          &    -                     &  -                       &    -                       &    -         &     -     \\
$\Delta d_{56}$(\%)  & -0.31            &    -0.25/ -0.31      &    -                     &  -                      &    -                       &    -         &      -     \\
$\Delta d_{67}$(\%)  &  0.00            &      -               &     -                    &  -                      &    -                       &     -        &      -     \\
\end{tabular}
\end{ruledtabular}
\label{tab:surfrx1}
\end{table*}%

\begin{table*}[ht]
\caption{Changes (\%) in the interlayer spacing, $\Delta d_{i,i+1}$,  of Bi(111) films with respect to those in bulk Bi upon atop adsorption of one monolayer of atomic H. Results are compared with our converged values (12 bilayers) in the absence of H (clean) reported in Table~\ref{tab:surfrx1}. The distance between the surface bismuth atom and H, $d(Bi-H)$ is  shown. The symbol (*) indicates the side of the film on which H was adsorbed}
 \begin{ruledtabular}
 \begin{tabular}{ccccc}
                 &  \multicolumn{3}{c}{H/Bi(111)}   & clean Bi(111)          \\
                 & 6 BL           & 10 BL            &12 BL & 12 BL     \\
\hline
$\Delta d_{12}$(\%)  &  -1.89          &  -1.83         & -1.89     &  -1.57      \\
$\Delta d_{23}$(\%)  &  +1.35          &  +1.40          &  +1.61    &  +1.31      \\
$\Delta d_{34}$(\%)  &  -1.13          &  -1.19          &  -1.25    &  -0.87     \\
$\Delta d_{45}$(\%)  &  +0.04         &  +0.35           &  +0.48    &  +0.13      \\
$\Delta d_{56}$(\%)  &  -0.31          &  -0.56            &  -0.69    &  -0.31     \\
$\Delta d_{67}$(\%)  &  -0.71         &  +0.13           &  +0.04    &   0.00      \\
$\Delta d_{78}$(\%)  &   +0.06        &   -0.31          &  -0.50    &  +0.06      \\
$\Delta d_{89}$(\%)  &  -2.21         &  +0.09           &  +0.18    &   0.00     \\
$\Delta d_{910}$(\%)  &  +0.62       &  -0.19          & -0.31     &   0.00      \\
$\Delta d_{1011}$(\%)  & -3.81      & +0.04          &+0.18      &   0.00      \\
$\Delta d_{1112}$(\%)  & -3.20*        &  -0.12         &-0.31      &   0.00      \\
$\Delta d_{1213}$(\%)  &               &  -0.22         &+0.26      &   0.00      \\
$\Delta d_{1314}$(\%)  &               &  +0.06         &-0.25      &   0.00      \\
$\Delta d_{1415}$(\%)  &               &  -0.76         &+0.57      &   0.00      \\
$\Delta d_{1516}$(\%)  &               &   +0.25        &-0.12      &   0.00      \\
$\Delta d_{1617}$(\%)  &               &  -1.94         &+0.57      &   0.00      \\
$\Delta d_{1718}$(\%)  &               &  +0.74         &+0.06      &  +0.06      \\
$\Delta d_{1819}$(\%)  &               &  -3.62         &+0.18      &   0.00      \\
$\Delta d_{1920}$(\%)  &               &   -3.13*        &+0.25      &   -0.31      \\
$\Delta d_{2021}$(\%)  &               &                &-1.39      &   +0.13      \\
$\Delta d_{2122}$(\%)  &               &                &+0.80     &   -0.87      \\
$\Delta d_{2223}$(\%)  &               &                &-3.24     &   +1.31      \\
$\Delta d_{2324}$(\%)  &               &                &-3.00*      &   -1.57      \\
$d(Bi-H)$          & 1.863         & 1.864          &1.865      &     -        \\
\end{tabular}
\end{ruledtabular}
\label{tab:surfrx2}
\end{table*}%

\newpage

\clearpage

\begin{figure*}
\includegraphics[width=0.9\textwidth]{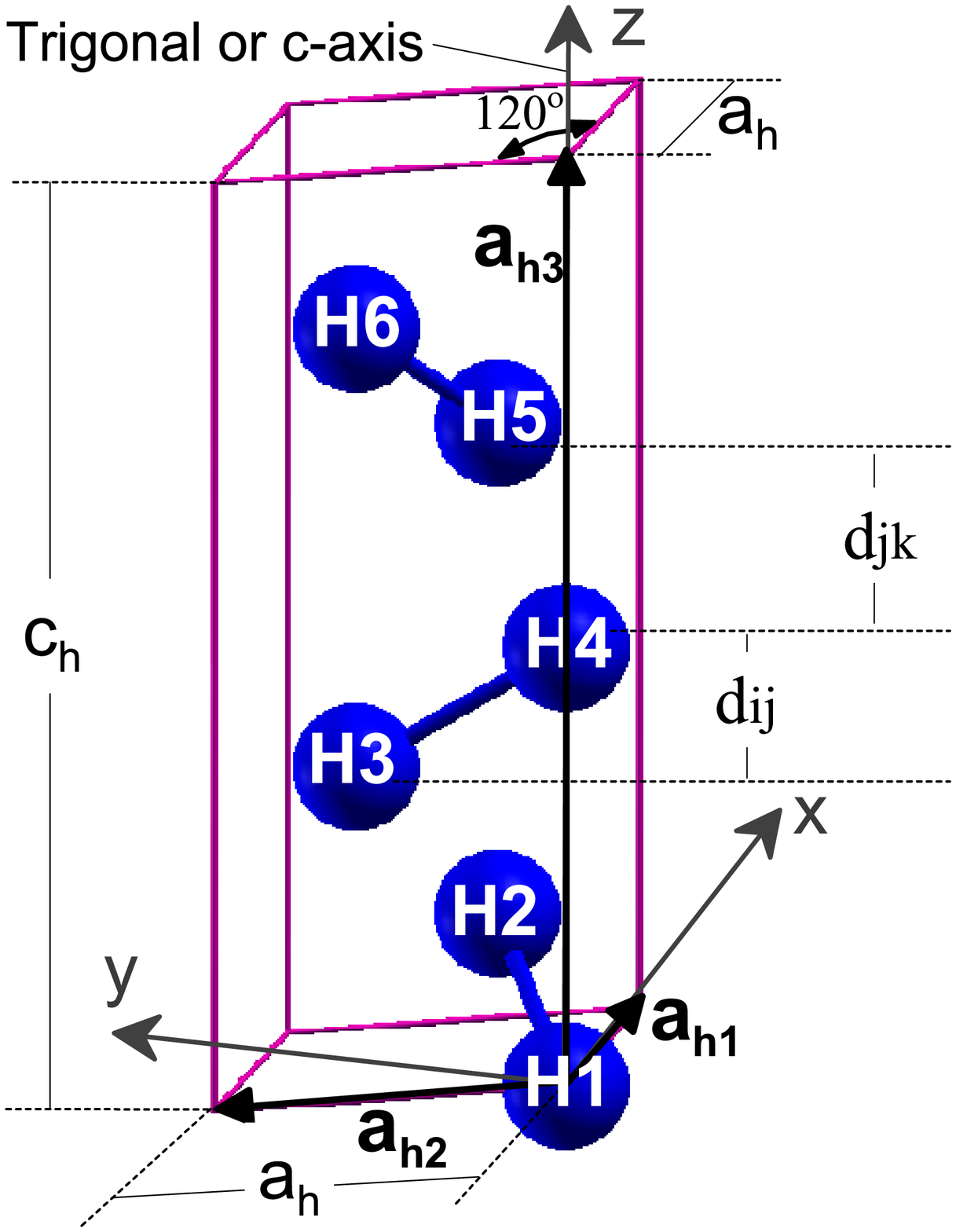}
\caption{\label{fig:bulkstrhex}
}
\end{figure*}

FIG.~\ref{fig:bulkstrhex} (Color online): Hexagonal representation of the structure of bulk Bi.
$a_h=\sqrt{3}b$ and
$c_h=3c_r $, where
$b = \frac{1}{\surd{3}}a_r \sqrt{2(1-cos\alpha)}$;
$c_r = \frac{1}{\surd{3}}a_r \sqrt{1+2 cos\alpha}$
($a_r$, $\alpha$, $u$, $a_h$, and $c_h$ are given in Table~\ref{tab:bulkstr}).
The position of the six atoms in the hexagonal unit cell are
H1= $\bm{0}$;
H2= $ \frac{1}{3} [ 2\bm{a_{h1}} +  \bm{a_{h2}} + (6u-1)\bm{a_{h3}}]  $;
H3= $ \frac{1}{3} [ \bm{a_{h1}} +  2\bm{a_{h2}} + \bm{a_{h3}}]  $;
H4= $2u \bm{a_{h3}}$;
H5= $ \frac{1}{3} [ 2\bm{a_{h1}} +  \bm{a_{h2}} + 2\bm{a_{h3}}]  $; and
H6= $ \frac{1}{3} [ \bm{a_{h1}} +  2\bm{a_{h2}} + (6u+1)\bm{a_{h3}}]  $.

\begin{figure*}
\includegraphics[width=0.6\textwidth]{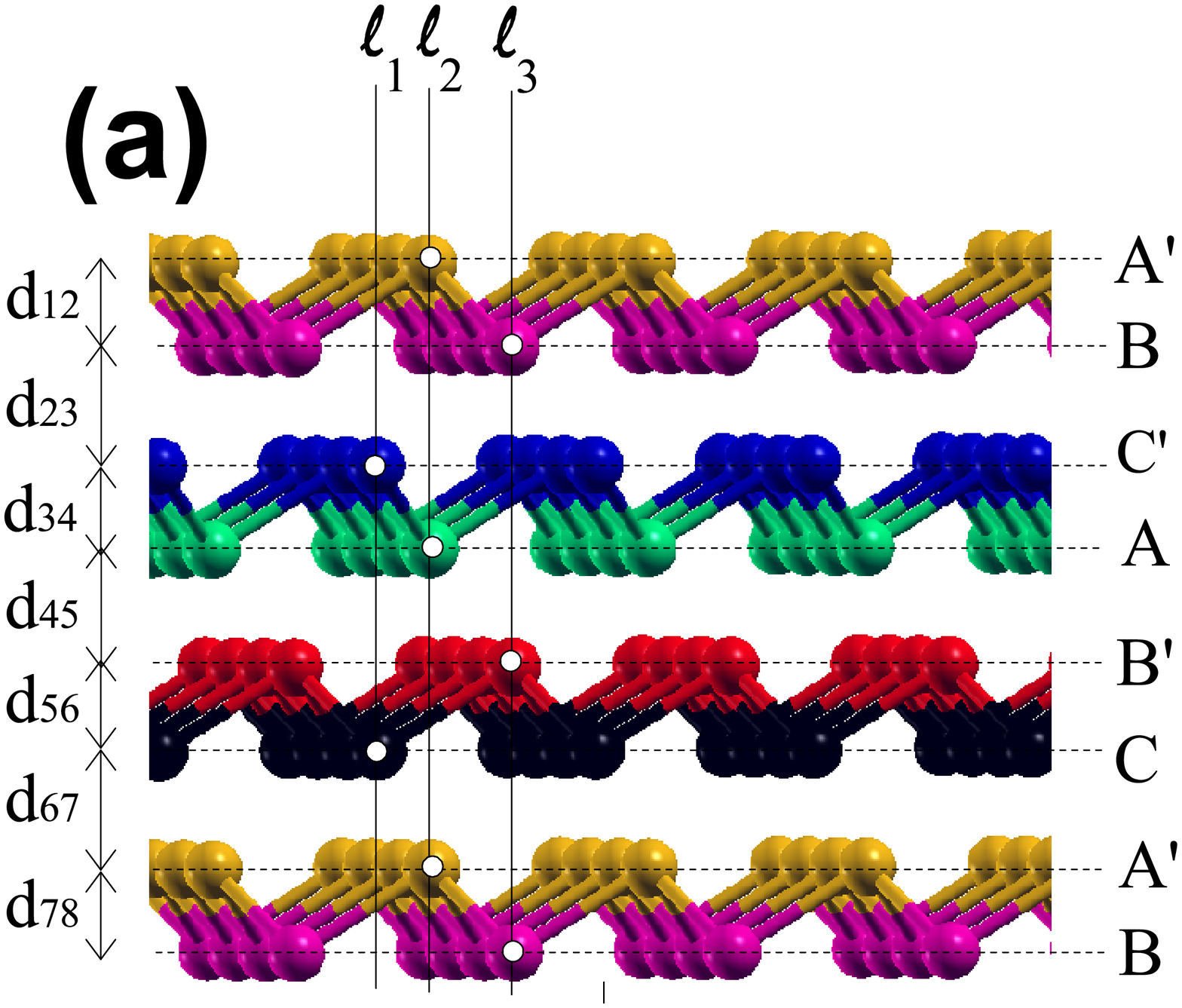}
\includegraphics[width=0.5\textwidth]{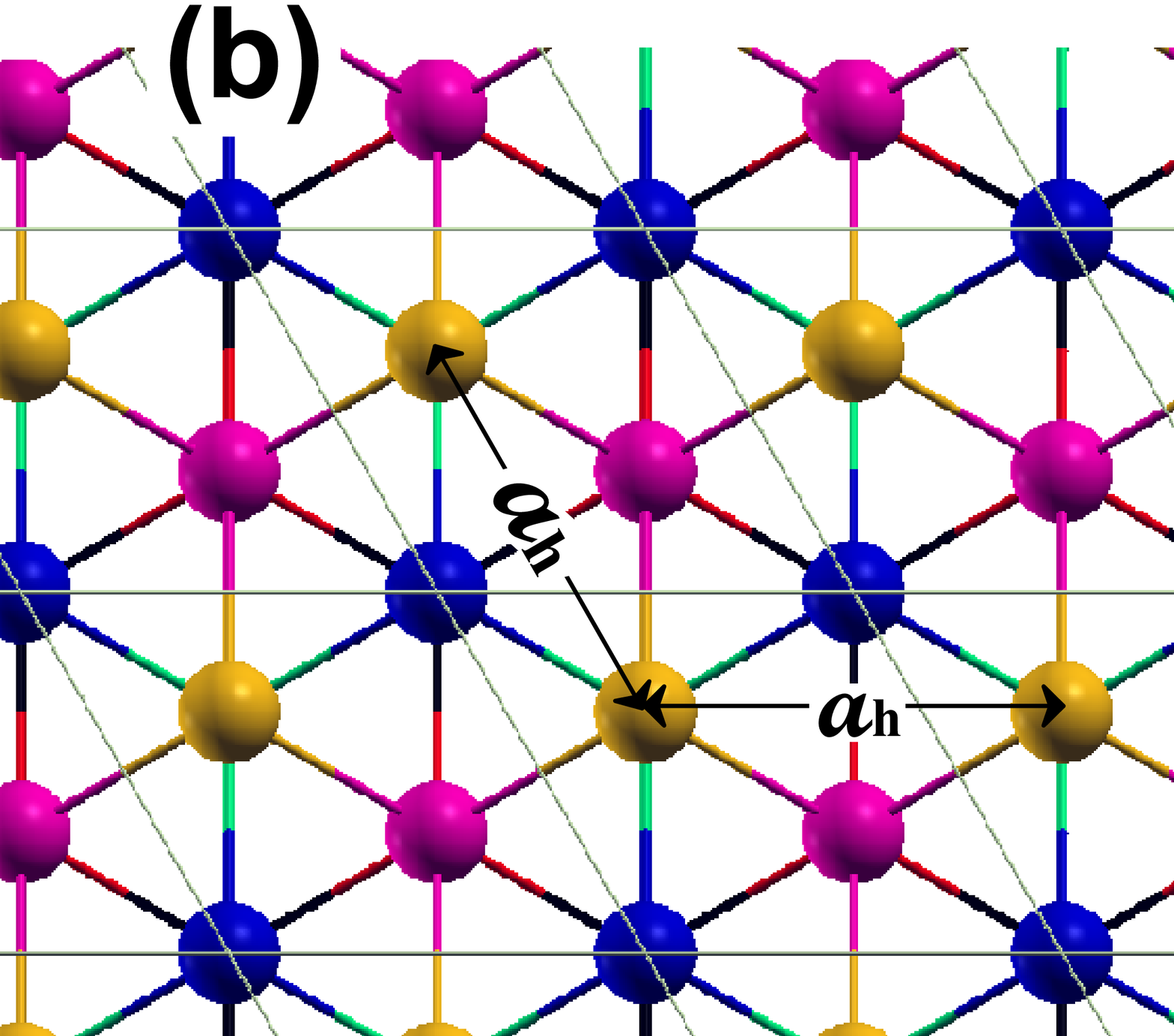}
\includegraphics[width=0.5\textwidth]{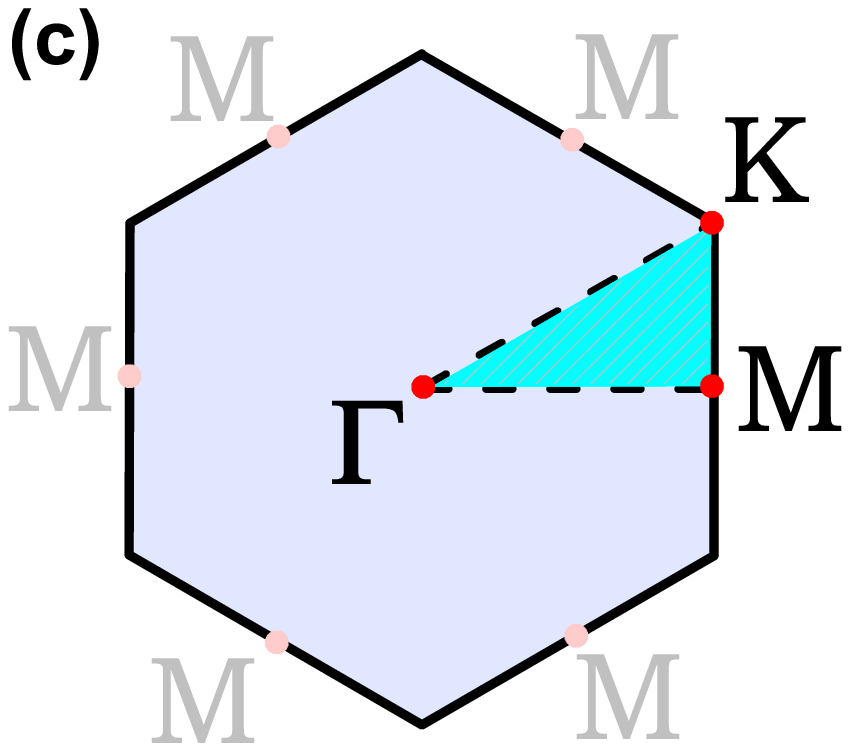}
\caption{\label{fig:surfstr}
}
\end{figure*}

FIG.~\ref{fig:surfstr} (Color Online) (a) Side view of a Bi slab underlying the Bi(111) surface. A, A', B, B', C, and C' represent the planes defined by each layer, which are perpendicular to the trigonal axis. The sticks connecting the atoms indicate bonds between first NNs  (between atoms on planes A' and B, C' and A, or B' and C), in order to  highlight the puckered bilayer-like structure of Bi(111). Lines $l_1$, $l_2$ and $l_3$ and the white dots show the large separation between an atom and the atom directly below (or above) along a direction parallel to the normal to  Bi(111). (b) Top view of the Bi(111) surface. The experimental and calculated values of $a_h$ are given in Table~\ref{tab:bulkstr}. The first light grey (yellow), grey (pink) and dark grey (blue) balls represent the first, second and third layers of Bi(111). (c) Surface Brillouin zone of Bi(111) indicating the high-symmetry points, including the six equivalent $M$ points around $\Gamma$, and the irreducible Brillouin zone (shaded).

\begin{figure*}
\includegraphics[width=0.5\textwidth]{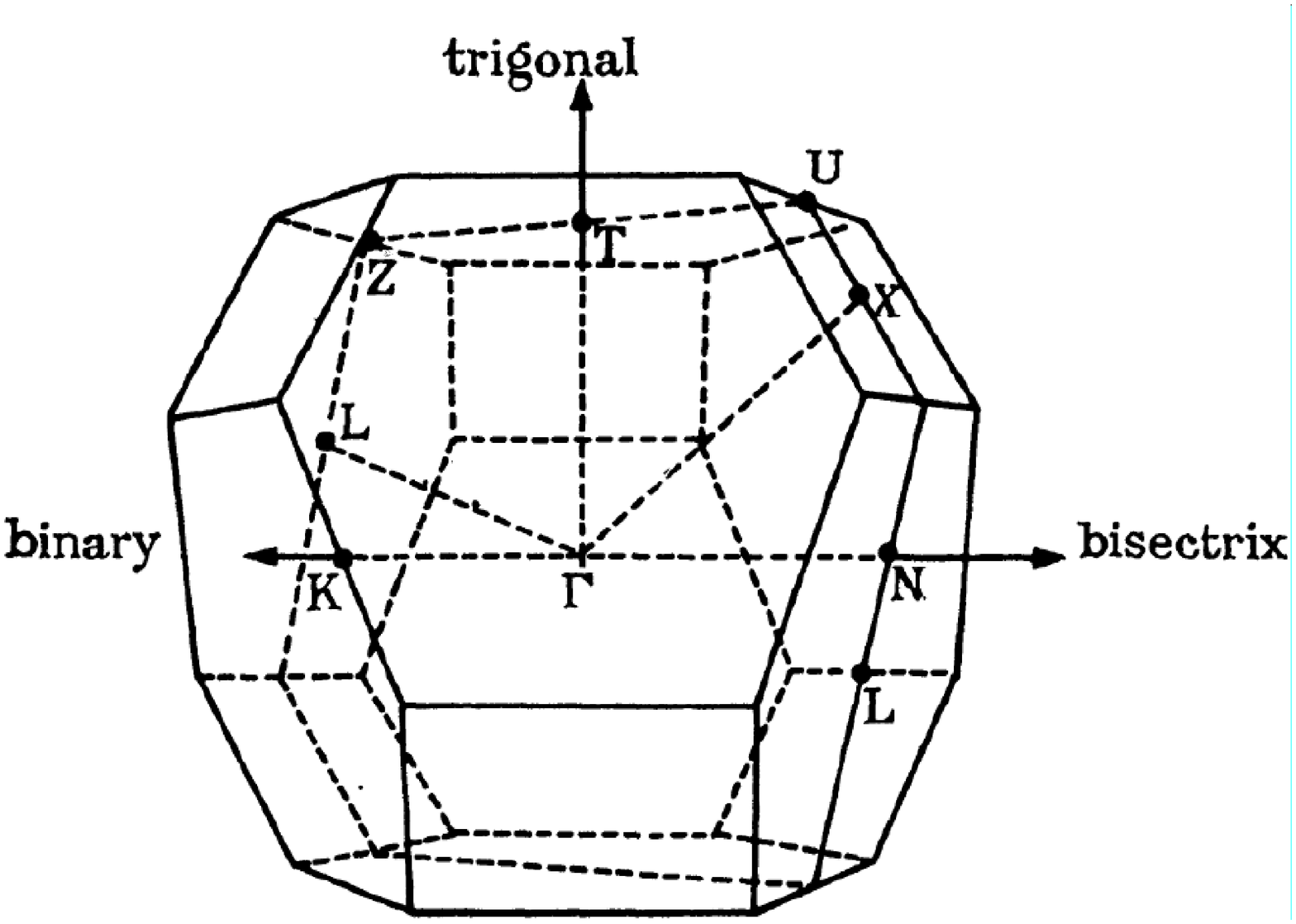}
\includegraphics[width=0.5\textwidth]{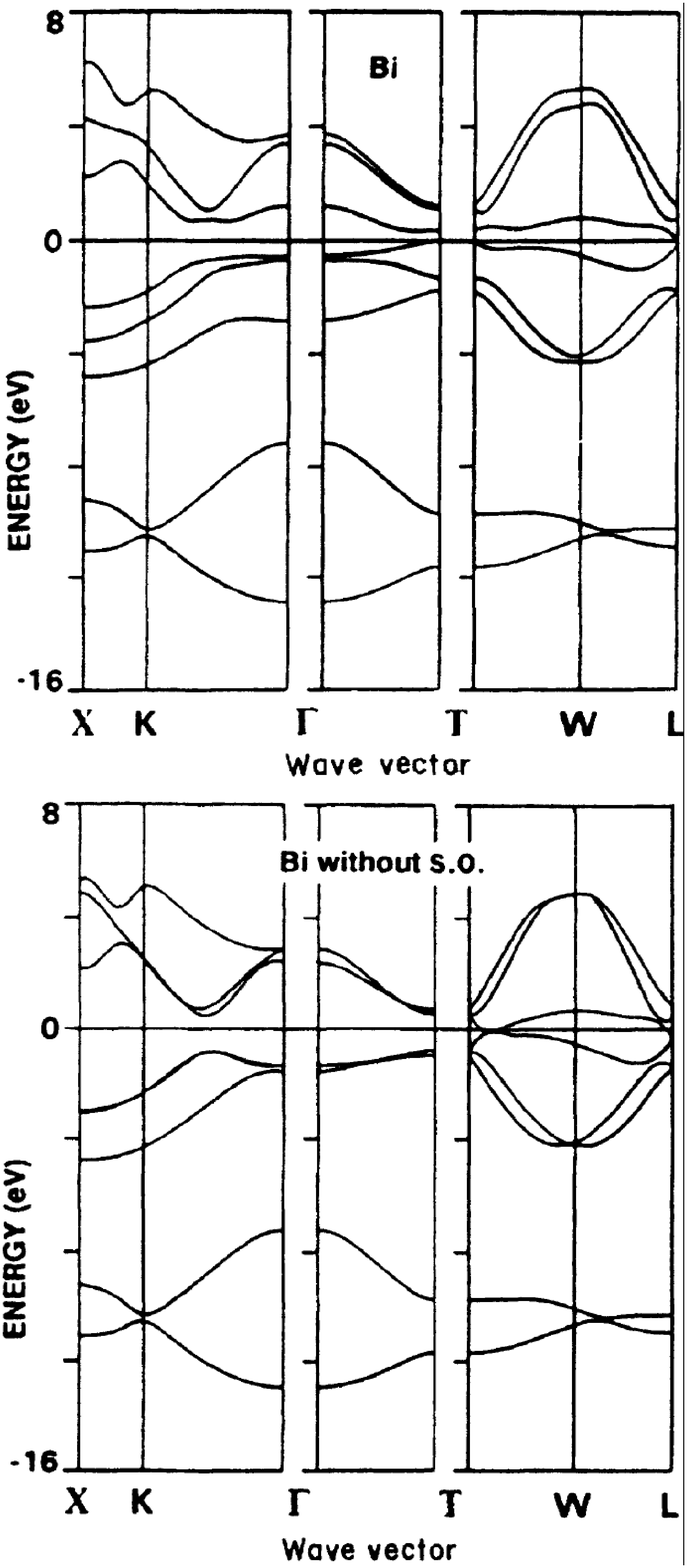}
\caption{\label{fig:bzelecstr}
}
\end{figure*}

FIG.~\ref{fig:bzelecstr}  (Top) The first Brillouin zone of the A7 structure taken from Ref.~\onlinecite{AsA7Bigood}, showing the binary, bisectrix and
trigonal axes and the notation for the high-symmetry points. (Bottom) The first Surface Brillouin zone of Bi(111) showing the high-symmetry points.
(Bottom) Electronic stricture of bulk Bi along some of the high-symmetry lines with and without taking into account the SOC by Gonze \emph{et al}.~\cite{gonze1990}

\begin{figure*}
\includegraphics[width=0.8\textwidth]{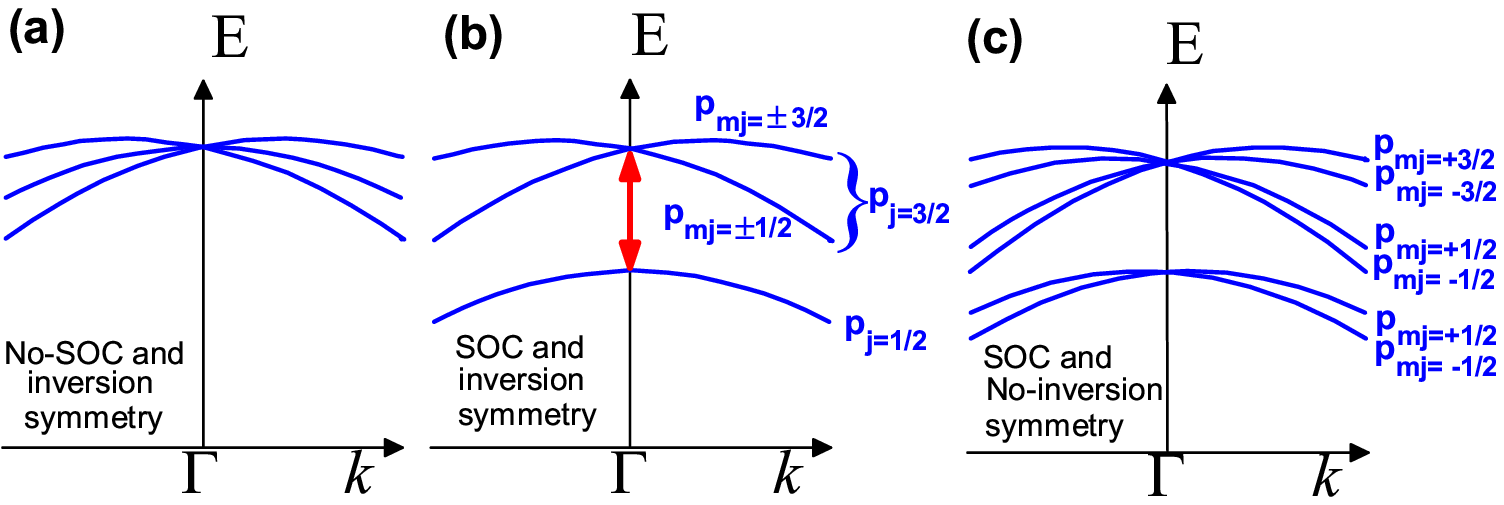}
\caption{\label{fig:soc}
}
\end{figure*}

FIG.~\ref{fig:soc} (Color Online) Schematic representation of \emph{p}-levels (a) without the influence of the SOC and inversion symmetry; (b) with the influence of the SOC and inversion symmetry; and (c) with the influence of the SOC and without inversion symmetry.~\cite{winkler,krupin2004}

FIG.~\ref{fig:bpdosss} (Color Online) (Top) Projected band-structure of bulk Bi on the SBZ of Bi(111) from first-principles and including the SOC along the high-symmetry lines of the surface Brillouin zone, $\overline{\Gamma M}$ and $\overline{\Gamma K}$. (Bottom) Band-structure of a 39-bilayer slab along the high-symmetry lines of the SBZ with and without SOC, indicating the surface states that arise only if the SOC (red arrows) is taken into account and  the pair of metallic surface states  in the "A7-distortion" gap  appearing regardless whether or not the SOC is taken into account (blue arrows).

FIG.~\ref{fig:surfelec} (Color Online) Band-structure of Bi(111) around $E_F$ for slabs of various thicknesses (7, 12, 21, 27, and 39 BLs) along $\overline{\Gamma M}$  comparing the calculated surface states against the measurements of Hengsberger \emph{et al.}~\cite{hengsberger} (Exp.1) and Ast and H\"ochst~\cite{AstMpoint2003} (Exp.2) and the calculated projected band-structure of bulk Bi on the SBZ of Bi(111).

FIG.~\ref{fig:convgap} (Color Online) Calculated gap $\Delta E$ between the two metallic bands  as a function of slab thickness and fitting curve $\Delta E = \frac{B}{N_B}-C$, where $B=2866.1$~meV and $C=40$~meV.

FIG.~\ref{fig:socsia} (Color Online) Evolution of the metallic branches (red lines) of a 24-BL Bi(111)  as a function of the strength of the SOC for films (Top) with inversion and (Bottom) without inversion symmetry. The black lines correspond all other states of the band-structure. The diameter of the dots represent how much the metallic band is localized on a single surface of the film. In the case of the slab without inversion symmetry, the red dots correspond to the unperturbed surface (see text).

Figure~\ref{fig:magmom_discon} (Color online) Metallic surface-state  branches localized at the "top" side of a 39-BL film with inversion symmetry indicating the polarization of the magnetic moment: Red for positive ("up") and black for negative ("down").

Figure~\ref{fig:magmom_evo}(Color online) Magnetic moment at the two surface bilayers (top and bottom) of a 39-BL film associated with the (a)  low-energy  and (b) high-energy metallic branches. The two degenerate bands composing the low (high) energy branch are denoted by  L1 and L2 (H1 and H2)  and their localization at either the "top" or "bottom" layer is indicated.

FIG.~\ref{fig:pldosani} (Color online) $p_{xy}$ (left) and $p_{z}$ (right) character of the two metallic branches of Bi(111) at the positions of the atoms of the surface bilayers.

FIG.~\ref{fig:magy} (Color online) Positive projection of the magnetization along the in-plane $y$-axis corresponding to the contribution of the two metallic branches of Bi(111) as projected onto the  two surface bilayers.

FIG.~\ref{fig:dosatom} (Color Online)  Local DOS of one bulk atom and one atom of the surface layer of a 39-BL slab.

FIG.~\ref{fig:dosfilm} (Color Online) Total DOS of a 3.2, 6, and 15~nm films (39 bilayers) versus the LDOS contribution of the topmost bilayer.

FIG.~\ref{fig:interintrabi} (Color Online) Band-structure of an inversion-symmetric Bi(111) slab of 12~BLs (black) around $E_F$ and along $\overline{\Gamma M}$ and $\overline{\Gamma K}$  compared against (Top) a 12~BLs (pink) inversion-asymmetric slab because of an artificial inter-bilayer perturbation that consisted of increasing  the inter-bilayer distance between the surface bilayer and the rest of the film by 0.3~{\AA} and (Bottom)  a 12~BLs (red) inversion-asymmetric slab because of an artificial inter-bilayer perturbation that consisted of increasing  the intra-bilayer distance between the surface layer and the rest of the film by 0.1~{\AA}.

FIG.~\ref{fig:surfelecH} (Color Online) Band-structure of H/Bi(111) around $E_F$ for slabs of various thicknesses (6, 12, and 21 BL) comparing the calculated surface states  against those  of clean Bi(111) for a slab of 39 BL and the projected band-structure of bulk Bi (black lines).



\clearpage

\begin{figure*}
\includegraphics[width=0.6\textwidth]{bi111_24bl_inv_new.eps}
\includegraphics[width=0.6\textwidth]{bi111_24bl_noinv_red_blue.eps}
\caption{\label{fig:socsia}
}
\end{figure*}

\begin{figure*}
\includegraphics[width=0.85\textwidth]{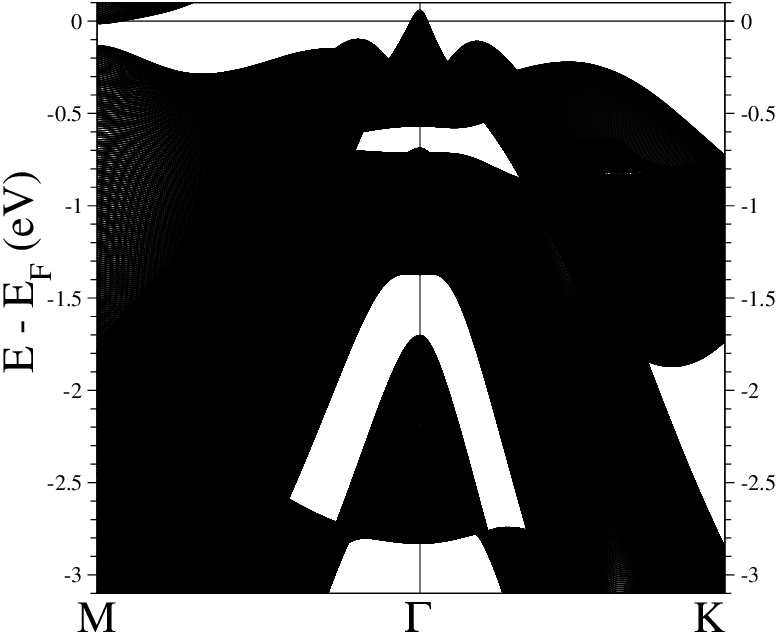}
\includegraphics[width=0.8\textwidth]{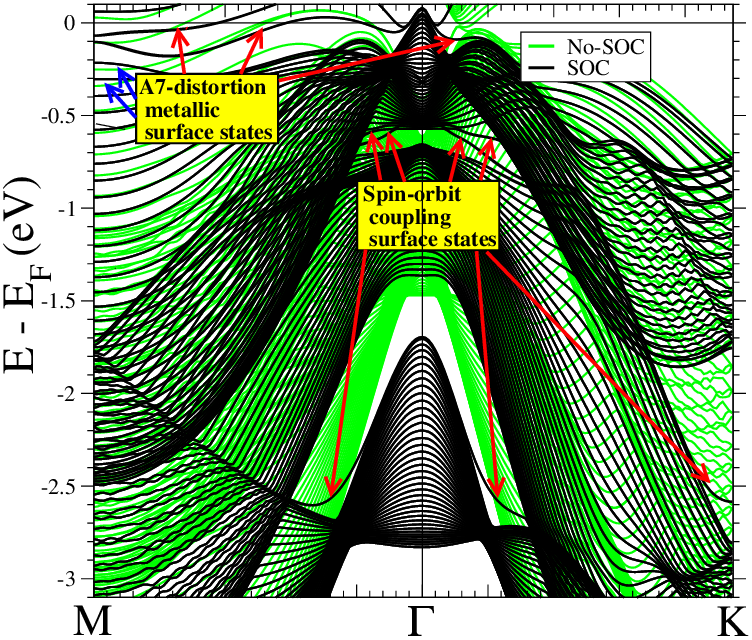}
\caption{\label{fig:bpdosss}
}
\end{figure*}


\begin{figure*}
\includegraphics[width=0.8\textwidth]{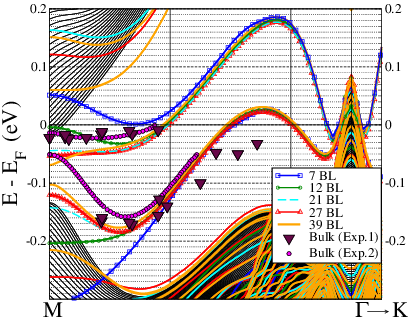}
\caption{\label{fig:surfelec}
}
\end{figure*}

\begin{figure*}
\includegraphics[width=0.8\textwidth]{surface_states_conv.eps}
\caption{\label{fig:convgap}
}
\end{figure*}

\begin{figure*}
\includegraphics[width=0.8\textwidth]{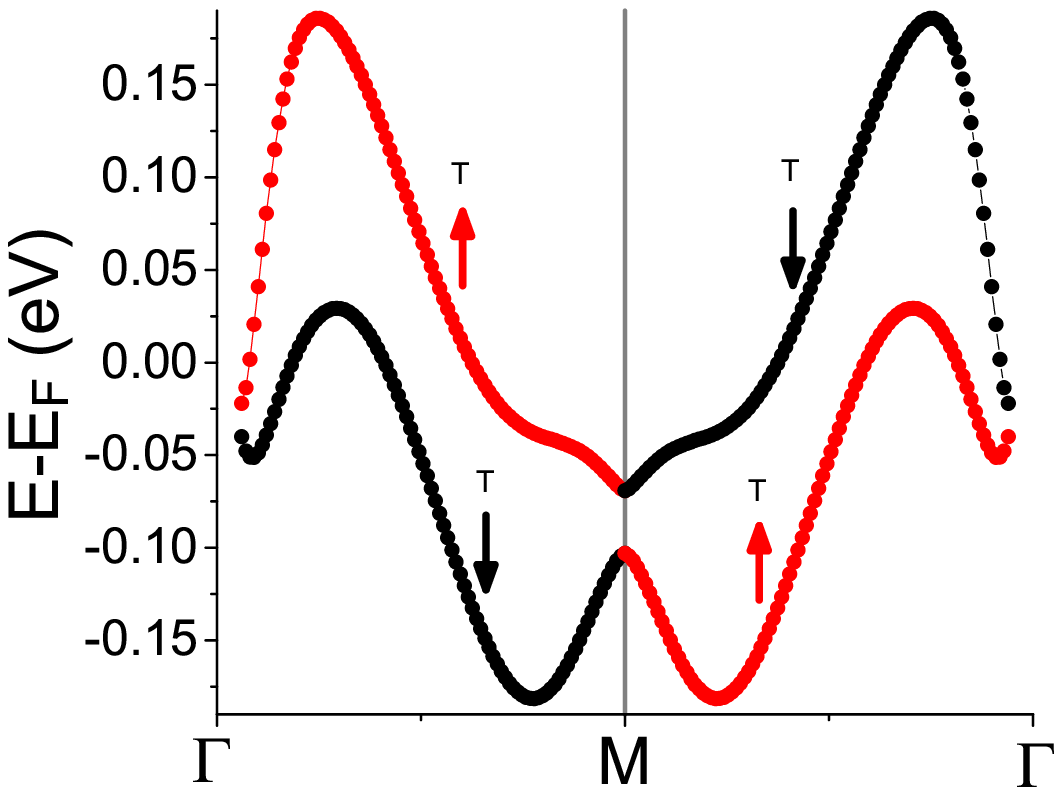}
\caption{\label{fig:magmom_discon}
}
\end{figure*}

\begin{figure*}
\includegraphics[width=0.6\textwidth]{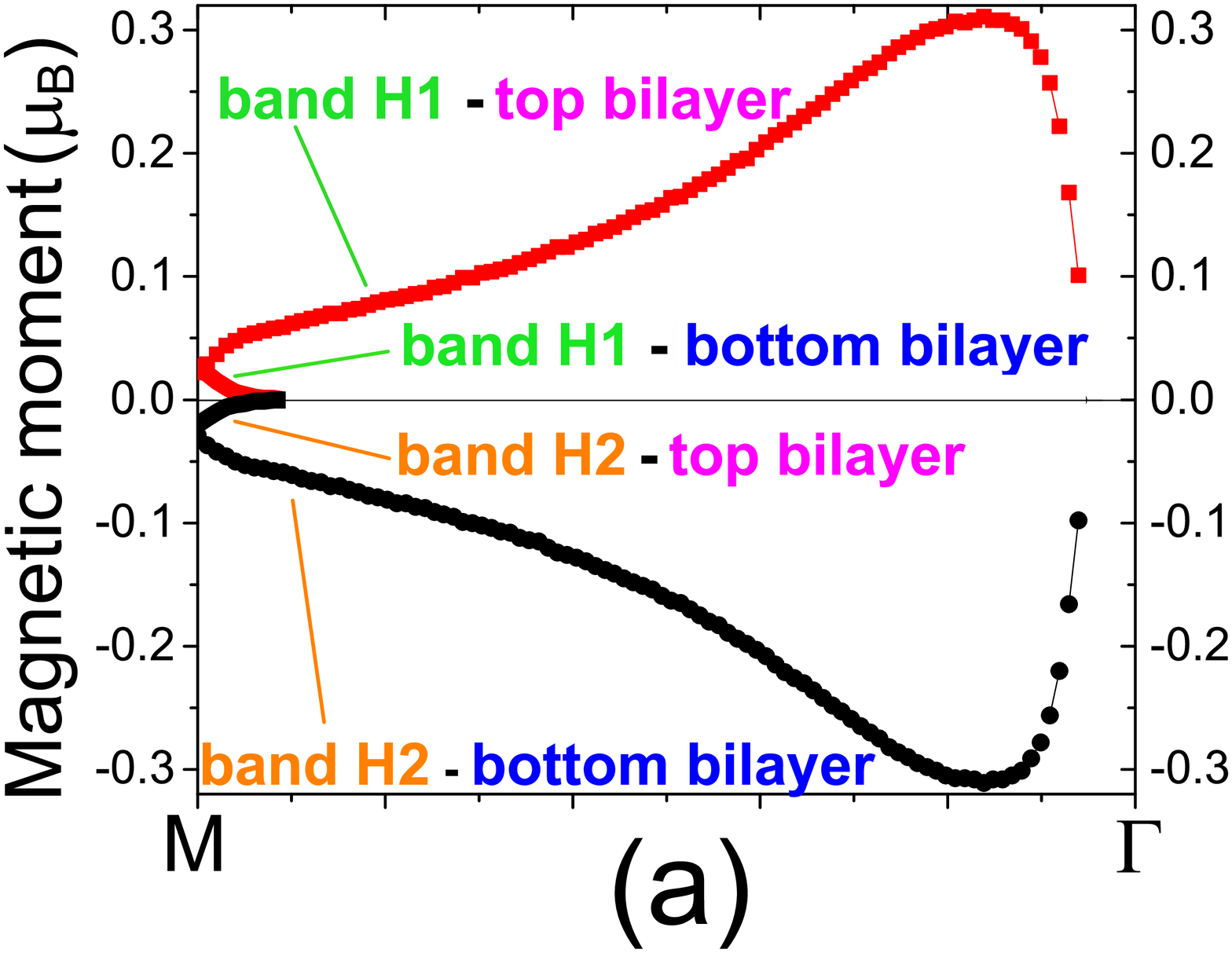}
\includegraphics[width=0.6\textwidth]{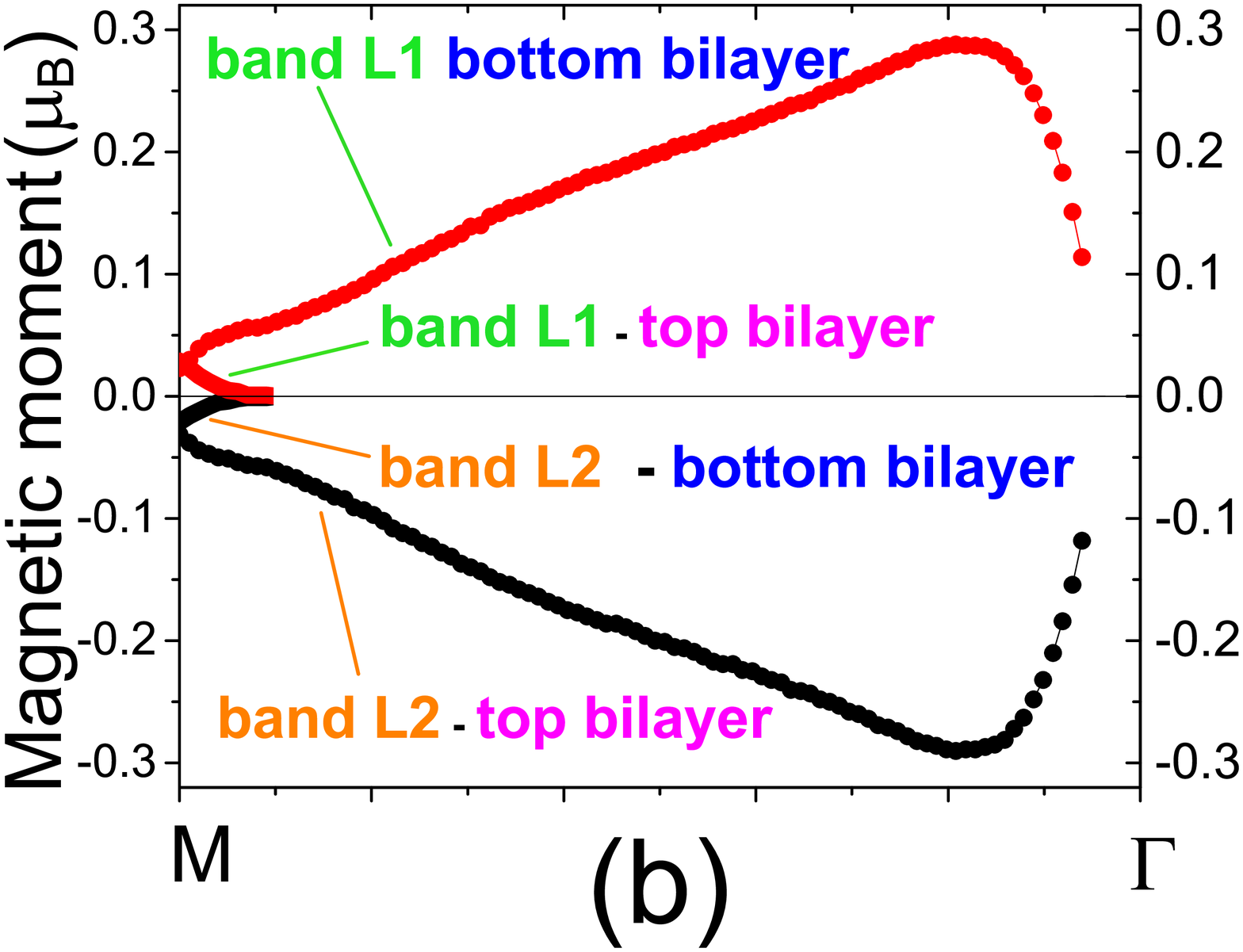}
\caption{\label{fig:magmom_evo}
}
\end{figure*}

\begin{figure*}
\includegraphics[width=0.8\textwidth]{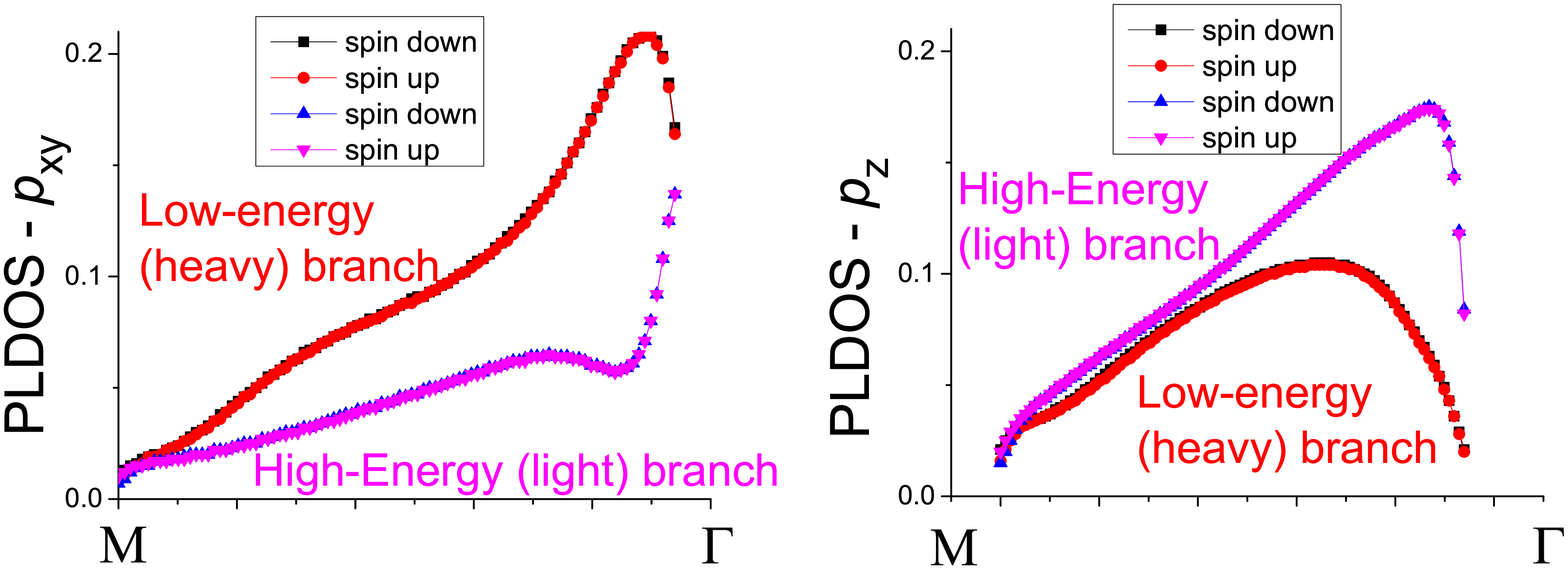}
\caption{\label{fig:pldosani}
}
\end{figure*}

\begin{figure*}
\includegraphics[width=0.5\textwidth]{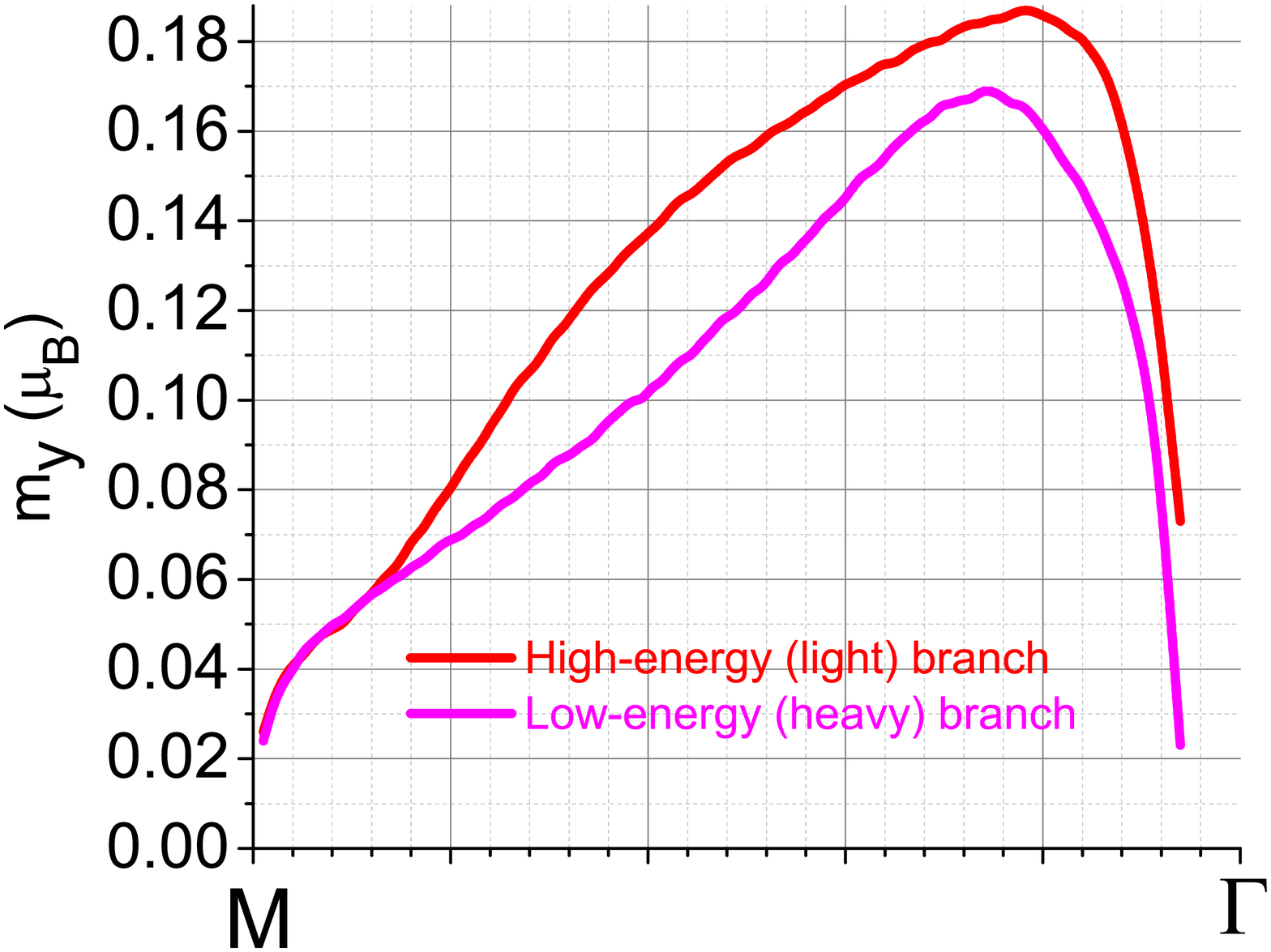}
\caption{\label{fig:magy}
}
\end{figure*}

\begin{figure*}
\includegraphics[width=0.8\textwidth]{bdsr_IS_vs_interb.eps}
\includegraphics[width=0.8\textwidth]{bdsr_IS_vs_interl.eps}
\caption{\label{fig:interintrabi}
}
\end{figure*}

\begin{figure*}
\includegraphics[width=0.8\textwidth]{bands_surf_bulk_and_Hsurf.eps}
\caption{\label{fig:surfelecH}
}
\end{figure*}

\begin{figure*}
\includegraphics[width=0.5\textwidth]{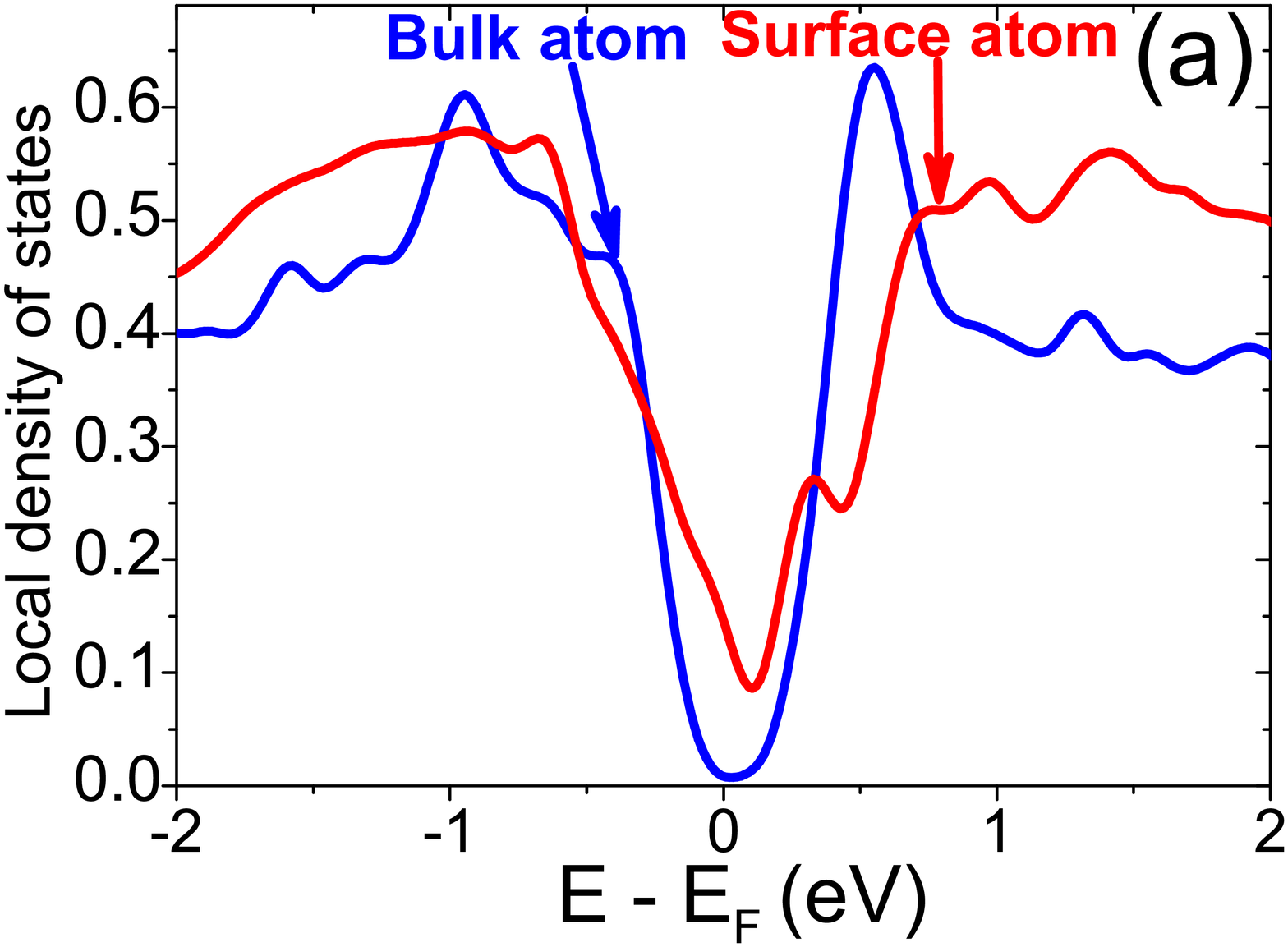}
\includegraphics[width=0.5\textwidth]{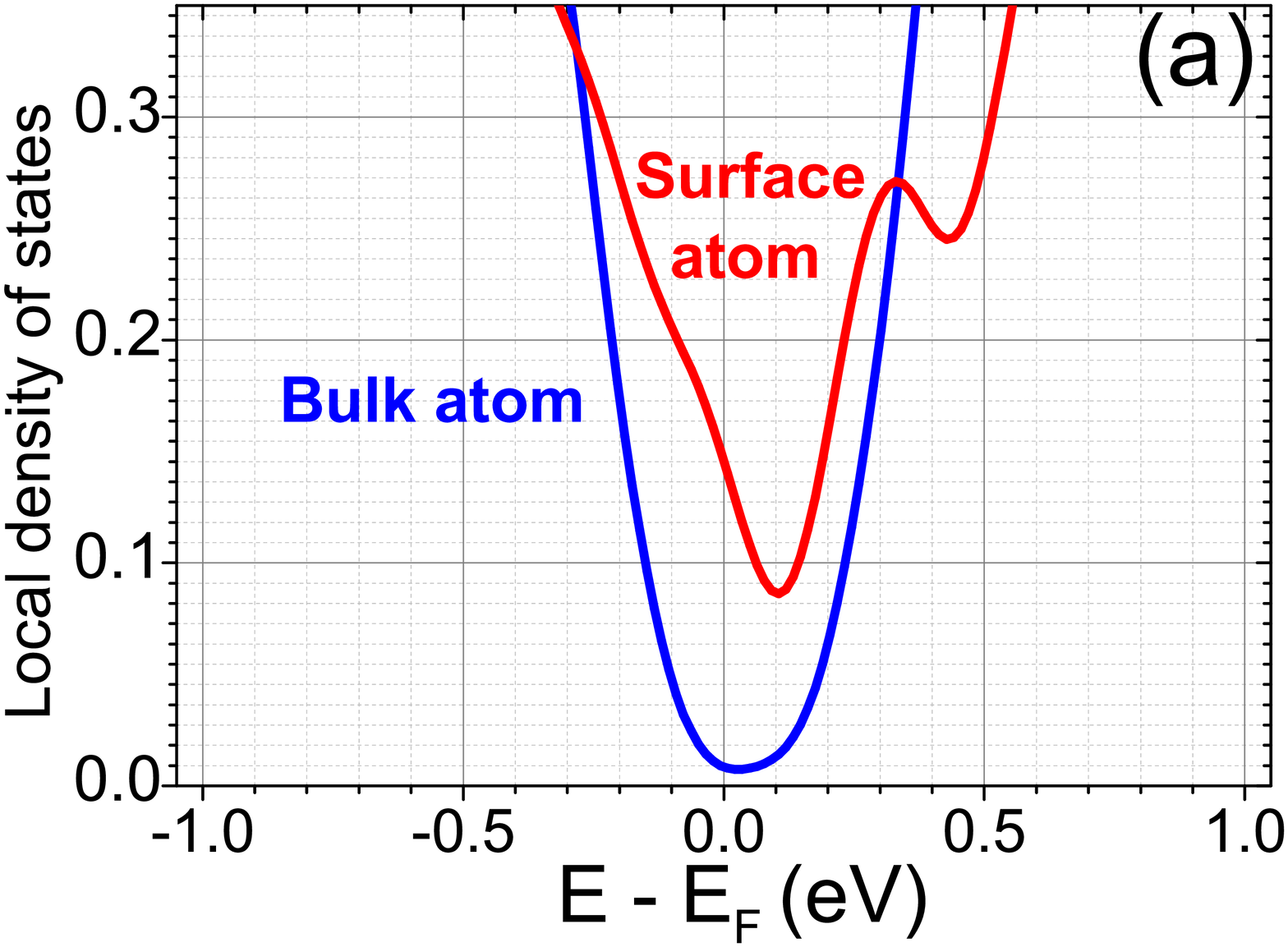}
\caption{\label{fig:dosatom}
}
\end{figure*}

\begin{figure*}
\includegraphics[trim=0cm 33mm 0cm 0cm, clip=true,width=0.5\textwidth]{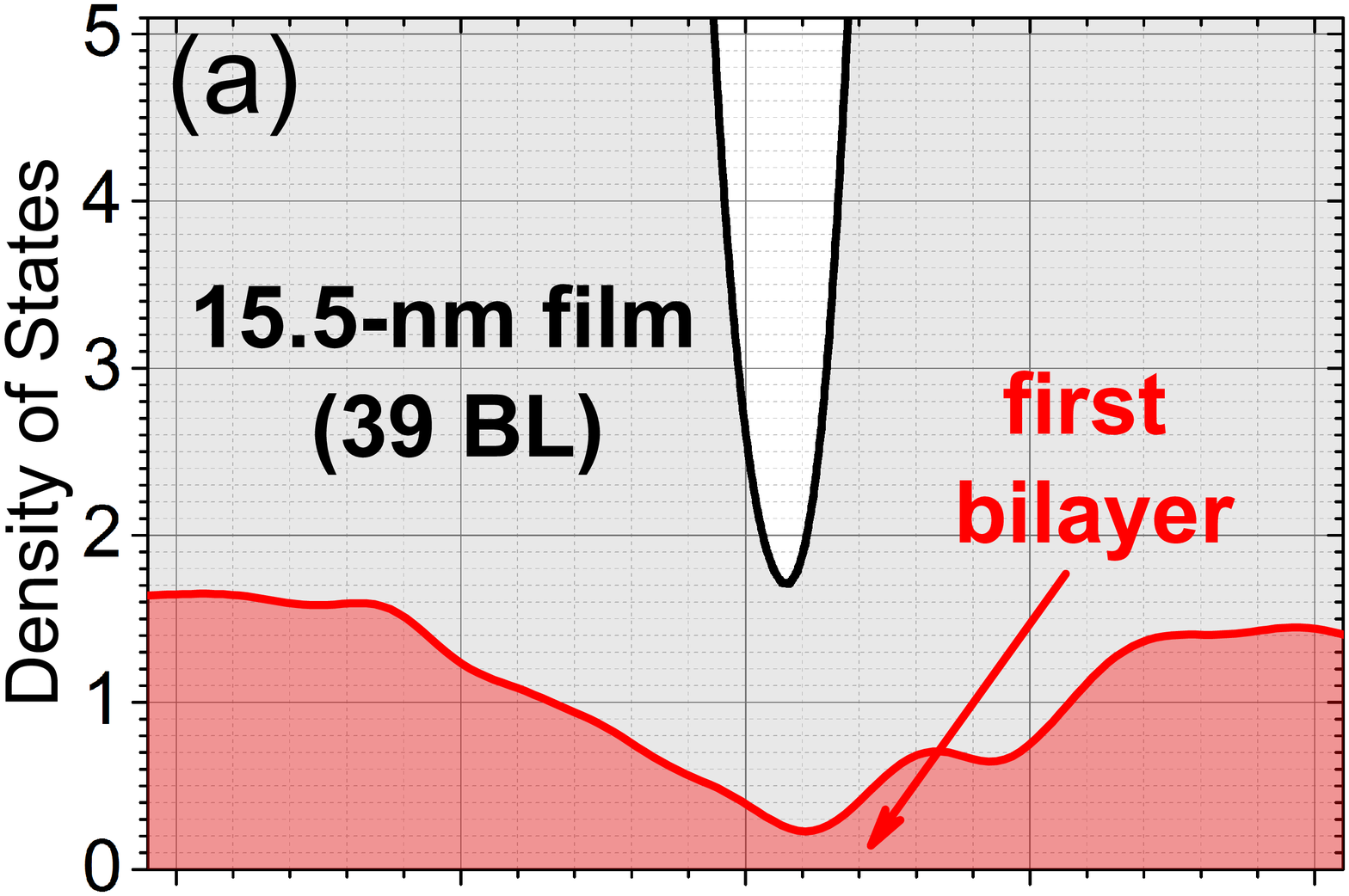}
\includegraphics[trim=0cm 33mm 0cm 2.5mm, clip=true,width=0.5\textwidth]{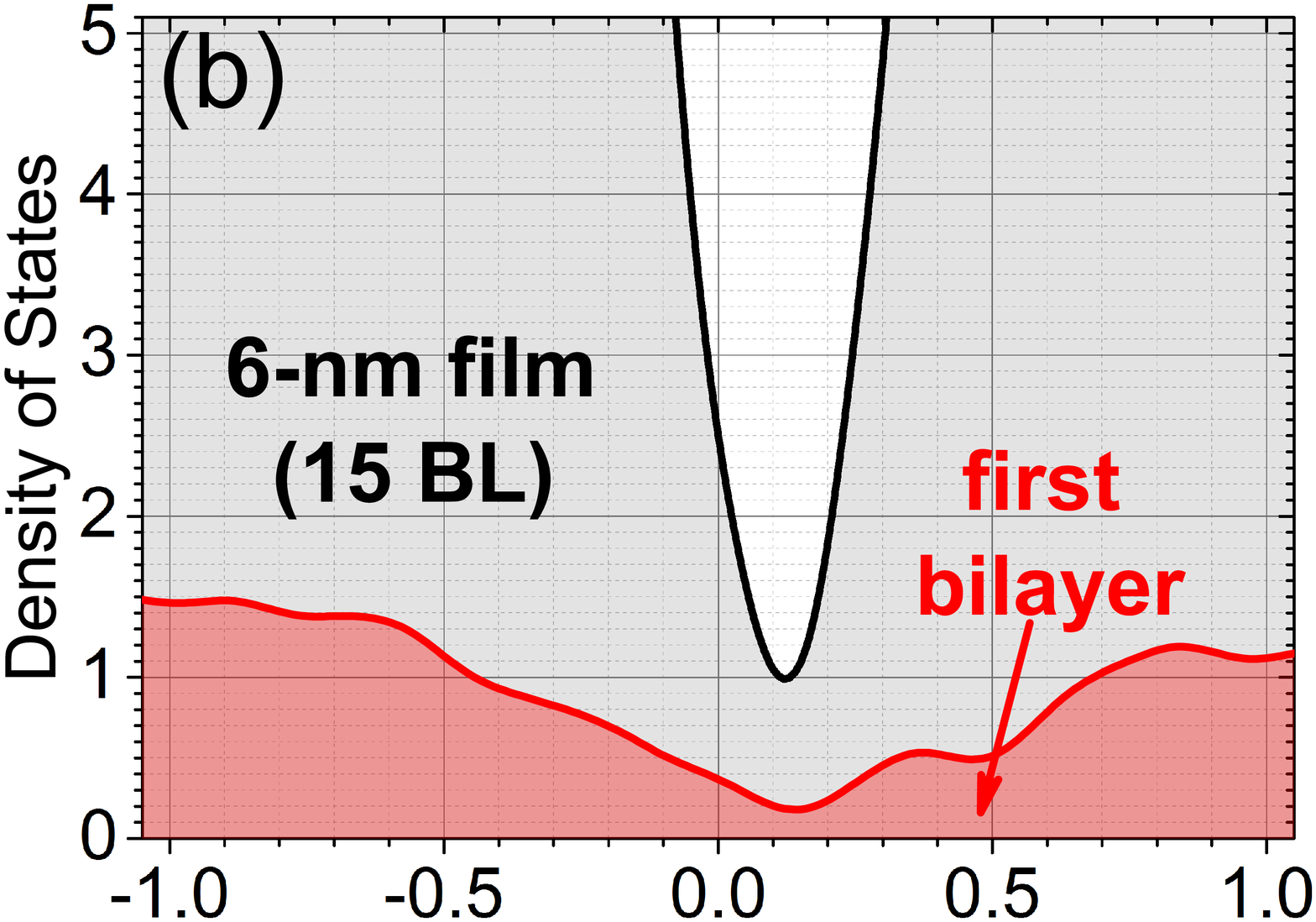}
\includegraphics[trim=0cm 0mm 0cm 0mm, clip=true,width=0.5\textwidth]{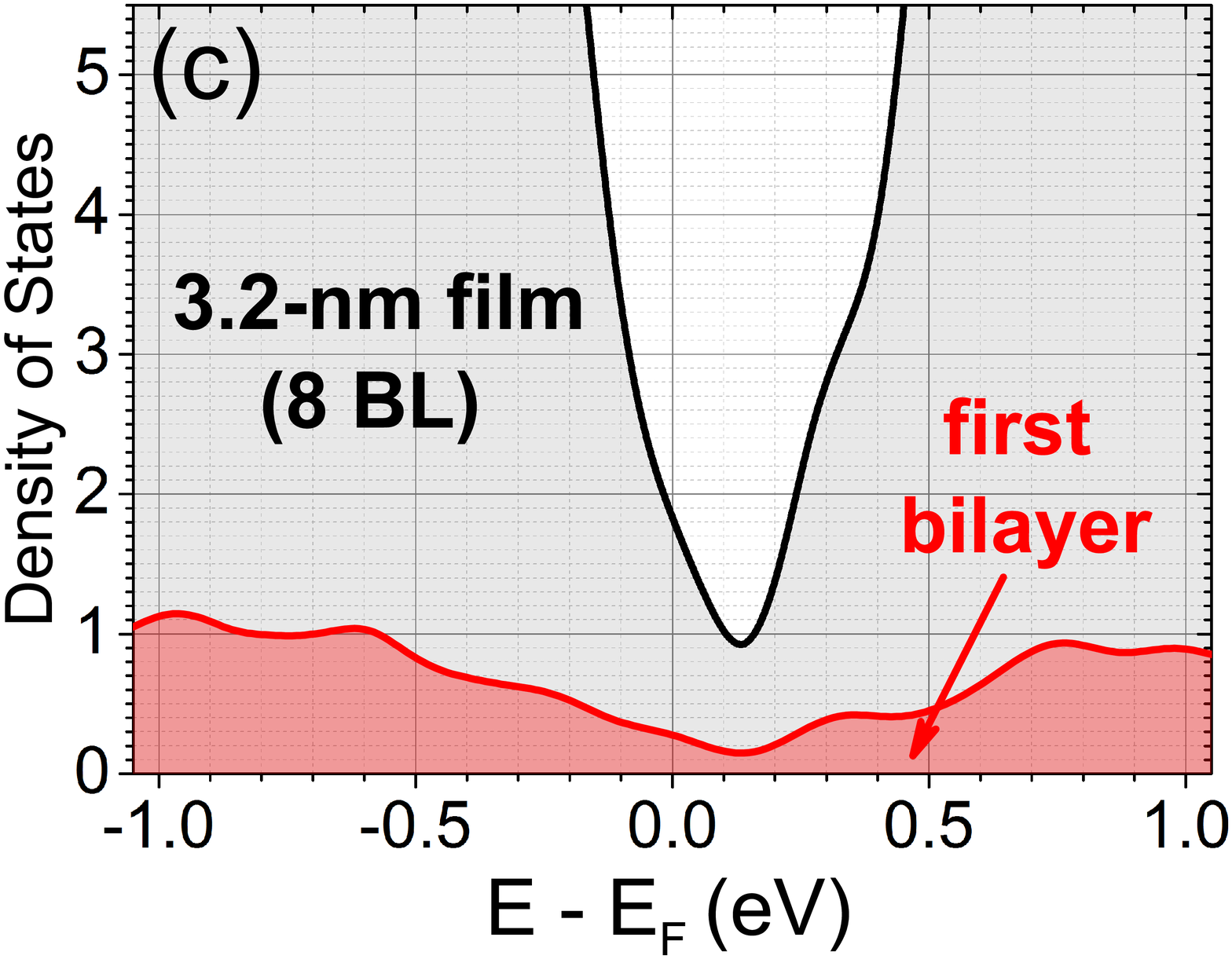}
\caption{\label{fig:dosfilm}
}
\end{figure*}


\end{document}